\documentclass[12pt,a4paper]{iopart}

\usepackage{graphicx}
\usepackage{color}
\usepackage{mathrsfs} 
\usepackage{dsfont}
\usepackage{iopams}  
\usepackage{caption}
\usepackage{tabularx}
\usepackage{multirow}
\usepackage[usenames,dvipsnames,table]{xcolor}
\usepackage{mathdots}

\usepackage[bookmarks=true,
            bookmarksopen=true,colorlinks=false]{hyperref}

\newcommand{\eqref}{\eref}
\newcommand{\dd}{\mathrm{d}}
\newcommand{\ee}{\mathrm{e}}
\newcommand{\ii}{\mathrm{i}}
\newcommand{\R}{\mathds R}

\newcommand{\Scri}{\mathscr{I}}

\renewcommand{\vec}{\boldsymbol}

\newcounter{mnotecount}

\newcommand{\mnotex}[1]
{\protect{\stepcounter{mnotecount}}$^{\mbox{\footnotesize $\bullet$\themnotecount}}$ 
\marginpar{
\raggedright\tiny\em
$\!\!\!\!\!\!\,\bullet$\themnotecount: #1} }

\newcommand{\bit}{\begin{itemize}}
\newcommand{\eit}{\end{itemize}}
\newcommand{\ben}{\begin{enumerate}}
\newcommand{\een}{\end{enumerate}}
\newcommand{\beq}{\begin{equation}}
\newcommand{\eeq}{\end{equation}}
\newcommand{\bea}{\begin{eqnarray}}
\newcommand{\eea}{\end{eqnarray}}
\newcommand{\nn}{\nonumber}

\newcommand{\bwt}{\begin{widetext}}
\newcommand{\ewt}{\end{widetext}}

\begin{document}

\title[]{Fully pseudospectral solution of the conformally invariant wave equation on a Kerr background}

\author{J\"org Hennig$^1$ and Rodrigo Panosso Macedo$^2$}
\address{$^1$Department of Mathematics and Statistics,
           University of Otago,
           PO Box 56, \phantom{$^1$}Dunedin 9054, New Zealand\\
		 $^2$School of Mathematical Sciences, Queen Mary, University of London, 
		   Mile End \phantom{$^2$}Road, London E1 4NS, United Kingdom}
\eads{\mailto{jhennig@maths.otago.ac.nz} and \mailto{r.panossomacedo@qmul.ac.uk}}

\begin{abstract}
We study axisymmetric solution to the conformally invariant wave equation on a Kerr background by means of numerical and analytical methods. Our main focus is on the behaviour of the solutions near spacelike infinity, which is appropriately represented as a cylinder. Earlier studies of the wave equation on a Schwarzschild background have revealed important details about the regularity of the corresponding solutions. It was found that, on the cylinder, the solutions generically develop logarithmic singularities at infinitely many orders. Moreover, these singularities also `spread' to future null infinity. However, by imposing certain regularity conditions on the initial data, the lowest-order singularities can be removed. Here we are interested in a generalisation of these results to a rotating black hole background and study the influence of the rotation rate on the properties of the solutions. To this aim, we first construct a conformal compactification of the Kerr solution which yields a suitable representation of the cylinder at spatial infinity. Besides analytical investigations on the cylinder, we numerically solve the wave equation with a fully pseudospectral method, which allows us to obtain highly accurate numerical solutions. This is crucial for a detailed analysis of the regularity of the solutions. In the Schwarzschild case, the numerical problem could effectively be reduced to solving $(1+1)$-dimensional equations. Here we present a code that can perform the full $2+1$ evolution as required for axisymmetric waves on a Kerr background.
\\[2ex]{}
\end{abstract}


\section{Introduction}

The idealised concept of an \emph{isolated system} is important in many areas of physics, because it allows us to extract properties of a particular physical system without taking the system's entire environment into account. In general relativity, this is particularly useful, for example, for studying the gravitational waves emitted by a well-defined source. Isolated systems in relativity are appropriately modelled as \emph{asymptotically flat spacetimes}, which obey certain fall-off conditions for the metric and curvature at large distances. Alternatively, an elegant geometric framework is provided by Penrose's conformal rescalings \cite{Penrose1963, Penrose1964a, Penrose1964b, Penrose1965}. In this approach, an asymptotically flat spacetime is characterised by existence of a conformal factor such that the rescaled spacetime admits the attachment of a smooth boundary representing events at infinity. Questions about the global behaviour of a spacetime can then be answered by studying local properties at the conformal boundary. For recent overviews of conformal methods, see \cite{Frauendiener2004, Kroon2016, BeyerFrauendienerHennig2020}.

The conformal boundary consists of two null hypersurfaces, namely past and future null infinity $\Scri^-$ and $\Scri^+$. The former emanates from past timelike infinity $i^-$ and focuses at spacelike infinity $i^0$, and the latter emanates from $i^0$ and focuses at future timelike infinity $i^+$. While $i^\pm$ and $i^0$ are regular points in Minkowski spacetime, they are generally singular whenever the ADM mass does not vanish. Clearly, any matter content of a spacetime will initially emerge from $i^-$ and finally end up at $i^+$, for which reason these `points' are singular. The singularity of $i^0$, on the other hand, is due to the gravitational field itself and is present even when the asymptotic region only contains vacuum.

A formulation of Einstein's field equations adapted to the setting of conformal spacetimes and the singular nature of $i^0$ is provided by Friedrich's generalised conformal field equations \cite{Friedrich1998}. In particular, the behaviour of field near spacelike infinity is appropriately resolved by representing the `point' $i^0$ by an entire cylinder $I$ of topology $S^2\times\R$. This cylinder connects $\Scri^-$ and $\Scri^+$ via the `critical sets' $I^-$ and $I^+$. 

When radiation enters the spacetime through $\Scri^-$, then it can influence the information available on $\Scri^+$ by first scattering (either at matter or through self-interaction) and then further propagating to $\Scri^+$. However, if the equations that govern the particular physical process can be extended to null infinity, then it can also happen that information is entirely travelling within the conformal boundary --- from $\Scri^-$ through the cylinder $I$ and further to $\Scri^+$. Clearly, this should be considered to be unphysical, for it does not involve the actual physical spacetime in any form. Hence this phenomenon was referred to as \emph{causality violation at infinity} in \cite{BeyerFrauendienerHennig2020}.

Since, the equations for the field variables and their normal derivatives reduce to intrinsic transport equations on $I$, the cylinder can indeed be thought of as a `bridge' through which information can travel from $\Scri^-$ to $\Scri^+$.  A particularly interesting feature (and mathematical difficulty) is that these transport equations degenerate at the critical sets $I^\pm$. 
If Cauchy initial data for the field equations are provided on a spacelike hypersurface, then they induce data for the intrinsic equations on the intersection $I^0$ of this hypersurface and the cylinder $I$.
The corresponding  solutions on the cylinder are then generally expected to develop singularities at the critical sets.

In order to better understand the subtle role that the cylinder plays in the structure of asymptotically flat spacetimes, it is useful to study the behaviour of fields at and near  spacelike infinity. Rather than considering the full (conformal) field equations, interesting insights are already possible by analysing the properties of solutions to simpler equations on a fixed background spacetime. In this paper, we continue investigations of the \emph{conformally invariant wave equation}, which was studied on a Minkowski background in \cite{FrauendienerHennig2014} and on a Schwarzschild spacetime in \cite{FrauendienerHennig2017, FrauendienerHennig2018}.
As examples for other related problems discussed in the literature, we also refer to numerical solutions of the \emph{spin-2 equation} on Minkowski \cite{BeyerDoulis2012, BeyerDoulis2014, DoulisFrauendiener2013, MacedoValienteKroon2018} and analytical investigations of the \emph{Maxwell equations} on Schwarzschild \cite{ValienteKroon2007, ValienteKroon2008}.

The most important observations from previous studies of the conformally invariant wave equation are the following. In the simplest case of  a Minkowski background, it was found in \cite{FrauendienerHennig2014} that we can impose a certain regularity condition on the data at $I^0$ which achieves that the resulting solutions are globally regular. In the Schwarzschild case \cite{FrauendienerHennig2017, FrauendienerHennig2018}, on the other hand, it turned out that only very simple solutions are globally regular, while general solutions have limited $C^k$ regularity at $I^+$ due to logarithmic singularities. In particular, instead of just a single regularity condition, there is a whole hierarchy of  conditions for the Schwarzschild wave equation. These determine the regularity of function values and higher normal derivatives of the solutions at the cylinder. By imposing sufficiently many of the conditions, arbitrary finite $C^k$ regularity at $I^+$ can be achieved.
The numerical simulations in \cite{FrauendienerHennig2018} also confirmed that, as expected, the singularities are not restricted to the cylinder, but propagate through $I^+$ to future null infinity. This detailed study of the regularity of the solutions was possible thanks to the highly accurate numerical solutions that were obtained with a \emph{fully pseudospectral scheme}, which utilises spectral expansions in space \emph{and} time directions. This approach was first developed in \cite{HennigAnsorg2009}, and various $(1+1)$-dimensional physical applications were later studied in \cite{AnsorgHennig2011} and \cite{Hennig2013}.

How do the properties of solutions to the wave equation change if we replace the spherically-symmetric Schwarzschild background by a rotating Kerr black hole? Is it still possible to enforce arbitrarily regular solutions by imposing suitable conditions on the initial data? Can the propagation of axisymmetric waves be numerically treated with the $(2+1)$-dimensional generalisation of the fully pseudospectral method that was presented in \cite{MacedoAnsorg2014}? These are the central questions that we address in this paper.

When departing from static backgrounds, the first challenge to tackle such questions is to find a suitable coordinate system representing the cylinder at spatial infinity. It is known that all stationary spacetimes admit a regular representation of the set $I$~\cite{AcenaKroon2011}, a result that clearly contains the Kerr solution as a special case. The generic proof starts from an initial value problem, and it relies on a specific conformal gauge and frame --- the so-called conformal Gauss coordinates (for instance, see~\cite{Paetz:2019sjd}). Even though the authors then showed that the essentials of this construction do not depend on such a gauge choice, an explicit coordinate system yielding a regular cylinder at spatial infinity for the Kerr spacetime is not available in the literature. Indeed, when constructing conformal Gauss coordinates at spatial infinity for the Kerr solution, ref.~\cite{Paetz:2018nbd} restricts itself to the leading order of a series expansion about the cylinder.

By relaxing the condition on the use of conformal Gauss coordinates, we first fill this gap and construct a conformal compactification of the Kerr solution in Sec.~\ref{sec:compactification}. Then we derive simple exact solutions to the conformally invariant wave equation in Sec.~\ref{sec:testsolutions}, which can be used for first tests of the numerical method. Afterwards, in Sec.~\ref{sec:BehaviourCylinder}, we study properties on the cylinder. In particular, we observe the coupling of different angular modes and investigate the regularity of the solutions. Sec.~\ref{sec:results} is dedicated to a comprehensive exposition of the numerical methods and results. We start with a short overview of spectral methods in Sec.~\ref{sec:specmeth}, where we discuss some main properties that are relevant for this work. Then we numerically solve the intrinsic equations on the cylinder in Sec.~\ref{sec:firstnumerics}, before we perform the full $2+1$ evolution in Sec.~\ref{sec:2plus1}. Finally, we discuss our results in Sec.~\ref{sec:discussion}.
Furthermore, three appendices provide some details on: (A) a different way to construct our conformal compactification, (B) the Schwarzschild limit of our equations and a comparison with the previous calculations on Schwarzschild, and (C) a summary of the fully pseudospectral method.

\section{Conformal compactification and the conformally invariant wave equation}
\label{sec:compactification}

\subsection{Coordinate system adapted to spacelike infinity}
In order to obtain highly accurate numerical solutions to the conformally invariant wave equation, we first need to construct suitable coordinates in which the solutions are expected to be sufficiently well-behaved. These coordinates should
allow us to obtain a regular conformally rescaled metric, they should blow up spacelike infinity $i^0$ to the cylinder $I$, and contain (a part of) future null infinity $\Scri^+$. This can be achieved as follows.

We start from the physical Kerr metric with mass $m$ and rotation parameter $a$ in Boyer--Lindquist coordinates $(\tilde r,\theta,\varphi,\tilde t)$,
\begin{equation}\fl
\label{eq:Kerr_BL}
 \tilde g = \tilde\Sigma\left(\frac{\dd\tilde r^2}{\tilde\Delta}+\dd\theta^2\right)
             +(\tilde r^2+a^2)\sin^2\theta\,\dd\varphi^2
             -\dd\tilde t^{\,2}+\frac{2m\tilde r}{\tilde\Sigma}(a\sin^2\theta\,\dd\varphi-\dd\tilde t\,)^2,
\end{equation}
where
\begin{equation}
 \tilde\Sigma=\tilde r^2+a^2\cos^2\theta,\quad
 \tilde\Delta=\tilde r^2-2m\tilde r+a^2.
\end{equation}
The function $\tilde\Delta$ can also be expressed as 
$\tilde\Delta=(\tilde r-\tilde r_+)(\tilde r-\tilde r_-)$ in terms of the coordinate positions
\begin{equation}
 \tilde r_\pm=m\pm\sqrt{m^2-a^2}
\end{equation}
of the event and Cauchy horizons.

The desired new coordinates are obtained in several steps. Firstly, we change the angular coordinate from $\varphi$ to $\phi$ via
\begin{equation}\label{eq:phitrans}
 \dd\varphi=\dd\phi+\frac{a}{\tilde\Delta}\,\dd\tilde r,
\end{equation}
which is motivated by the transformation from Boyer--Lindquist to outgoing Kerr coordinates. This leads to the metric
\begin{eqnarray}
 \tilde g &=& \tilde\Sigma\left(\frac{\dd\tilde r^2}{\tilde\Delta}+\dd\theta^2\right)
             +(\tilde r^2+a^2)\sin^2\theta\,\left(\dd\phi+\frac{a}{\tilde\Delta}\,\dd\tilde r\right)^2 - \dd\tilde t^{\,2}\nonumber\\
           && +\frac{2m\tilde r}{\tilde\Sigma}\left[a\sin^2\theta\,\left(\dd\phi+\frac{a}{\tilde\Delta}\,\dd\tilde r\right)-\dd\tilde t\right]^2.
\end{eqnarray}
Note that the relation \eqref{eq:phitrans} can be solved exactly, even though the explicit formula is not required in the following. The result is (after a suitable choice of integration constants)
\begin{equation}
 \varphi=\left\{\begin{array}{ll}
          \displaystyle
          \phi+\frac{a}{2\sqrt{m^2-a^2}}\ln\frac{\tilde r-\tilde r_+}
          {\tilde r-\tilde r_-},&
           |a|<m,\\[2ex]
		 \displaystyle
          \phi-\frac{a}{\tilde r-m},& |a|=m.
         \end{array}\right. 
\end{equation}

Secondly, we compactify\footnote{Following \cite{Macedo2020}, the compactification \eqref{eq:compact_dimensionless} corresponds to the \emph{radial fixing gauge}, in which the limit $|\kappa| \rightarrow 1$ yields the extremal Kerr metric. An alternative option is the \emph{Cauchy horizon fixing gauge} $\tilde r = \tilde r_+ \frac{1-r(1-\kappa^2)}{r}$. Here,   the extremal limit leads to a discontinuous transition into the near horizon geometry of the Kerr spacetime.} 
the radial coordinate $\tilde r$ and, for simplicity, rescale in terms of the length scale provided by the event horizon coordinate $\tilde r_+$ to obtain dimensionless coordinates $r$ and $t$,
\begin{equation}
\label{eq:compact_dimensionless}
 \tilde r = \frac{\tilde r_+}{r},\quad \tilde t=\tilde r_+ t.
\end{equation}
We also introduce the new parameter
\begin{equation}\label{eq:Defkappa}
 \kappa:=\frac{a}{\tilde r_+}\in[-1,1],
\end{equation}
where $\kappa=0$ corresponds to the Schwarzschild solution and $\kappa=\pm1$ to extreme Kerr.
In the following, we take $\kappa$ and $\tilde r_+$ as the fundamental parameters, in terms of which we have
\begin{equation}
 m=\frac{1+\kappa^2}{2} \tilde r_+,\quad
 a=\kappa \tilde r_+,\quad
 r_-=\kappa^2 \tilde r_+.
\end{equation}
Moreover, we replace $\tilde\Sigma$ and $\tilde\Delta$ by dimensionless quantities $\Sigma$ and $\Delta$ as follows,
\begin{equation}\fl
 \tilde\Sigma = \frac{\tilde r_+^2}{r^2}+\kappa^2 \tilde r_+^2\cos^2\theta
              = \frac{\tilde r_+^2}{r^2}\Sigma,\quad
 \tilde\Delta = \frac{\tilde r_+^2}{r^2}-\frac{1+\kappa^2}{r}\tilde r_+^2
                +\kappa^2 \tilde r_+^2
              = \frac{\tilde r_+^2}{r^2}\Delta
\end{equation}
with
\begin{equation}\fl
\label{eq:ConfKerrFunc}
 \Sigma:=1+\kappa^2 r^2\cos^2\theta,\quad 
 \Delta:=1-(1+\kappa^2)r+\kappa^2 r^2
       \equiv (1-r)(1-\kappa^2 r).
\end{equation}
In terms of $\kappa$, $\tilde r_+$, $\Sigma$, $\Delta$ and the new coordinates $(r,\theta,\phi,t)$, the metric takes the form
\begin{eqnarray}
 \tilde g &=& \frac{\tilde r_+^2}{r^2}\left[\Sigma\left(\frac{\dd r^2}
              {r^2\Delta}+\dd\theta^2\right)
             +\left(1+\kappa^2 r^2\right)\sin^2\theta\,\left(\dd\phi-\frac{\kappa}{\Delta}\,\dd r\right)^2 - r^2\dd t^2\right.\nonumber\\
           && \qquad \left.+\frac{(1+\kappa^2) r^3}{\Sigma}
              \left[\kappa \sin^2\theta\,\left(\dd\phi-\frac{\kappa}{\Delta}\,\dd r\right)-\dd t\right]^2\right].
\end{eqnarray}
Hence we can express the physical metric as
\begin{equation}
 \tilde g=\Theta^{-2} g
\end{equation}
with the conformal metric
\begin{eqnarray}
 g &=& \Sigma\left(\frac{\dd r^2}
              {r^2\Delta}+\dd\theta^2\right)
             +\left(1+\kappa^2 r^2\right)\sin^2\theta\,\left(\dd\phi-\frac{\kappa}{\Delta}\,\dd r\right)^2 - r^2\dd t^2\nonumber\\
           && +\frac{(1+\kappa^2) r^3}{\Sigma}
              \left[\kappa \sin^2\theta\,\left(\dd\phi-\frac{\kappa}{\Delta}\,\dd r\right)-\dd t\right]^2
\end{eqnarray}
and the conformal factor
\begin{equation}\label{eq:ConformalFactor}
 \Theta=\frac{r}{\tilde r_+}.
\end{equation}
Note that $\tilde r_+$ only appears in $\Theta$, but not in $g$, i.e.\ it is an insignificant scaling parameter. 

Finally, we compactify the time coordinate and blow up $i^0$ to the cylinder $I$. This can be achieved by introducing coordinates adapted to a family of null curves in the $r$-$t$ plane. In terms of a parameter $\lambda$, such a curve is given by functions $r=r(\lambda)$, $t=t(\lambda)$ with
\begin{equation}
\label{eq:func_t}
 \frac{\dd t}{\dd\lambda}=-\frac{1}{F(r)}\,\frac{\dd r}{\dd\lambda},\quad
 F(r):=\frac{r^2\Delta}{1+\kappa^2 r^2}.
\end{equation}
Remarkably, the function $F$ is independent of $\theta$, which is only the case thanks to the earlier transformation $\varphi\mapsto\phi$, see \eqref{eq:phitrans}. A similar construction based on the original angle $\varphi$ would not work as the $\theta$-dependence of the analogous function $F$ in that case spoils the intended coordinate transformation.

Now we perform the transformation $(r,\theta,\phi,t)\mapsto (\rho,\theta,\phi,\tau)$ with
\begin{equation}
\label{eq:i0_blowup}
 r=\rho(1-\tau),\quad
 t=\int_r^\rho \frac{\dd s}{F(s)}.
\end{equation}
Explicitly, $t$ is given by
\begin{equation}\fl
\label{eq:t_of_tau_rho}
 t=\left\{\begin{array}{ll}
           \displaystyle
           \frac{\tau}{r}-(1+\kappa^2)\ln(1-\tau)+\frac{1+\kappa^2}{1-\kappa^2}\left(\kappa^2\ln\frac{1-\kappa^2\rho}{1-\kappa^2r}-\ln\frac{1-\rho}{1-r}\right), & |\kappa|<1\\[2ex]
           \displaystyle
           \frac{\tau}{r}-2\ln(1-\tau)-2\ln\frac{1-\rho}{1-r}+\frac{2\rho\tau}{(1-\rho)(1-r)}, & |\kappa|=1
          \end{array}\right. ,
\end{equation}
but similarly to the explicit formula for $\varphi$, this is not required in the following.

In the new coordinates $(\rho,\theta,\phi,\tau)$, the cylinder is located at $\rho=0$. Moreover, $\tau=0$ corresponds to the initial hypersurface $\tilde t=t=0$, where we want to prescribe initial data. Finally, (a part of) $\Scri^+$ is located at $\tau=1$, and the critical set $I^+$ is at $\rho=0$, $\tau=1$. Note that all curves of the form $\rho=\mathrm{constant}$, $\theta=\mathrm{constant}$, $\phi=\mathrm{constant}$, $t=\lambda$ are null curves, i.e.\ $\rho$ is a null coordinate. In fact, these curves are null \emph{geodesics} of the conformal metric, because the geodesic equation reduces to $\Gamma^i_{\tau\tau}=0$, and all these Christoffel symbols do indeed vanish.\footnote{Interestingly, if we consider the same curves with respect to the \emph{physical} metric, then we have $\tilde\Gamma^i_{\tau\tau}=0$ for $i=\rho,\theta,\phi$, but $\tilde\Gamma^\tau_{\tau\tau}=\frac{2}{1-\tau}\neq 0$. Hence these curves are only autoparallel curves with respect to $\tilde g$, and a suitable change of the parameter $\lambda$ is required to obtain an affinely parametrised geodesic.}

In terms of the new coordinates $(\rho,\theta,\phi,\tau)$, the conformal metric takes the following form,
\begin{eqnarray}\label{eq:confmetric}
 \fl
 g=\left(\begin{array}{cccc}
   \!\!\frac{1}{F(\rho)}\left(\frac{(1+\kappa^2)r-\Sigma}{\Sigma F(\rho)}r^2+2(1-\tau)\right)
    & 0
    & -(1-\tau+\frac{(1+\kappa^2)r^3}{\Sigma F(\rho)})\kappa\sin^2\theta
    & -\frac{\rho}{F(\rho)}\\
    \!\!0 & \Sigma & 0 & 0\\
    \!\!-(1-\tau+\frac{(1+\kappa^2)r^3}{\Sigma F(\rho)})\kappa\sin^2\theta
    & 0
    & \left(1+\kappa^2 r^2+\frac{(1+\kappa^2)r^3 \kappa^2\sin^2\theta}{\Sigma}\right)\sin^2\theta
    & \kappa\rho\sin^2\theta\\
    \!\!-\frac{\rho}{F(\rho)}
    & 0
    & \kappa\rho\sin^2\theta
    & 0
   \end{array}\right)\nonumber\\
\end{eqnarray}
and the corresponding inverse metric is
\begin{equation}\fl
 g^{-1} = \left(\begin{array}{cccc}
           \kappa^2\frac{F^2(\rho)}{\Sigma}\sin^2\theta
          & 0
          & \kappa\frac{F(\rho)}{\Sigma}
          & g^{\rho\tau}\\    
          0 & \frac{1}{\Sigma} & 0 & 0\\
          \kappa\frac{F(\rho)}{\Sigma}    
          & 0
          & \frac{1}{\Sigma\sin^2\theta}
          & \frac{\kappa}{\rho\Sigma}[(1-\tau)F(\rho)-r^2]\\
          g^{\rho\tau}
          & 0
          & \frac{\kappa}{\rho\Sigma}[(1-\tau)F(\rho)-r^2]
          & g^{\tau\tau}
       \end{array}\right),  
\end{equation}
where the `more involved' components are given by
\begin{eqnarray}
 g^{\rho\tau} &=& -\frac{F(\rho)}{\rho\Sigma}\left[1+\kappa^2 r^2-(1-\tau)\kappa^2 F(\rho)\sin^2\theta\right],\\
 g^{\tau\tau} &=& \frac{1}{\rho^2\Sigma}\left[r^2\Delta-2(1-\tau)(1+\kappa^2r^2)F(\rho)+(1-\tau)^2\kappa^2 F^2(\rho)\sin^2\theta\right].
\end{eqnarray}
We also calculate the determinant of the metric and the Ricci scalar. The results are
\begin{equation}
 \det(g)=-\frac{\rho^2\Sigma^2}{F^2(\rho)}\,\sin^2\theta,\quad
 R = \frac{6r}{\Sigma}[1+\kappa^2 (1-2r)].
\end{equation}

Note that the same coordinates can also be introduced in a more geometric way by constructing a suitable null tetrad basis. Details on this procedure can be found in \ref{sec:null_tetrad}.

\begin{figure}[ht]
\begin{center}
\includegraphics[width=13cm]{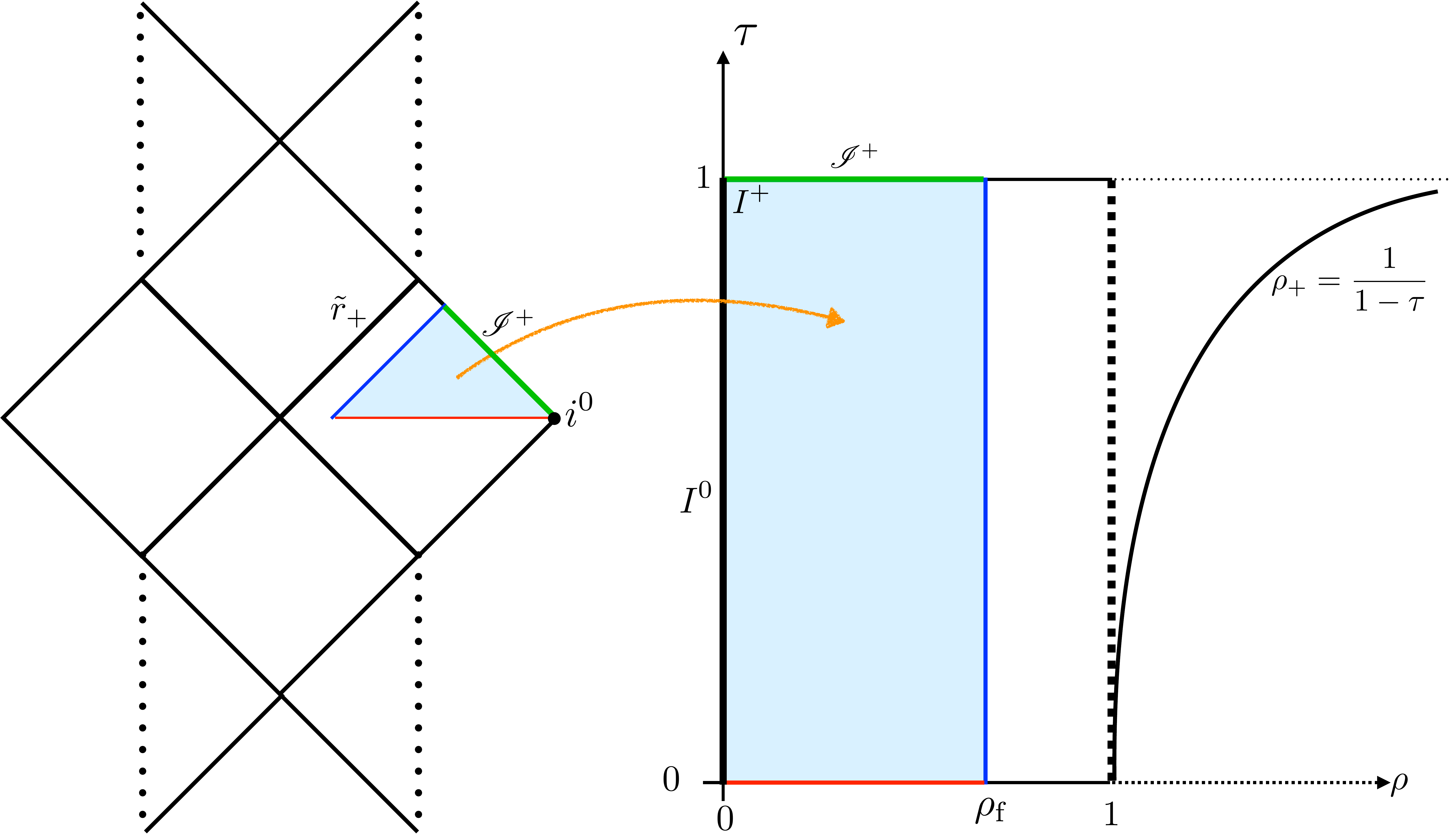}
\end{center}
\caption{{\em Left panel:} Carter--Penrose diagram for the Kerr spacetime (at $\theta=\pi/2$, $\phi=\mathrm{constant}$). Highlighted is the region near spacelike infinity $i^0$, where we construct the new coordinates $\{\tau,\rho,\theta,\phi\}$. {\em Right panel:} In the new coordinates, the cylinder $I^0$ at spacelike infinity is located at $\rho=0$, with the critical set $I^+$ at $\rho=0$, $\tau=1$. Future null infinity $\Scri^+$ is mapped to $\tau =1$. The coordinate transformation \eqref{eq:i0_blowup} and \eqref{eq:t_of_tau_rho} is valid only in the exterior black-hole region. We study the solutions to the conformal wave equation in the domain $(\tau, \rho)\in [0,1]\times[0, \rho_{\rm f}]$ with $\rho_{\rm f}<1$. 
}
\label{fig:Coordinates}
\end{figure}

Finally, notice that the event and inner Cauchy horizons located at $\Delta=0$, i.e.\ at $r_+=1$ and $r_-=\kappa^{-2}$, are parametrised by the curves $\rho_+ = 1/(1-\tau)$ and $\rho_- = \kappa^{-2}/(1-\tau)$, respectively, in the new coordinates $\{\rho, \theta, \phi, \tau \}$. However, we emphasise that our new coordinates only cover a region \emph{outside} the black hole event horizon. Indeed, the coordinate transformation \eqref{eq:i0_blowup} and \eqref{eq:t_of_tau_rho} introduces further coordinate singularities at $F(\rho) = 0 \leftrightarrow \Delta(\rho)=0$ --- cf.~\eqref{eq:func_t} --- where $\det(g) \rightarrow \infty$. In particular, the smallest root of $F(\rho)$ is $\rho = 1 \leq \rho_+ $. Thus, we restrict ourselves to a region $\rho\in [0, \rho_{\rm f}]$, with $\rho_{\rm f} < 1$, making the horizon and the interior of the black hole not accessible in our coordinates, see Fig.~\ref{fig:Coordinates}. Similarly to the considerations for a Schwarzschild background in \cite{FrauendienerHennig2017}, it should be possible to modify the coordinate transformation in order to obtain horizon-penetrating coordinates, which avoid these coordinate singularities. However, since our main interest here is in the behaviour at infinity, this is not an issue for our present discussion.

\subsection{Conformally invariant wave equation}

With the previous preparations, we are now in a position to formulate the wave equation 
\begin{equation}
 0=\Box f-\frac{R}{6} f
  \equiv \frac{1}{\sqrt{-\det(g)}}\left(\sqrt{-\det(g)} g^{ij}f_{,i}\right)_{,j}-\frac{R}{6} f
\end{equation}
in our coordinates. We restrict ourselves to axisymmetric solutions $f=f(\rho,\theta,\tau)$, which are independent of the angle $\phi$. Then the wave equation reads
\begin{eqnarray}\label{eq:CWE}
 \fl
 0 &=&  \kappa^2 \sin^2\theta \left[\rho F(\rho) f_{,\rho}\right]_{,\rho}
       +\frac{\rho}{F(\rho)\sin\theta}\left(\sin\theta f_{,\theta}\right)_{,\theta}
       \nonumber\\
 \fl
   &&  +\frac{1}{\rho F(\rho)}\Big[\left[r^2\Delta-2(1-\tau)
       (1+\kappa^2r^2)F(\rho)+(1-\tau)^2\kappa^2 F^2(\rho)\sin^2\theta\right] 
       f_{,\tau}\Big]_{,\tau}
       \nonumber\\
 \fl
   &&  -\Big[\left[1+\kappa^2 r^2-(1-\tau)\kappa^2 F(\rho)\sin^2\theta\right] f_{,\rho}\Big]_{,\tau}
        \nonumber\\
 \fl
   &&  -\Big[\left[1+\kappa^2 r^2-(1-\tau)\kappa^2 F(\rho)\sin^2\theta\right] f_{,\tau}\Big]_{,\rho}
       -\frac{\rho r}{F(\rho)} [1+\kappa^2 (1-2r)] f.
\end{eqnarray}

An important observation is that $\theta$ is explicitly present not only in the $\theta$-derivative term, but in four other terms as well. These additional appearances only vanish in the Schwarzschild limit $\kappa=0$. As a consequence (unlike in the Schwarzschild case as discussed in \cite{FrauendienerHennig2018}), the angular dependence cannot be separated and we need to consider the full equation in $2+1$ dimensions.

Finally, we observe that the Schwarzschild limit $\kappa\to0$ of the conformal metric \eqref{eq:confmetric} and the wave equation \eqref{eq:CWE} are slightly different from the expressions used in the numerical investigations in \cite{FrauendienerHennig2017, FrauendienerHennig2018}. With our present approach, we interestingly obtain somewhat simpler equations. The relationship between the different coordinate systems is discussed in \ref{sec:Schwarzschild}.

\section{Simple test solutions}\label{sec:testsolutions}

In order to test the pseudospectral code and to get some first insights into the behaviour of solutions, it is useful to construct simple but nontrivial exact solutions to the conformally invariant wave equation. To this end, we consider the wave equation with respect to the \emph{physical} metric \eqref{eq:Kerr_BL}, for which the determinant and Ricci scalar are given by
\begin{equation}
 \det(\tilde g)=-\tilde\Sigma^2\sin^2\theta,\quad
 \tilde R=0.
\end{equation}
We look for solutions $\tilde f=\tilde f(\tilde r,\theta)$ that are independent of the Boyer--Lindquist time coordinate $\tilde t$, but will later obtain a nontrivial time-dependence with respect to $\tau$ due to the transformation 
$\tilde r=\tilde r_+/[\rho(1-\tau)]$, cf.~\eqref{eq:compact_dimensionless} and \eqref{eq:i0_blowup}. The corresponding conformally invariant wave equation is
\begin{equation}\fl
 0    = \frac{1}{\sqrt{-\det(\tilde g)}}\left(\sqrt{-\det(\tilde g)} 
      \tilde g^{ij}\tilde f_{,i}\right)_{,j}-\frac{\tilde R}{6}\tilde f
   = \frac{1}{\tilde\Sigma}(\tilde\Delta\tilde f_{,\tilde r})_{,\tilde r}
     +\frac{1}{\tilde\Sigma\sin\theta}(\sin\theta\tilde f_{,\theta})_{,\theta}.
\end{equation}
The simple structure of this equation allows us to separate the $\theta$-dependence with an expansion in terms of Legendre polynomials, 
\begin{equation}\label{eq:ExactSolLegendre}
 \tilde f(\tilde r,\theta) = \sum_{\ell=0}^\infty \tilde\psi_\ell(\tilde r) P_\ell(\cos\theta).
\end{equation}
Due to the Legendre differential equation
\begin{equation}
 \frac{1}{\sin\theta}[\sin\theta\ P_\ell(\cos\theta)_{,\theta}]_{,\theta}=-\ell(\ell+1)P_\ell(\cos\theta),
\end{equation}
the functions $\tilde\psi_\ell(\tilde r)$ have to satisfy the ODEs
\begin{equation}\label{eq:ODE}
 (\tilde\Delta \tilde\psi_{\ell,\tilde r})_{,\tilde r}-\ell(\ell+1)\tilde\psi_\ell=0,\quad
 \ell=0,1,2,\dots
\end{equation}

Once we have found $\tilde f$, we can obtain the corresponding solution $f$ to the wave equation with respect to $g$ in terms of the new coordinates $(\rho,\theta,\phi,\tau)$ via
\begin{equation}\label{eq:ftransform}
 f(\rho,\theta,\tau)=\Theta^{-1}\tilde f(\tilde r,\theta)
 \equiv \frac{\tilde r_+}{r} \tilde f(\tilde r,\theta)
 \equiv \frac{\tilde r_+}{\rho(1-\tau)}
		\tilde f\left(\frac{\tilde r_+}{\rho(1-\tau)},\theta\right),
\end{equation}
since the wave equation has the conformal weight $-1$ in 4 spacetime dimensions.

Considering sub-extremal black holes first (for which $-1<\kappa<1$), some particular solutions with contribution from just one value of $\ell$ are as follows,
\begin{eqnarray}
 \fl
 \label{eq:exact_sol-l0}
 \ell=0:\quad && \displaystyle f = \frac{P_0(\cos\theta)
						}{r}\ln\frac{1-\kappa^2 r}{1-r},\\
 \fl
 \ell=1:      && \displaystyle f = \frac{P_1(\cos\theta)
					 }{r^2}\left[\left(1-\frac{1+\kappa^2}{2} r\right)\ln\frac{1-\kappa^2 r}{1-r}
              -(1-\kappa^2)r\right],\\
 \fl
 \ell=2:      && f=\frac{P_2(\cos\theta)
		}{r^3}
              \left[\left(1-(1+\kappa^2)r+\frac{1+4\kappa^2+\kappa^4}{6}r^2\right) 
              \ln\frac{1-\kappa^2r}{1-r}\right.\nonumber\\
 \fl       && \qquad \left.-(1-\kappa^2)\left(r-\frac{1+\kappa^2}{2}r^2\right)\right],\\
 \fl
  \label{eq:exact_sol-l3}
 \ell=3:      && f=\frac{P_3(\cos\theta)
		}{r^4}  
              \left[\left(1-\frac32(1+\kappa^2)r
              +\frac35(1+3\kappa^2+\kappa^4)r^2\right.\right.\nonumber\\
           && \qquad\left.\left.
              -\frac{1+9\kappa^2+9\kappa^4+\kappa^6}{20}r^3 
              \right)\ln\frac{1-\kappa^2 r}{1-r}\right.\nonumber\\
           && \qquad\left.
              -(1-\kappa^2)\left(r-(1+\kappa^2)r^2
              +\frac{11+38\kappa^2+11\kappa^4}{60}r^3\right)\right],
\end{eqnarray}
where $r=\rho(1-\tau)$ as before.
Note that the integration constants for all examples have been chosen such that $f$ is regular at $r=0$, and such that the constant term in the polynomial in front of the logarithm is equal to one. The solution for arbitrary $\ell$ can be given in terms of hypergeometric functions.

In the case of extremal black holes ($\kappa=\pm1$), particular solutions (with suitably chosen integration constants) are
\begin{equation}
\label{eq:ExactSolution_extrem}
 f = \frac{r^\ell}{(1-r)^{\ell+1}} P_\ell(\cos\theta), \quad \ell=0,1,2,\dots
\end{equation}

How do the above solutions behave  in the limit $r\rightarrow 0$, i.e.\ near the cylider ($\rho=0$) or near future null infinity ($\tau=1$)?
 In the extremal case, one can easily read off from \eqref{eq:ExactSolution_extrem} that we have 
\begin{equation}\label{eq:ExactSolLimit}
 f\sim r^\ell P_\ell(\cos\theta)\quad\mbox{as}\quad r\to0.
\end{equation}
In the sub-extremal case, an expansion about $r=0$ reveals that the solutions \eqref{eq:exact_sol-l0}-\eqref{eq:exact_sol-l3} do have exactly the same asymptotic behaviour \eqref{eq:ExactSolLimit}. Moreover, we can verify that this is not limited to the  above four explicit examples, but also holds in general for \emph{all} solutions that do only depend on $r$ (or $\tilde r$) and are regular at $i^0$. This follows by only considering the leading terms in \eqref{eq:ODE} for large $\tilde r$, which gives
\begin{equation}
 \tilde r^2\tilde\psi_{\ell,\tilde r\tilde r}+2\tilde r\tilde\psi_{\ell,\tilde r}-\ell(\ell+1)\tilde\psi_\ell=0. 
\end{equation}
This equation has the bounded solutions $\tilde\psi_\ell\sim\tilde r^{-\ell-1}\sim r^{\ell+1}$. With \eqref{eq:ftransform} and \eqref{eq:ExactSolLegendre} this indeed confirms that, for any $\ell$, the corresponding mode has the asymptotic behaviour \eqref{eq:ExactSolLimit}.

Note that the sub-extremal solutions \eqref{eq:exact_sol-l0}-\eqref{eq:exact_sol-l3} vanish in the extremal limit $|\kappa| \rightarrow 1$. 
Nevertheless, if we appropriately rescale the solutions first, then we can recover the appropriate extremal solution in \eqref{eq:ExactSolution_extrem}. One possibility is a rescaling via $f\mapsto f/(1-\kappa^2)^{2\ell+1}$. The extremal limit does then produce solutions proportional to \eqref{eq:ExactSolution_extrem}.
Another possibility, which is even more convenient for our numerical experiments, is to rescale in such a way that $f = 1$ holds at a fixed point $(\rho^*, \theta^*, \tau^*)$ regardless of $\kappa$. This can be achieved with the transformation  $f\mapsto f/f^*$, $f^* := f(\rho^*, \theta^*, \tau^*)$, which also yields a well-defined extremal limit.

\section{General behaviour near the cylinder}
\label{sec:BehaviourCylinder}
 
We leave the realm of the simple exact solutions from the previous section and now discuss the behaviour of general solutions near spacelike infinity.
In particular, we identify a coupling of the angular modes, which is caused by the spin parameter $\kappa$ in the Kerr solution. We also derive conditions on the initial data that ensure a certain degree of regularity at $I^+$ ($\rho=0$, $\tau=1$).
  
For that purpose, we assume that $f$ near $\rho=0$ has an expansion in $\rho$ of the form
\begin{equation}\label{eq:expansion}
 f(\rho,\theta,\tau)=f_0(\theta,\tau)+\rho f_1(\theta,\tau)+\rho^2 f_2(\theta,\tau)+\dots
\end{equation}
Plugging this into the conformally invariant wave equation \eqref{eq:CWE}, we obtain an equation for $f_n$,
\begin{equation}\label{eq:cylinderEQ}
\fl
 (1-\tau^2) f_{n,\tau\tau}+2(n-\tau)f_{n,\tau}
 -\frac{1}{\sin\theta}(\sin\theta f_{n,\theta})_{,\theta}=R_n, \quad n=0,1,2,\dots
\end{equation}
The inhomogeneity $R_n$ can be given explicitly, but is a rather lengthy expression. It  depends linearly on up to eight previous functions $f_{n-1}$, $f_{n-2}$, $\dots$, $f_{n-8}$ and their derivatives.\footnote{Note that the corresponding equation for the Schwarzschild wave equation has the same left-hand side, but $R_n$ is different and, in particular, ``only'' requires up to six previous functions \cite{FrauendienerHennig2018}. Hence the rotation of the black hole couples even more of the functions $f_n$.}

 \subsection{Angular mode coupling}\label{sec:ModeCoupling}
An interesting feature of the wave equation on a rotating Kerr background is the coupling of different angular modes, which is not present in the Schwarzschild case discussed in~\cite{FrauendienerHennig2018}. This can be observed as follows.

The left-hand side of \eqref{eq:cylinderEQ} has the property that, if $f_n$ is proportional to $P_\ell(\cos\theta)$ for some $\ell$, then the entire left-hand side is again proportional to $P_\ell(\cos\theta)$. However, this is different for the right-hand side. The inhomogeneity $R_n$ can be written as 
\begin{equation}\label{eq:Rdecomposition}
 R_n=\tilde R_n+\cos^2\theta\,\hat R_n,
\end{equation}
where $\hat R_n$ vanishes for $n=0$ and $n=1$, but is present for $n\ge 2$ if $\kappa\neq 0$. The expressions $\tilde R_n$ and $\hat R_n$ also have the property that, if all relevant functions $f_{n-1}$, $f_{n-2}$, etc.\ are proportional to $P_\ell(\cos\theta)$, then they are proportional to $P_\ell$ as well. The same does not apply to $R_n$. Due to the $\cos^2\theta$ multiplier, $R_n$ generally also contains terms proportional to $P_{\ell-2}(\cos\theta)$ and $P_{\ell+2}(\cos\theta)$. Hence the different $\ell$-modes are coupled.
 
In order to see this more explicitly, we expand $f_n$ and $R_n$ in a Legendre polynomial basis via
\bea\label{eq:Legendre}
 f_n(\theta,\tau)&=\sum_{\ell=0}^\infty \psi_{n\ell}(\tau)P_\ell(\cos\theta) &\quad\longrightarrow\quad \psi_{n\ell} = \frac{2\ell+1}{2} \langle f_n, P_\ell \rangle,\\
 R_n(\theta,\tau)&=\sum_{\ell=0}^\infty R_{n\ell}(\tau)P_\ell(\cos\theta)  
 &\quad\longrightarrow\quad R_{n\ell} = \frac{2\ell+1}{2} \langle R_n, P_\ell \rangle,
\eea
where the scalar product for functions of $x=\cos\theta$ is defined by 
\begin{equation}
 \langle f,g\rangle=\int_{-1}^1 f(x)g(x)\,\dd x=\int_0^\pi f(\cos\theta)g(\cos\theta)\sin\theta\,\dd\theta. 
\end{equation}
In particular, we have $\langle P_{\ell'}, P_\ell\rangle = \frac{2}{2\ell+1}\delta_{\ell'\ell}$.
Then we obtain the following equation for $\psi_{n\ell}$ from \eqref{eq:cylinderEQ},
\begin{equation}
\label{eq:Eq_Cylider_mode_nl}
 (1-\tau^2)\ddot\psi_{n\ell} + 2(n-\tau)\dot\psi_{n\ell} + \ell(\ell+1)\psi_{n\ell}=R_{n\ell},
\end{equation}
where a dot denotes differentiation with respect to $\tau$. 
By expressing the factor $\cos^2\theta$ in \eqref{eq:Rdecomposition} in terms of Legendre polynomials, $\cos^2\theta=(P_0+2P_2)/3$, we obtain the projection 
$\langle \cos^2\theta\ P_{\ell'} , P_\ell  \rangle = (\langle P_{\ell'},P_\ell \rangle + 2 \langle P_2 P_{\ell'} ,P_\ell \rangle)/3$. This allows us to reformulate the right-hand side of \eqref{eq:Eq_Cylider_mode_nl} as\footnote{Applying formulae for the integration over three Legendre polynomials from Appendix D of~\cite{Alcubierre2009}, we have 
$$
\langle P_2 P_{\ell'} ,P_\ell \rangle = \frac{2}{2\ell+1}\Bigg(\frac{3\ell(\ell-1)}{2(2\ell-3)(2\ell-1)} \delta_{\ell'\ell-2} + \frac{\ell(\ell+1)}{(2\ell-1)(2\ell+3)}\delta_{\ell'\ell}+ \frac{3(\ell+1)(\ell+2)}{2(2\ell+3)(2\ell+5)}\delta_{\ell' \ell+2}\Bigg).
$$ 
} 
\bea
\label{eq:ModeCouple_rhs}
R_{n\ell} &=& \tilde{R}_{n\ell}  +  \frac{2\ell(\ell+1) -1}{(2\ell-1)(2\ell+1)} \hat{R}_{n\ell}  \nn \\
&+& \frac{\ell(\ell-1)}{(2\ell-3)(2\ell-1)}\hat{R}_{n\ell-2} + \frac{(\ell+1)(\ell+2)}{(2\ell+3)(2\ell+5)} \hat{R}_{n\ell+2}
\eea

From \eqref{eq:ModeCouple_rhs} and the similar expressions for $R_{n\ell\pm2}$ we see that, if any of the relevant previous functions $f_{n-1}$, $f_{n-2}$, $\dots$ contains an angular mode $\ell$, then we find source terms on the right-hand side of \eqref{eq:Eq_Cylider_mode_nl} not only for $\ell$, but for $\ell\pm2$ as well. Consequently, $f_n$ will also contain the modes $\psi_{n\ell\pm2}$ in addition to $\psi_{n\ell}$.

\begin{figure}[ht]
\begin{center}
\includegraphics[width=13cm]{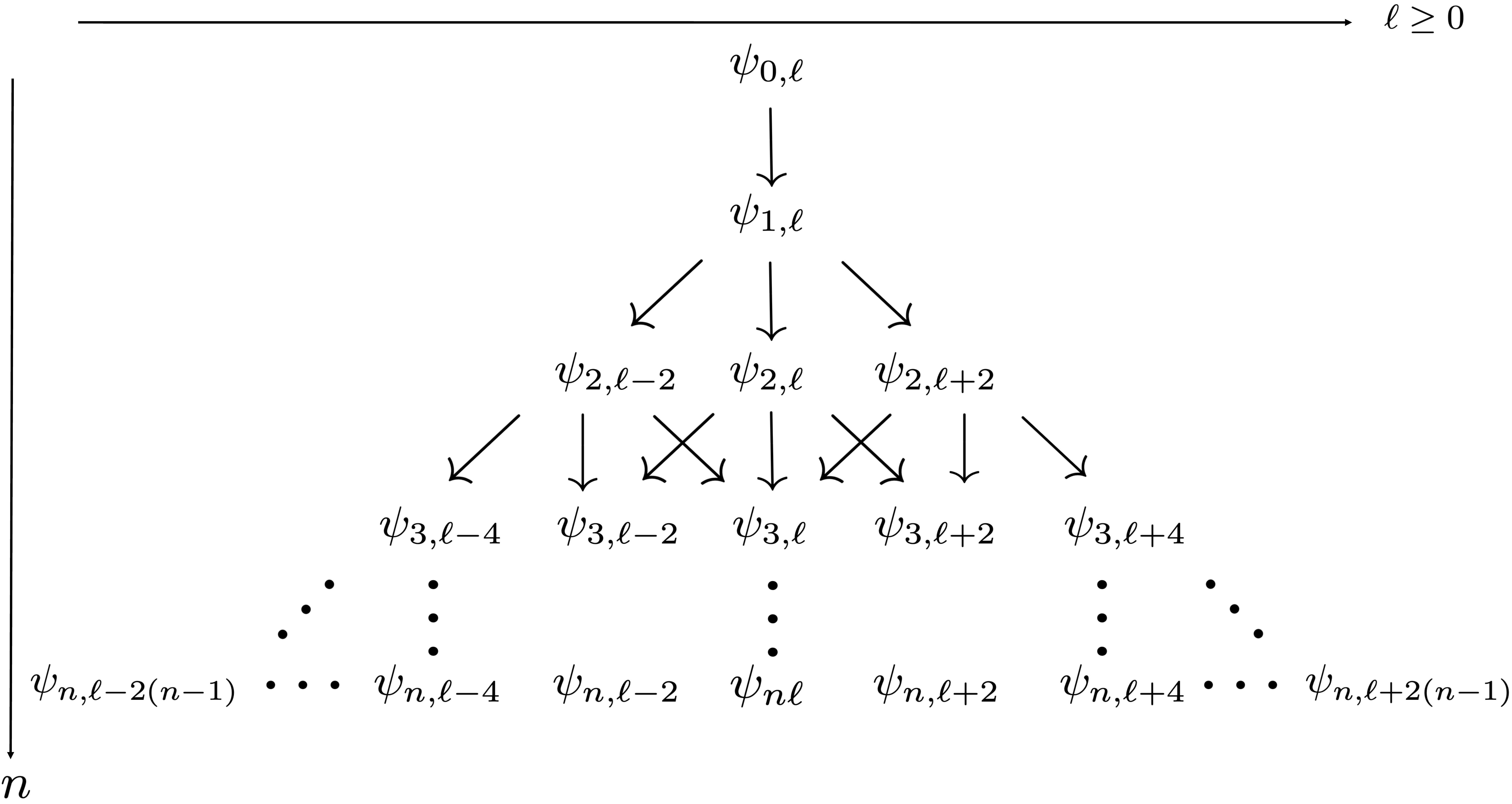}
\end{center}
\caption{Schematic representation of the angular mode coupling at spatial infinity. (Of course, only functions with nonnegative $\ell$-index are practically relevant.)
}
\label{fig:ModeCouplingSpatialInft}
\end{figure}

Therefore, any mode $\psi_{0\ell}$ in the lowest-order function $f_0$ will successively excite more and more modes in higher-order functions $f_n$ as illustrated in 
Fig.~\ref{fig:ModeCouplingSpatialInft}.

\subsection{Regularity conditions}\label{sec:Regularity}
Since Eq.~\eqref{eq:Eq_Cylider_mode_nl}  has exactly the same form as the corresponding equation in the Schwarzschild case (see Eq.~(9) in \cite{FrauendienerHennig2018}), we also arrive at the same conclusion: Generally, the functions $\psi_{n\ell}$ are regular for $0\le\tau<1$, but they contain terms of the form $(1-\tau)^n\ln(1-\tau)$ and hence have logarithmic singularities at $\tau= 1$. However, for a suitable choice of the integration constants, those terms disappear. This corresponds to an appropriate fine-tuning of the initial data and their spatial derivatives at $\rho=0$, $\tau=0$, i.e.\ at $I^0$, where the initial slice $\tau=0$ approaches the cylinder.

With the explicit form of the inhomogeneity $R_{n\ell}$ in \eqref{eq:Eq_Cylider_mode_nl}, we can derive the corresponding regularity conditions for the first few values of $n$ and $\ell$.

\subsubsection{Conditions for regularity of $f_0$.}

For $n=0$, the inhomogeneity vanishes, i.e.\ we have $R_0=0$ and hence $R_{0\ell}=0$. Therefore, \eqref{eq:Eq_Cylider_mode_nl} becomes
\begin{equation}
 \frac{\dd}{\dd\tau}[(1-\tau^2)\dot\psi_{0\ell}]+\ell(\ell+1)\psi_{0\ell}=0.
\end{equation}
The corresponding solutions are simply given in terms of Legendre polynomials $P_\ell$ and Legendre functions of the second kind $Q_\ell$,
\begin{equation}
\label{eq:Sol_n0}
 \psi_{0\ell}(\tau) = c_\ell P_\ell(\tau)+d_\ell Q_\ell(\tau)
\end{equation}
with integration constants $c_\ell$ and $d_\ell$. Since $Q_\ell(\tau)\sim\ln(1-\tau)$ near $\tau=1$, these solutions are generally singular. However, we can enforce regularity by restricting the initial data. For that purpose, since $P_\ell(0)=0$ for odd $\ell$ and $\dot P_\ell(0)=0$ for even $\ell$, we can impose the conditions 
\begin{equation}\label{eq:cond0}
 \left\{
 \begin{array}{ll}
  \psi_{0\ell}(0)=0,    & \mbox{for odd $\ell$},\\
  \dot\psi_{0\ell}(0)=0,& \mbox{for even $\ell$}.
 \end{array}\right.
\end{equation}
This ensures that $d_\ell=0$ so that the singular terms vanish. 
Note that, if we decide to impose these conditions for \emph{all} $\ell$, this corresponds to choosing initial data for which $f(0,\theta,0)$ is an even function of $\cos\theta$, and $\dot f(0,\theta,0)$ is an odd function of $\cos\theta$, cf.~\eqref{eq:expansion} and \eqref{eq:Legendre}.

As a practical application, we consider here and in the following subsections a particular family of initial data. We assume that the initial function values and $\tau$-derivatives at $\tau=0$ contain only angular contributions corresponding to $\ell=0,1,\dots, 4$, but no modes with $\ell>4$. Hence we choose data of the form
\begin{eqnarray}\label{eq:data1}
 f(\rho,\theta,0) &=& g_0(\rho)P_0(\cos\theta)+g_1(\rho)P_1(\cos\theta)
                 +\dots+g_4(\rho)P_4(\cos\theta),\\             
 \label{eq:data2}
 \frac{\partial f}{\partial\tau}(\rho,\theta,0) 
             &=& h_0(\rho)P_0(\cos\theta)+h_1(\rho)P_1(\cos\theta)
                 +\dots+h_4(\rho)P_4(\cos\theta),
\end{eqnarray}
which depend on the ten free functions $g_\ell(\rho)$, $h_\ell(\rho)$, $\ell=0,1,\dots, 4$. 
 
Comparing these data with the expansion \eqref{eq:expansion} in the radial direction, the initial data for $f_n(\theta, \tau)$, $n=0,1,2,\dots$, as solutions to \eqref{eq:cylinderEQ} take the form
\begin{equation}
\fl
 f_n(\theta,0)=\frac{1}{n!}\sum_{\ell=0}^4 g_\ell^{(n)}(0)P_\ell(\cos\theta),\quad
 \dot f_n(\theta,0)=\frac{1}{n!}\sum_{\ell=0}^4 h_\ell^{(n)}(0)P_\ell(\cos\theta),
\end{equation}
where $g^{(n)}_\ell(0) = \frac{\dd^n\, g_\ell}{\dd\rho^n}\big|_{\rho=0}$ and $h^{(n)}_\ell(0) = \frac{\dd^n\, h_\ell}{\dd\rho^n}\big|_{\rho=0}$. 
The corresponding initial data for the modes $\psi_{n\ell}(\tau)$ as solutions to \eqref{eq:Eq_Cylider_mode_nl} are then given by  
\begin{equation}\label{eq:psi-g-h-relations}
\fl
  \psi_{n\ell}(0)=
   \left\{\begin{array}{ll}
		   \frac{g_\ell^{(n)}(0)}{n!}, & \ell=0,\dots,4,\\
           0, & \ell>4,
          \end{array}\right.\qquad
  \dot\psi_{n\ell}(0)=
   \left\{\begin{array}{ll}
		   \frac{h_\ell^{(n)}(0)}{n!}, & \ell=0,\dots,4,\\
           0, & \ell>4.
          \end{array}\right.         
\end{equation}

Applying the conditions \eqref{eq:cond0}, we obtain the following requirements for regularity of $\psi_{0\ell}(\tau)$ for $\ell=0,1,\dots,4$,
\begin{eqnarray}
\label{eq:RegCond_n0l0}
 \ell=0:\quad & h_0(0)=0\quad&\Leftrightarrow\quad \dot\psi_{00}(0)=0,\\
 \ell=1:\quad & g_1(0)=0\quad&\Leftrightarrow\quad \psi_{01}(0)=0,\\
 \ell=2:\quad & h_2(0)=0\quad&\Leftrightarrow\quad \dot\psi_{02}(0)=0,\\
 \label{eq:RegCond_n0l3}
 \ell=3:\quad & g_3(0)=0\quad&\Leftrightarrow\quad \psi_{03}(0)=0,\\
 \ell=4:\quad & h_4(0)=0\quad&\Leftrightarrow\quad \dot\psi_{04}(0)=0.
\end{eqnarray}

Higher modes $\psi_{0\ell}(\tau)$ with $\ell>4$ are identically equal to zero --- they vanish initially due to our choice of data, and they still vanish for $\tau>0$. Note that the same does not generally hold for $\psi_{n\ell}(\tau)$ with larger values of $n$, where modes can appear during the time evolution that are not present in the initial data. This will be demonstrated in subsection \ref{sec:f2cond} below.

\subsubsection{Conditions for regularity of $f_1$.}

For $n=1$, the inhomogeneity $R_1$ is nonzero, but does not yet contain the term that introduces the mode coupling mentioned above. 
 
 Considering the same family \eqref{eq:data1}, \eqref{eq:data2} of initial data introduced in the previous subsection, we obtain the following regularity conditions (assuming the conditions for $n=0$ are already satisfied),
\begin{eqnarray}
\label{eq:RegCond_n1l0}
 \ell=0:\quad && \mbox{NA ($\psi_{10}$ is always regular)},\\
 \ell=1:\quad && g_{1,\rho}(0)+h_{1,\rho}(0)+2(1+\kappa^2)h_1(0)=0,\\
 \ell=2:\quad && g_{2,\rho}(0)+\frac92 (1+\kappa^2)g_{2}(0)=0,\\
 \ell=3:\quad && g_{3,\rho}(0)+h_{3,\rho}(0)+\frac{16}{3}(1+\kappa^2)h_{3}(0)=0,\\
 \ell=4:\quad && g_{4,\rho}(0)+\frac{137}{18} (1+\kappa^2)g_{4}(0)=0.
\end{eqnarray}
For initial data subject to these conditions, the terms $\sim (1-\tau)\ln(1-\tau)$ on the cylinder are eliminated.
Higher modes $\psi_{1\ell}(\tau)$ with $\ell>4$ are again identically equal to zero.

Similarly to the previous subsection, an equivalent formulations of the above regularity conditions in terms of $\psi_{n\ell}(0)$ and $\dot\psi_{n\ell}(0)$ can easily be obtained by virtue of \eqref{eq:psi-g-h-relations}.
 
\subsubsection{Conditions for regularity of $f_2$.}\label{sec:f2cond}

Next we consider conditions for regularity of $\psi_{2\ell}(\tau)$ for our family of initial data. Again assuming the lower-order conditions are satisfied (i.e.\ those for $n=0$ and $n=1$), we obtain,
\begin{eqnarray}
 \fl
 \ell=0:\quad && g_{0,\rho}(0)+h_{0,\rho}(0)=0,\\
 \fl
 \label{eq:RegCond_n2l1}
 \ell=1:\quad && \mbox{NA ($\psi_{21}$ is always regular)},\\
 \fl
  \label{eq:RegCond_n2l2}
 \ell=2:\quad && \textstyle g_{2,\rho\rho}(0)+\frac12 h_{2,\rho\rho}
              +(1+\kappa^2)\left(\frac{2827}{1890} g_{2,\rho}(0)+\frac{17}{6} h_{2,\rho}(0)\right)=0,\\
 \fl
  \label{eq:RegCond_n2l3}
 \ell=3:\quad && \textstyle g_{3,\rho\rho}(0)+\frac{1+\kappa^2}{315}[4159 g_{3,\rho}(0)
               +169 h_{3,\rho}(0)]=0,\\
 \fl
 \label{eq:RegCond_n2l4}
 \ell=4:\quad && \textstyle g_{4,\rho\rho}(0)+\frac12 h_{4,\rho\rho}(0)
              +(1+\kappa^2)\left(\frac{173706}{52745}g_{4,\rho}(0)+\frac{34}{5}h_{4,\rho}(0)\right)=0.
\end{eqnarray}
Interestingly, the next mode $\psi_{25}(\tau)$ is not identically equal to zero. Instead, it is given by
\begin{equation}
\label{eq:psi_25}
\psi_{25}(\tau) = \frac{100}{189} h_3(0)\kappa^2\tau^3(1-\tau)^2. 
\end{equation}
Obviously, this function, together with its $\tau$-derivative, vanishes at $\tau=0$, but it is nonzero for $0<\tau<1$ (provided the $\ell=3$ mode in the initial data satisfies $h_3(0)\neq 0$ corresponding to $\dot\psi_{03}(0)\neq 0$). This is an example of a mode that is not present in the initial data, yet it still appears in the solution. It also demonstrates the coupling between modes, since the $\ell=3$ mode in the data [in form of the value $h_3(0)$ or $\dot\psi_{03}(0)$] determines the amplitude of $\psi_{25}(\tau)$, which is a contribution to the $\ell=5$ mode in the solution near the cylinder.

Another example is the $\ell=6$ mode, which is
\begin{equation}
\label{eq:psi_26}
\psi_{26}(\tau) = \frac{100}{33} g_4(0)\kappa^2\tau^2(1-\tau)^2(1-3\tau^2) 
\end{equation}
and depends on the $\ell=4$ mode in the initial data. Note that the mode coupling disappears in the Schwarzschild case $\kappa=0$, where we have $\psi_{25}(\tau)=\psi_{26}(\tau)\equiv 0$.

\subsubsection{Conditions for regularity of $f_3$.}

Finally, we consider regularity conditions for the two lowest modes in $f_3$. We obtain
\begin{eqnarray}
 \fl
 \label{eq:RegCond_n3l0}
 \ell=0:\quad && 12g_{0,\rho\rho}(0)+2(1+\kappa^2)g_{0,\rho}(0)
             +\frac{48}{15}\kappa^2g_{2}(0)-(9\kappa^4+10\kappa^2+9)g_{0}(0)=0,\\
 \fl
  \label{eq:RegCond_n3l1}
 \ell=1:\quad && (5\kappa^4+13\kappa^2+5)h_1(0)+\frac67\kappa^2h_3(0)
             +\frac52\left[2g_{1,\rho\rho}(0)+h_{1,\rho\rho}(0)\right]=0. 
\end{eqnarray}
These conditions are the first two examples where, for $\kappa\neq 0$, regularity of one mode (here $\ell=0$ or $\ell=1$) also requires information about a different mode in the initial data [here $\ell=2$ or $\ell=3$, respectively, in form of the values $g_2(0)$ or $\psi_{02}(0)$, and $h_3(0)$  or $\dot\psi_{03}(0)$]. For $\kappa=0$, on the other hand, those terms in the previous two equations vanish, i.e.\ the mode coupling disappears.

As a simple test for the above regularity conditions, we can easily verify that the everywhere regular exact solutions from Sec.~\ref{sec:testsolutions} do, of course, satisfy all of these conditions. Moreover, the exact solutions are special cases in which only a single angular mode $\ell$ is present. Hence all terms corresponding to mode coupling vanish. For example, the right-hand side of \eqref{eq:psi_25} is zero due to $h_3(0)=0$ for the $\ell=3$ solution in \eqref{eq:exact_sol-l3}. 

\section{Numerical Results}
\label{sec:results}
In this section, we numerically solve the conformally invariant wave equation near the cylinder on the Kerr background. Thereby, a particular technical challenge for the time integrator is resulting from the singular structure of the critical set. A robust approach to deal with this problem are fully pseudospectral codes~\cite{HennigAnsorg2009,Hennig2013,MacedoAnsorg2014,FrauendienerHennig2014,FrauendienerHennig2017,MacedoValienteKroon2018}, in which spectral methods are employed in both spatial \emph{and} time directions. Besides providing highly-accurate numerical solutions, this method also allows us to infer the underlying regularity structure of the solutions. Thus, it can be exploited to confirm and expand beyond the analytical studies. The next section briefly summarises a few necessary concepts in the theory of spectral methods underpinning the upcoming considerations. Refs.~\cite{Boyd,Canuto} are standard textbooks on the topic. A deeper discussion with a focus on problems in general relativity can be found in \cite{Grandclement:2007sb}. Further details on our  numerical algorithm are available in \ref{sec:SpecCode_2+1}.
\subsection{Spectral methods}
\label{sec:specmeth}
In order to introduce the basic idea of spectral methods, we focus here on a one-dimensional problem. Consider approximating a real valued function $f(\xi)$, defined on the interval $[-1,1]$. The function $f$ can be exactly expressed as
\beq
\label{eq:f_specmeth}
f(\xi) = \sum_{i=0}^{\infty} c_i T_i(\xi)
\eeq 
in terms of a basis given by the Chebyshev polynomials of the first kind, 
\begin{equation}
 T_i(\xi)=\cos(i\, \arccos \xi)
\end{equation}
with suitable Chebyshev coffiecients $c_i$.
We now employ the collocation methods to determine finitely many $c_i$ for an approximation $f_N$ of $f$. For a given truncation order $N$, we reformulate \eqref{eq:f_specmeth} as
\beq
f(\xi) = f_N(\xi) + R_N(\xi),\quad f_N(\xi)=\sum_{i=0}^N  c_i T_i(\xi),
\eeq
where the residual $R_N(\xi)$ encodes the error when expressing the function $f(\xi)$ in the Chebyshev basis up to order $N$ only. Then, given a discrete set of grid points $\{ \xi_k \}_{k=0}^N$, the coefficients $c_i$ are obtained by imposing the conditions $R_N(\xi_k)=0$, i.e.\ the residual vanishes at the grid points and the function is exactly represented there by the discrete basis. With the collocation method, the Chebyshev coefficients follow from inverting the equations
\beq
f(\xi_k) = f_N(\xi_k), \quad k=0,\dots,N.
\eeq
The choice of the grid points $\{ \xi_k \}_{k=0}^N$ may depend on particular properties of the problem, but the most common ones are the Chebyshev-Gauss grid $\xi_k = \cos(\pi \frac{k+1/2}{N+1})$ and the Chebyshev-Lobatto grid $\xi_k = \cos(\pi \frac{k}{N})$. These are based, respectively, on the zeros and the extremes of the function $T_N(\xi)$.

Of particular importance is the asymptotic behaviour of the Chebyshev coefficients, since they determine the rate at which the residual $||R_N(\xi)||_\infty$ decays to zero as $N\rightarrow \infty$. It can be shown (see  \cite{Boyd,Canuto,Grandclement:2007sb} and references therein) that we have the following properties.
\bit
\item Analytic functions: If $f(\xi)$ is of class $C^\omega([-1,1])$, then
\beq
c_i \sim {\cal C}^{-i} \Longrightarrow \|R_N\|_\infty\sim \bar{\cal{C}}^{-N},
\eeq
with some constants $\cal C$ and $\bar{\cal C}$.
\item Differentiable functions: If $f(\xi)$ is of class $C^k([-1,1])$, then
\beq
c_i \sim i^{-\varkappa} \Longrightarrow \|R_N\|_\infty \sim N^{-\varkappa+1},
\eeq
where the constant $\varkappa$ depends on the type and location of singularities in higher-order derivatives of $f$. Specifically, if $f$ is in $C^k([-1,1])$ due to logarithmic singularities of the form $\sim(1-\xi)^{k+1}\ln(1-\xi)$, but sufficiently regular in the open interval $(-1,1)$, then we have $\varkappa = 2k+3$.
\eit
Thus, from the decay rate of the Chebyshev coefficients, we can infer the regularity class of the fields. These properties will be exploited while analysing the numerical results. In particular, the analytical studies have already anticipated regularity loss of the solutions along the time direction due to appearance of terms of the form $(1-\tau)^\eta \ln (1-\tau)$, i.e.\ we expect to encounter functions of class $C^{\eta-1}$, which shall be identified by the Chebyshev coefficients algebraically decaying at the rate $c_i\sim i^{2\eta+1}$. One the other hand, coefficients decaying exponentially provide a strong indication for the solutions' analyticity. As special cases, polynomial solutions of order $p$ are easily identified since the coefficients $c_i$ vanish for $i>p$.

So far we have assumed that that an approximation $f_N$ is constructed for a known function $f$. In practice, $f$ will not be known {\it a priori}, but is supposed to satisfy some differential equation. In that case, the approximation $f_N$ is constructed from the condition that $f_N$, together with the resulting approximations for the derivatives\footnote{Spectral differentiation is discussed in refs.~\cite{Boyd,Canuto}.} $f_N'$, $f_N''$, etc., satisfy the differential equation (or boundary condition) exactly at the collocation points $\{\xi_k\}_{k=0}^N$.

\subsection{First numerical studies: $1+1$ evolution on the cylinder}\label{sec:firstnumerics}

As a first numerical test, we perform a $1+1$ evolution by considering the coupled system of intrinsic equations on the cylinder, fed with non-vanishing initial data (ID) for all fields with $n=0, 1, 2$.
These experiments provide a powerful tool to practically demonstrate the main features of the solutions on the cylinder that we derived in the previous section, namely the coupling of different angular modes and the finite $C^k$ regularity, where $k$ depends on how many regularity conditions are satisfied by the ID. 
Afterwards, in Sec.~\ref{sec:2plus1} below, we implement the full $2+1$ evolution and study the solution in a domain $(\rho,\theta,\tau)\in[0,\rho_{\rm f}]\times[0,\pi]\times[0,1]$.

We have seen that we can reduce the problem on the cylinder $\rho=0$ to the ODEs \eqref{eq:Eq_Cylider_mode_nl} for the modes $\psi_{n\ell}(\tau)$. For our first numerical calculations, however, we opt to explicitly include the angular dependence. Specifically, we focus on the system of coupled hyperbolic PDEs \eqref{eq:cylinderEQ} for the Taylor coefficients $f_n(\theta,\tau)$ with $n=0, 1, 2$. We perform the projection $\psi_{n\ell}(\tau) = \frac{2\ell +1}{2} \langle f_n, P_\ell\rangle$ numerically only {\em after} obtaining the solution $f_n$.

The reason for this approach is twofold. On the one hand, we want to be oblivious to the formal derivation of the mode coupling and regularity, i.e.\ we want to check that we can numerically reproduce these results. On the other hand, this $1+1$ setup provides us with a testbed for the full axisymmetric code that we describe in the next section. In particular, compared to previous numerical studies of the wave equation on spherically symmetric backgrounds, the critical set at $\rho=0$, $\tau=1$ now has a more intricate structure due to the angular dependence. Hence an investigation of the equations at the cylinder in $1+1$ dimensions directly tests the code's capability to numerically resolve the critical set. 

We use the fully spectral code developed in~\cite{HennigAnsorg2009,Hennig2013,MacedoAnsorg2014} to numerically solve the PDE \eqref{eq:cylinderEQ}. For that purpose, we introduce a new spatial variable $x=\cos\theta$, which replaces the angular coordinate $\theta$. For $n=0, 1, 2$, we provide Eq.~\eqref{eq:cylinderEQ} with ID at $\tau=0$ that correspond to a pure angular mode $\ell'\in\{0,\dots,4\}$, given by
\begin{equation}\label{eq:1+1ID}
 f_{n,\ell'}(\theta,0) = \alpha_{n,\ell'} P_{\ell'}(\cos\theta),\quad
 \dot{f}_{n,\ell'}(\theta,0) = \beta_{n,\ell'} P_{\ell'}(\cos\theta).
\end{equation}
Here, we are free to choose the constants $\alpha_{n,\ell'}$ and $\beta_{n,\ell'}$, which are related to the above free functions $g_\ell(\rho)$ and $h_\ell(\rho)$ in \eqref{eq:data1}, \eqref{eq:data2} by
\begin{equation}\label{eq:1+1ID2}
 \alpha_{n,\ell'} = \frac{1}{n!} \frac{\dd^n g_{\ell'}}{\dd\rho^n}\Big|_{\rho=0},\quad
 \beta_{n,\ell'} = \frac{1}{n!} \frac{\dd^n h_{\ell'}}{\dd\rho^n}\Big|_{\rho=0}.
\end{equation}

We denote the particular solutions to Eq.~\eqref{eq:cylinderEQ} that arise from the above pure $\ell'$-mode ID by $f_{n,\ell'}(\theta,\tau)$, and their projections in the Legendre basis by $\psi_{n\ell,\ell'}(\tau)=\frac{2\ell+1}{2}\langle f_{n,\ell'}, P_\ell\rangle$. 

\subsubsection{Mode coupling.}\label{sec:1+1coupling}

We begin by studying the effect of {\em each individual angular mode} $\ell'$ in the ID on the mode coupling. For example, extending the discussion from Sec.~\ref{sec:ModeCoupling}, we anticipate that ID for a single mode $\ell'$ also excite the following additional modes with $\ell = \ell'\pm 2$: 
\bea
\fl
\label{eq:ID02_psi20_psi24}
&& \textrm{ID mode } (n,\ell')=(0,2) \textrm{ excites} \left\{
\begin{array}{l}
\displaystyle
\psi_{20}(\tau) = \frac{2 \tau^2 (3-\tau^2)}{15(1+\tau)^2}\, \kappa^2  \alpha_{02},
\\
\displaystyle
\psi_{24}(\tau) = \frac{36 \tau^2(1-\tau)^2}{35}\, \kappa^2  \alpha_{02},
\end{array}
\right. \\
\fl
\label{eq:ID03_psi21_psi25}
&& \textrm{ID mode } (n,\ell')=(0,3) \textrm{ excites} \left\{
\begin{array}{l}
\displaystyle
\psi_{21}(\tau) = \frac{2 \tau^3(5-3\tau^2)}{35(1+\tau)^2}\, \kappa^2  \beta_{03},
\\
\displaystyle
\psi_{25}(\tau) = \frac{100 \tau^3(1-\tau)^2}{189}\, \kappa^2 \beta_{03},
\end{array}
\right. \\
\fl
\label{eq:ID04_psi22_psi26}
&& \textrm{ID mode } (n,\ell')=(0,4) \textrm{ excites} \left\{
\begin{array}{l}
\displaystyle
\psi_{22}(\tau) = \frac{40 \tau^2(1-\tau)^2}{21}\, \kappa^2  \alpha_{04},
\\[1ex]
\displaystyle
\psi_{26}(\tau) = \frac{100 \tau^2(1-\tau)^2(1-3\tau^2)}{33}\, \kappa^2 \alpha_{04}.
\end{array}
\right. 
\eea

Now we make a particular choice of initial data \eqref{eq:1+1ID}, \eqref{eq:1+1ID2}.
For that purpose, we consider $g_{\ell'}(0)$ ($\ell'$ even) or $h_{\ell'}(0)$ ($\ell'$ odd), as well as $h_{\ell'}{}_{,\rho}(0)$ and $h_{\ell'}{}_{,\rho\rho}(0)$ as freely specifiable data, and fix the remaining data --- i.e.\ $h_{\ell'}(0)$ ($\ell'$ even) or $g_{\ell'}(0)$ ($\ell'$ odd), as well as $g_{\ell'}{}_{,\rho}(0)$ and $g_{\ell'}{}_{,\rho\rho}(0)$ --- according the regularity conditions from Sec.~\ref{sec:Regularity}. 

As a representative example, we choose
\bea
\fl
& n=0: \quad \left\{
\begin{array}{ll}
 g_{\ell'}(0) = P_{\ell'}(0), & \mbox{for even $\ell'$}, \\  
 h_{\ell'}(0) = \ell'\, P_{\ell'-1}(0), & \mbox{for odd $\ell'$}, 
 \end{array}
 \right.
 \nn \\
& n=1: \quad h_{\ell'}{}_{,\rho}(0) = 10, \quad \label{eq:ID_1+1D} \\
& n=2: \quad h_{\ell'}{}_{,\rho\rho}(0) = -2. \nn
\eea
Note that the conditions at order $n=0$ are chosen such that the corresponding exact solutions have the projections $\psi_{0\ell,\ell'}(\tau)=P_\ell(\tau)\delta_{\ell\ell'}$.

\begin{figure}[ht!]\centering
\includegraphics[width=0.49\textwidth]{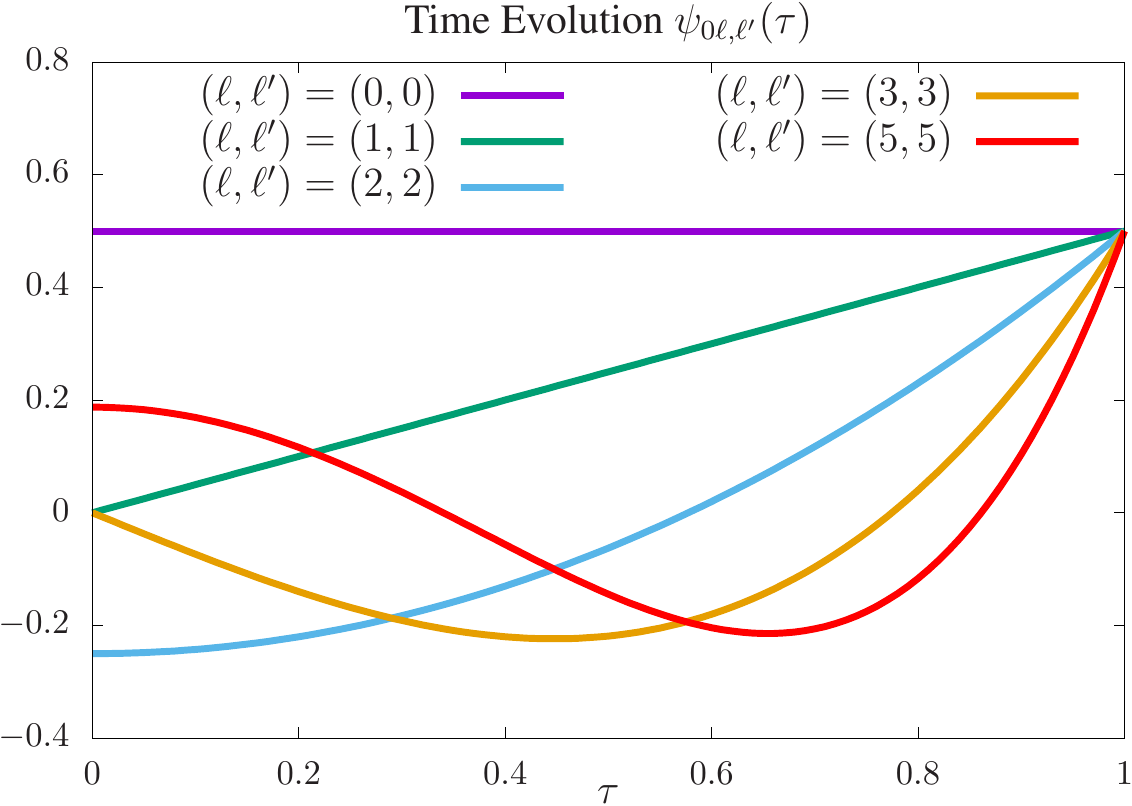}
\includegraphics[width=0.49\textwidth]{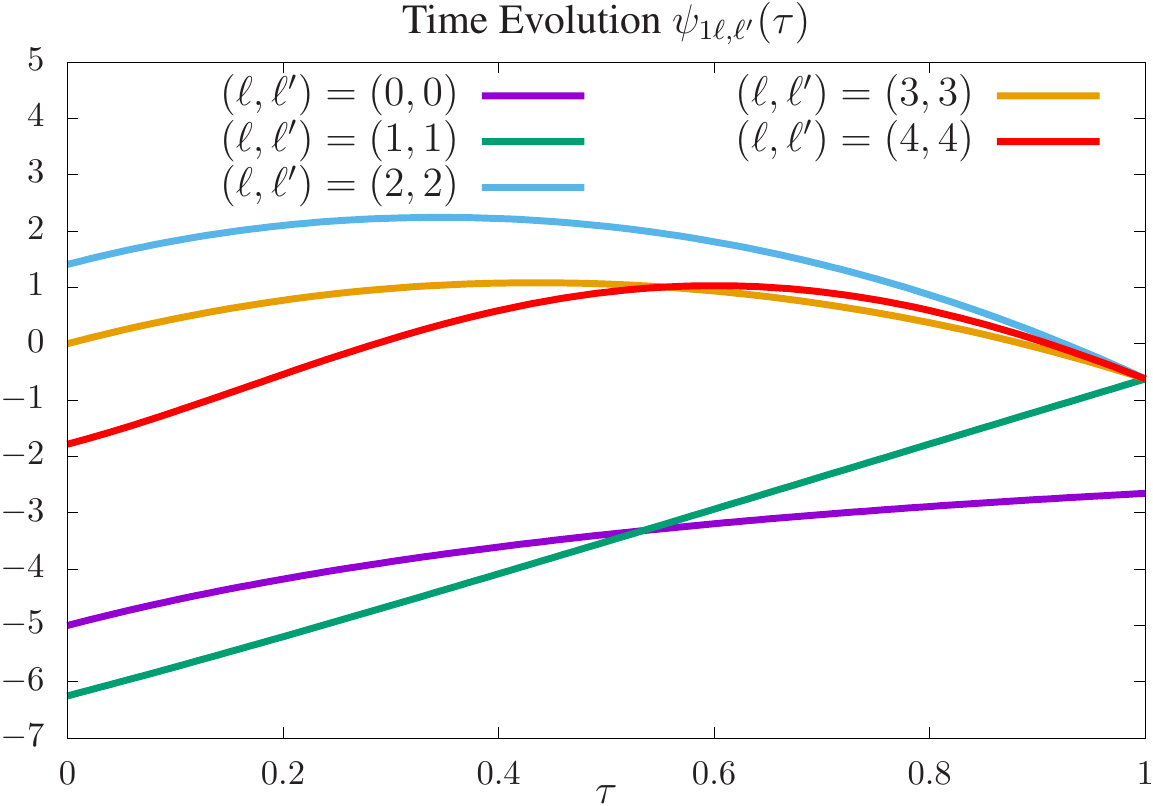}\\[2ex]
\includegraphics[width=0.49\textwidth]{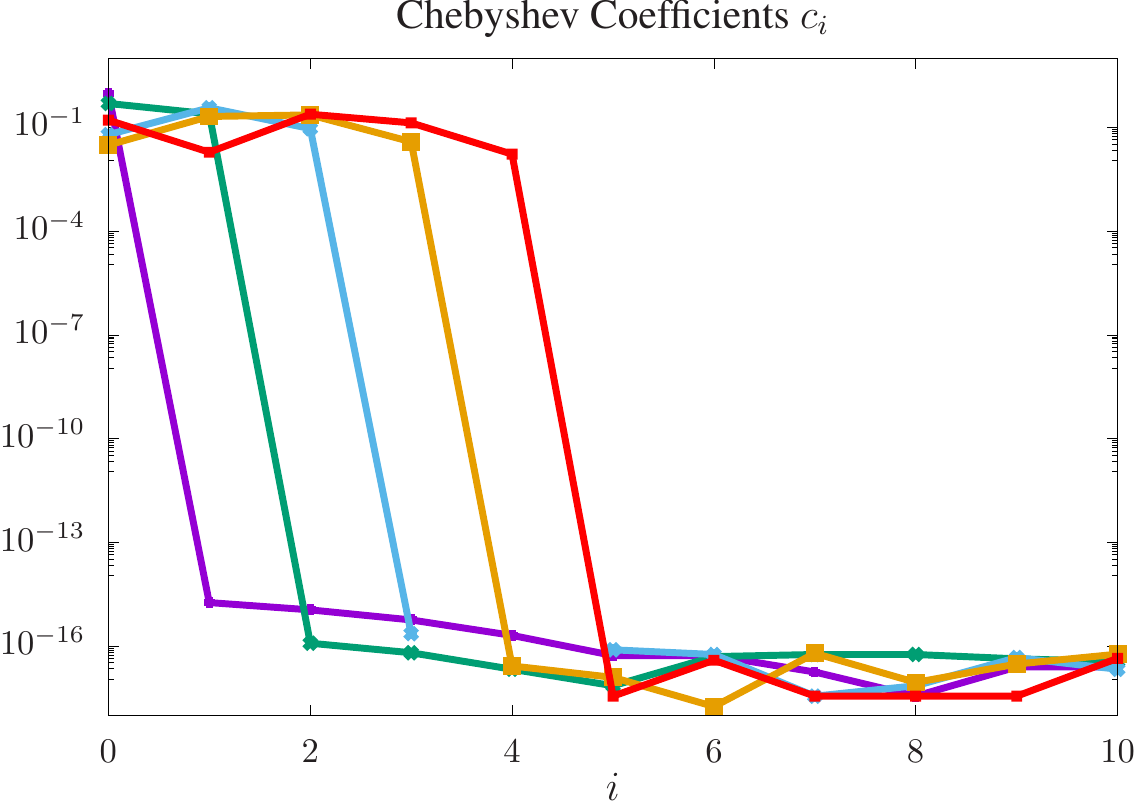}
\includegraphics[width=0.49\textwidth]{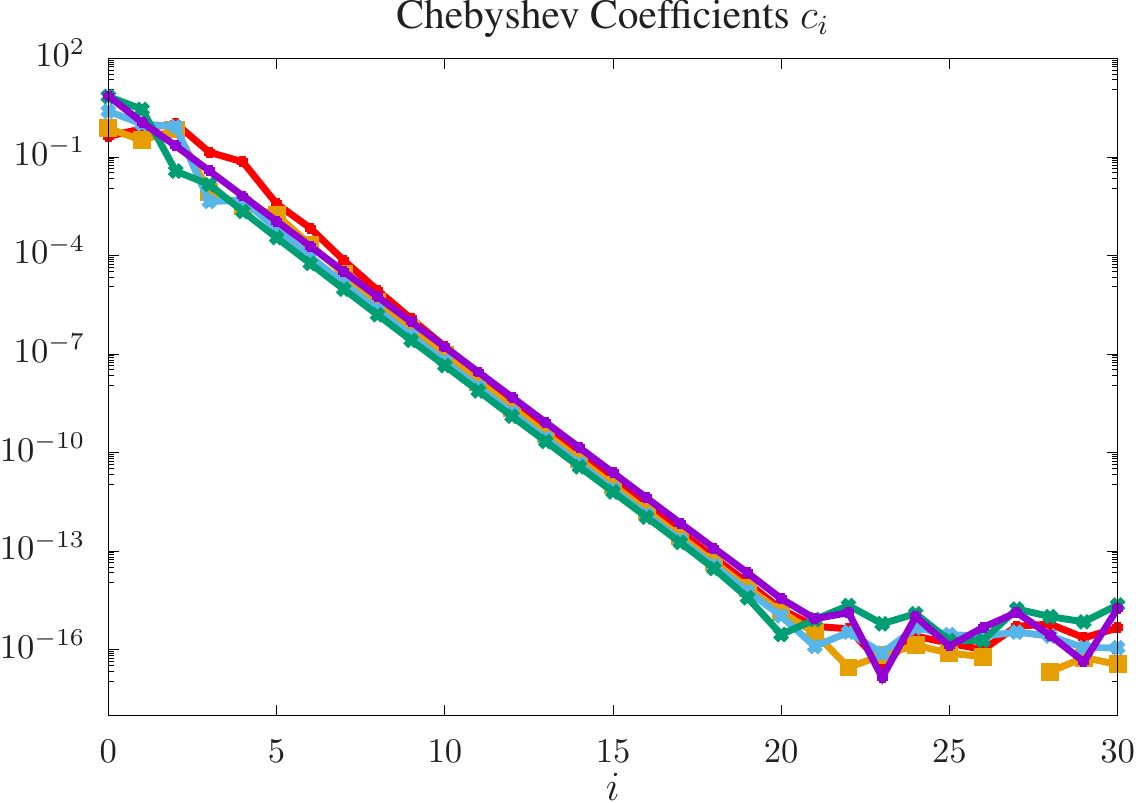}
 \caption{Time evolution on the cylinder for projected modes with $\ell=\ell'$ (top) and their Chebyshev coefficients (bottom) at orders $n=0$ (left) and $n=1$ (right). {\em Left panels:} The finite number of non-vanishing coefficients (up to numerical error $\sim 10^{-15}$)  confirms that $\psi_{0\ell,\ell'}$ are polynomials. {\em Right panels:} The coefficients' exponential decays show that the solutions $\psi_{1\ell,\ell'}$ are analytic. Modes with $\ell\neq\ell'$ are not present at orders $n=0, 1$.}
\label{fig:1+1_f0f1}
\end{figure}

\begin{figure}[ht!]\centering
\includegraphics[width=0.49\textwidth]{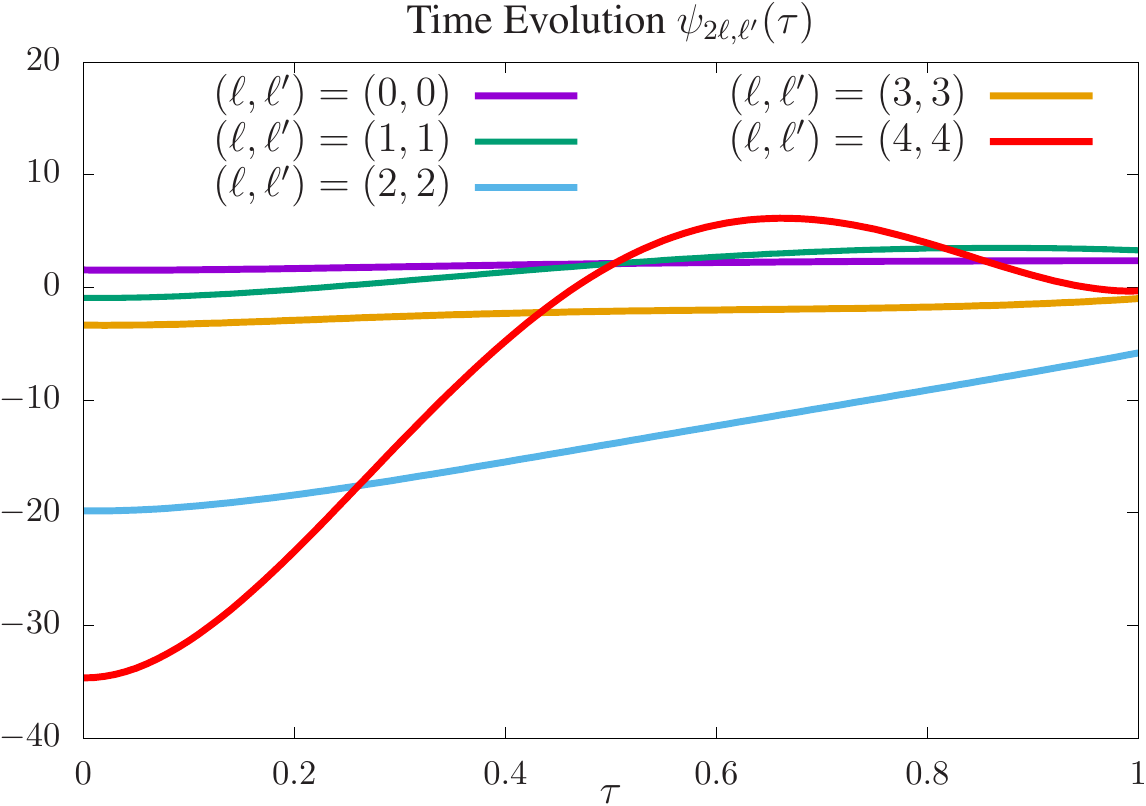}
\includegraphics[width=0.49\textwidth]{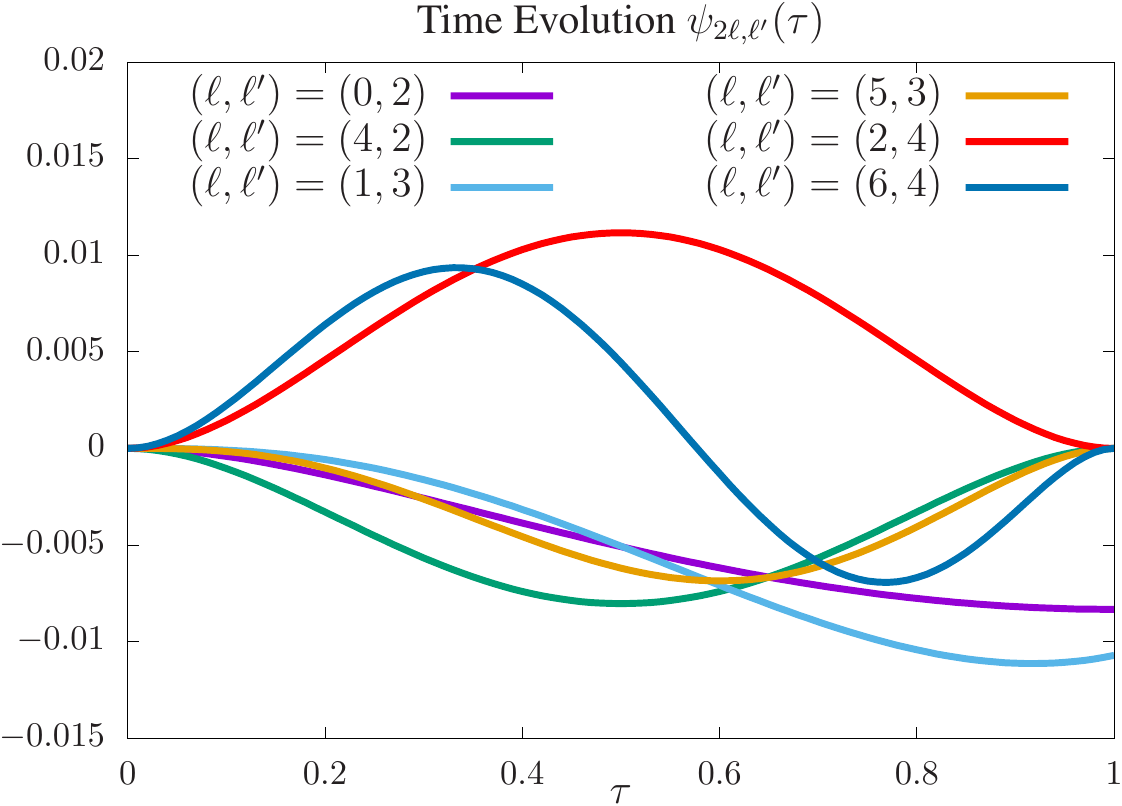}\\[2ex]
\includegraphics[width=0.49\textwidth]{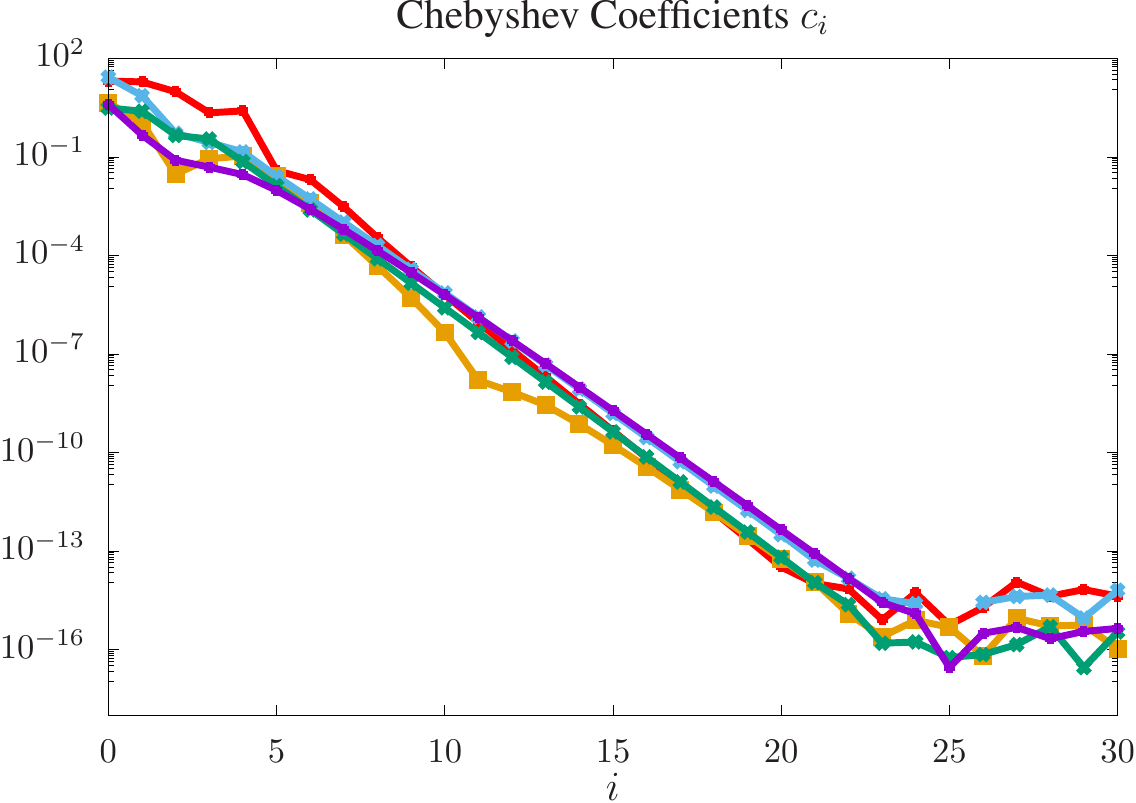}
\includegraphics[width=0.49\textwidth]{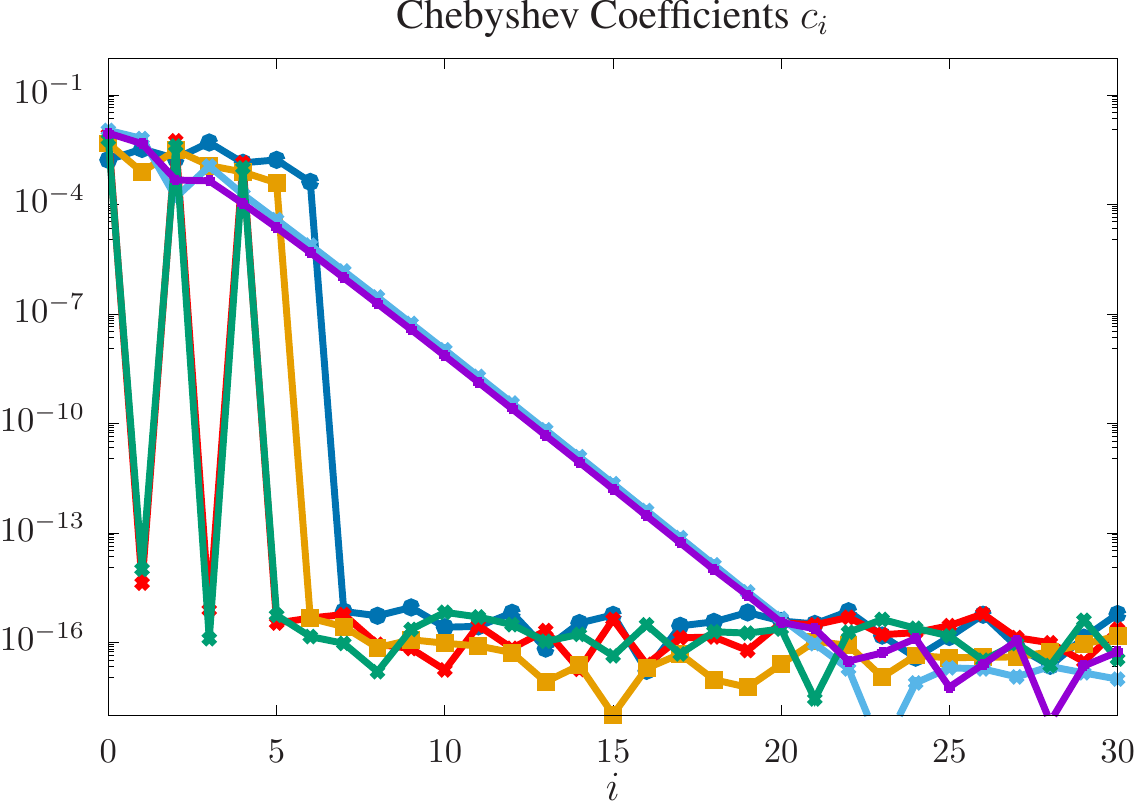}
\caption{Mode coupling on the cylinder at order $n=2$. Time evolution (top panels) and their Chebyshev coefficients (bottom panels) for the projections $\psi_{2\ell,\ell'}$ with modes $\ell=\ell'$ (left) and $\ell=\ell' \pm 2$ (right).   {\em Left Panel:} The coefficients' exponential decays demonstrate that the solutions $\psi_{2\ell,\ell'}$ with $\ell=\ell'$ are analytic.  {\em Right Panel:} The solutions initially vanish at $\tau=0$, but modes with $\ell=\ell' \pm 2$ are excited along the evolutions. The exponential decay/finite number of non-vanishing Chebyshev coefficients confirms the regular/polynomials character of the solutions --- cf.~\eqref{eq:ID02_psi20_psi24}-\eqref{eq:ID04_psi22_psi26}.}
  \label{fig:1+1_f2}
\end{figure}

The numerical results are shown in Figs.~\ref{fig:1+1_f0f1} and \ref{fig:1+1_f2}, which display the time evolution of each non-vanishing mode $\psi_{n\ell,\ell'}(\tau)$ together with their respective Chebyshev coefficients $c_i$ along the time direction. The diagrams correspond to the Kerr spin parameter $\kappa = 1/2$, but we found no difficulties to solve the equations in the whole range $|\kappa| \in [0,1]$. The numerical resolution is  $N_\theta=10$, $N_{\tau}=30$.
 
Fig.~\ref{fig:1+1_f0f1} focuses on the evolution of the fields with $n=0$ (left panels) and $n=1$ (right panels). As expected, the only non-vanishing $\ell$-modes are the ones that are already present in the ID, namely those with $\ell=\ell'$. The behaviour of the Chebyshev coefficients also confirms the regularity of the solutions. In particular, the polynomial behaviour in $\tau$ for $\psi_{0\ell,\ell'}(\tau)$ --- cf.~Eq.~\eqref{eq:Sol_n0} --- is evident (bottom left panel), since the coefficients vanish (up to numerical round off errors) for $i>\ell$. Moreover, the exponential decay of the Chebyshev coefficients for $n=1$ (bottom right panel) demonstrates that the solution $\psi_{n\ell,\ell'}(\tau)$ is analytic everywhere up to $\tau=1$. 
 
Next we discuss the order $n=2$. Fig.~\ref{fig:1+1_f2} displays the corresponding time evolution and Chebyshev coefficients for the fields $\psi_{2\ell,\ell'}(\tau)$. The left panels depict the results for the modes with $\ell = \ell'$, which are already contained in the ID. Once again, the exponential decay of the Chebyshev coefficients (bottom left panel) indicates that the solutions are analytic up to $\tau=1$. For $n=2$, however, we also observe the effect of mode coupling due to a non-vanishing spin parameter $\kappa$. The right panels in Fig.~\ref{fig:1+1_f2} confirm the excitation of modes with $\ell=\ell' \pm 2$ ($\ell'\geq 2$), as derived in Eqs.~\eqref{eq:ID02_psi20_psi24}-\eqref{eq:ID04_psi22_psi26}. In particular, the Chebyshev coefficients (bottom right panel) confirm the regularity of the fields $\psi_{20,2}(\tau)$ and $\psi_{21,3}(\tau)$, as well as the polynomial character of $\psi_{24,2}(\tau)$, $\psi_{25,3}(\tau)$, $\psi_{22,4}(\tau)$, and $\psi_{26,4}(\tau)$ --- cf.~\eqref{eq:ID02_psi20_psi24}-\eqref{eq:ID04_psi22_psi26}.

\subsubsection{Solutions with lower regularity.}

In our next numerical experiment, we consider ID that do \emph{not} satisfy all regularity conditions \eqref{eq:RegCond_n0l0}-\eqref{eq:RegCond_n2l4}. For that purpose, we introduce the following small modification of the ID \eqref{eq:ID_1+1D},
\bea
 n=0: \qquad 	& g_{\ell',\rho}(0) = g^{\rm reg}_{\ell',\rho}(0), \quad  
				& g_{\ell',\rho\rho}(0) = g^{\rm reg}_{\ell',\rho\rho}(0),\\
 n=1: \qquad 	& g_{\ell',\rho}(0) = g^{\rm reg}_{\ell',\rho}(0)(1+\epsilon), \quad  
				& g_{\ell',\rho\rho}(0) = g^{\rm reg}_{\ell',\rho\rho}(0), \\
 n=2: \qquad 	& g_{\ell',\rho}(0) = g^{\rm reg}_{\ell',\rho}(0), \quad  
				& g_{\ell',\rho\rho}(0) = g^{\rm reg}_{\ell',\rho\rho}(0)(1+\epsilon),
\eea
with $\epsilon = 1/10$, where $g^{\rm reg}_{\ell',\rho}(0)$ and $g^{\rm reg}_{\ell',\rho\rho}(0)$ refer to the previous data that obey the regularity conditions.
Note that we only modify the ID for $n=1, 2$, but not for $n=0$, because  a singularity of the form $\ln(1-\tau)$ would otherwise develop at $\tau=1$ and spoil the applicability of the current code. Moreover, the particular feature that $\psi_{21,1}(\tau)$ is always regular --- cf.~\eqref{eq:RegCond_n2l1} --- assumes the regularity at the previous order. 

\begin{figure}[ht!]\centering
    \includegraphics[width=0.49\textwidth]{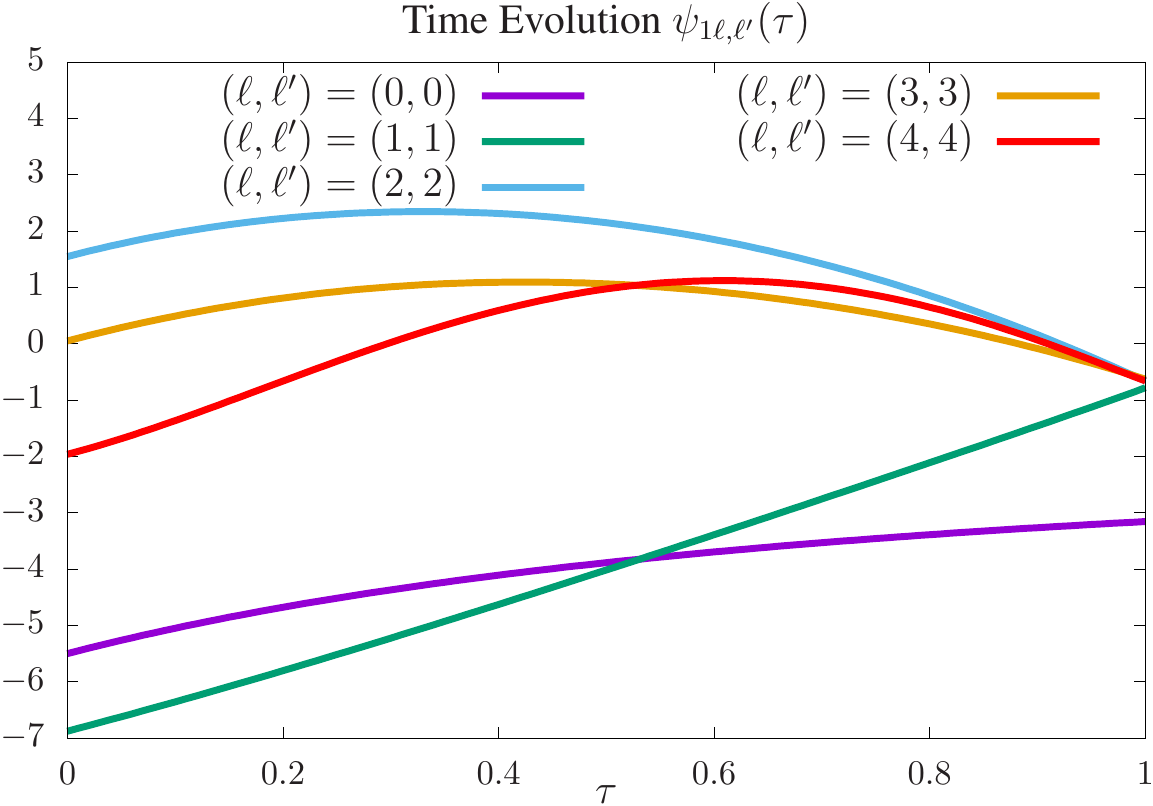}
     \includegraphics[width=0.49\textwidth]{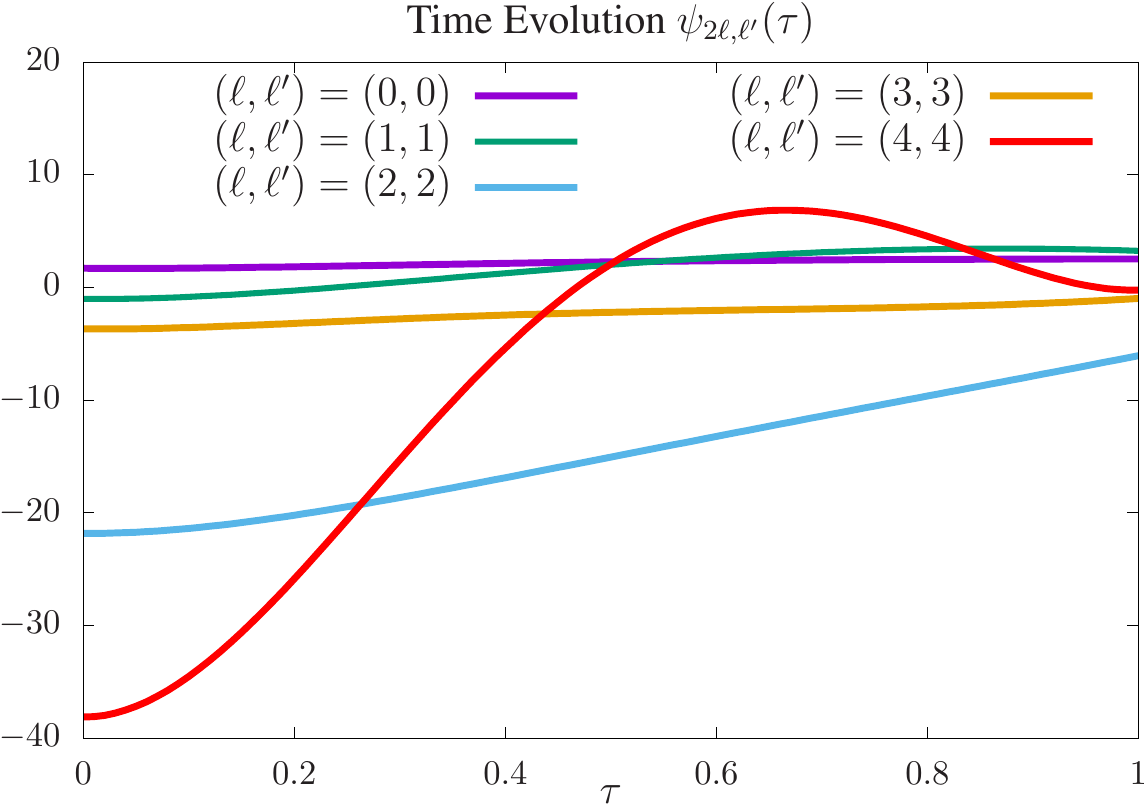}\\[2ex]
      \includegraphics[width=0.49\textwidth]{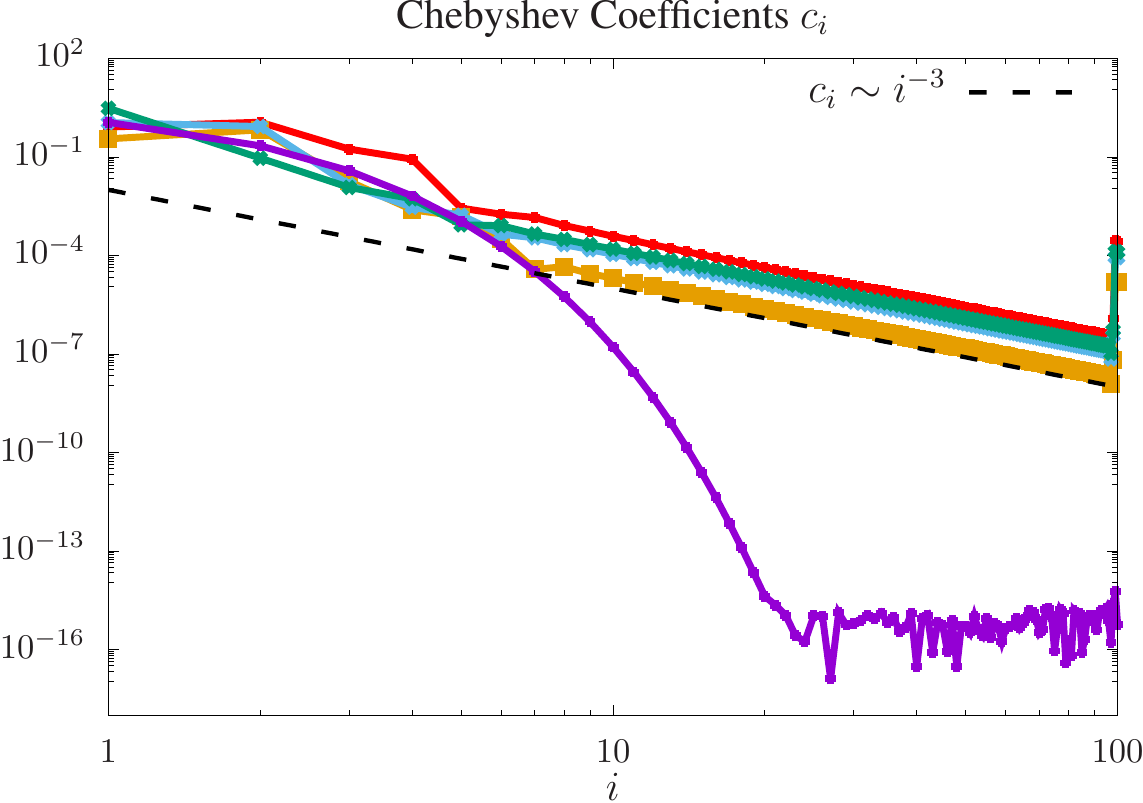}
     \includegraphics[width=0.49\textwidth]{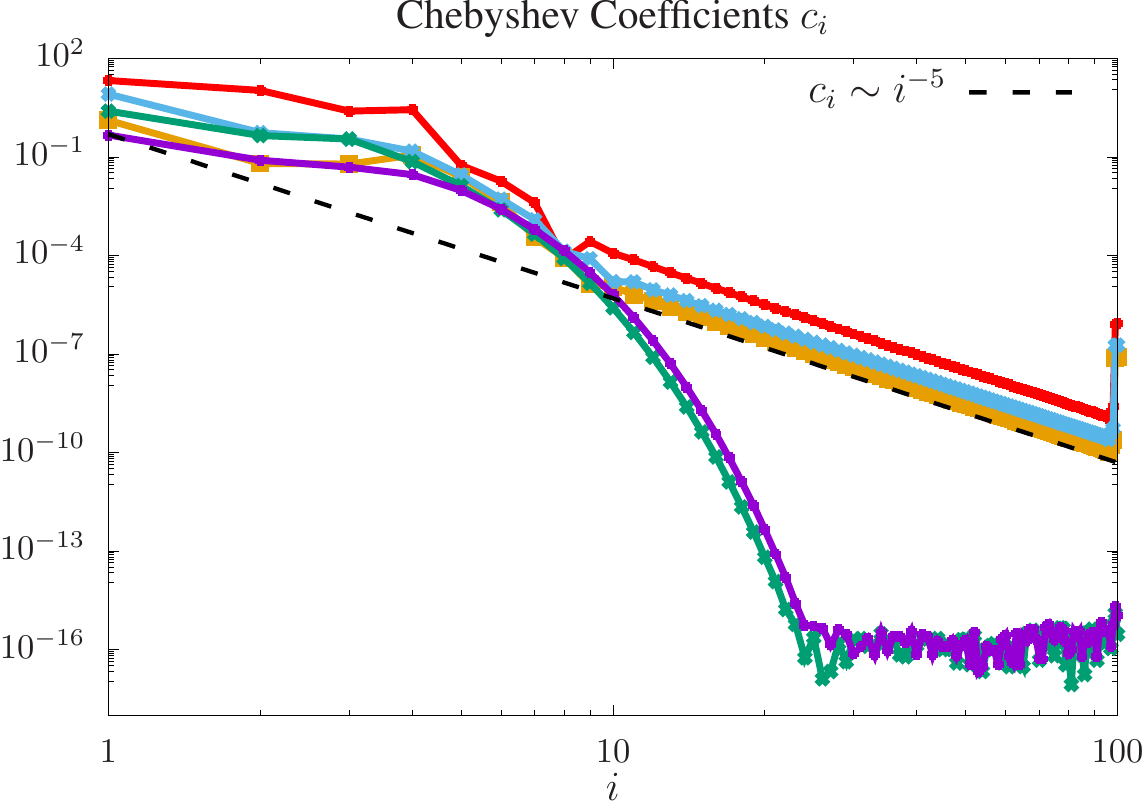}
  \caption{Time evolution on the cylinder for ID that do not satisfy the regularity conditions (top) and their respective Chebyshev coefficients (bottom) at orders $n=1$ (left) and $n=2$ (right).  The loss of regularity is not directly apparent in the time evolutions (cf.~top panels in Figs.~\ref{fig:1+1_f0f1} and \ref{fig:1+1_f2}). However, the algebraic decay of some Chebyschev coefficients (highlighted by the log-log scale) confirms the irregular character of the solutions. {\em Left panels:} The algebraic decay $c_i \sim i^{-3}$ indicates a singularity of the form $(1-\tau)\ln(1-\tau)$. {\em Right panels:} The algebraic decay $c_i \sim i^{-5}$ indicates a singularity of the form $(1-\tau)^2\ln(1-\tau)$.}
  \label{fig:1+1_f1f2_Log}
\end{figure}

The time evolution of these data is shown in Fig.~\ref{fig:1+1_f1f2_Log} for a resolution  $N_\theta=7$, $N_{\tau}=100$, where we focus on the orders $n=1$, $2$.

Since the modification in the initial data is relatively small, the fields $\psi_{1\ell,\ell'}(\tau)$ and $\psi_{2\ell,\ell'}(\tau)$ (top panels) do not show a perceptive difference compared to the regular solutions in Figs.~\ref{fig:1+1_f0f1}, \ref{fig:1+1_f2}. The singular behaviour at $\tau=1$, though, is apparent in the behaviour of the Chebyshev coefficients $c_i$ (bottom panels). Indeed, they clearly show an algebraic decay $c_i\sim i^{-\varkappa}$, as opposed to the exponential decay of the regular solutions. For $n=1$ (left panel), it is evident that the mode $\psi_{10,0}(\tau)$ is still regular, as anticipated in \eqref{eq:RegCond_n1l0}. The coefficients for the other modes, however, decay algebraically with the rate $c_i\sim i^{-3}$. The same analysis is valid for $n=2$ (right panel). In accordance with \eqref{eq:RegCond_n2l1}, the function $\psi_{21,1}(\tau)$ is still regular, whereas the other modes have coefficients decaying as $c_i\sim i^{-5}$. 

We emphasise that the algebraic decay is a direct consequence of the underlying theory of spectral methods (as discussed in Sec.~\ref{sec:specmeth}), and not an artefact of the chosen numerical resolution. This shows that $f_{1}(\tau,\theta)$ belongs to $C^0$, while $f_{2}(\tau,\theta)$ is in $C^1$. This result is in accordance with the theoretical prediction of singular terms of the form $(1-\tau)^n \ln(1-\tau)$ at order $n$, which  means that the Taylor coefficients $f_n(\tau,\theta)$ in the expansion about the cylinder $\rho=0$ are in $C^{n-1}$. Hence, in terms of the order $n$, the Chebyshev coefficients in the time direction should indeed decay as $c_i\sim i^{2n+1}$ whenever the ID are \emph{not} fine-tuned to remove the logarithmic terms. 

\subsection{Full $2+1$ evolution}\label{sec:2plus1}

The previous numerical studies on the cylinder have demonstrated that the fully pseudospectral scheme can successfully handle the angular dependence of the functions and the resulting more intricate structure of the critical set. Hence we are now in a position to tackle the full $(2+1)$-dimensional problem and solve the wave equation~\eqref{eq:CWE} for the unknown function $f(\rho,\theta,\tau)$. 

Since \eqref{eq:CWE} is a second-order PDE, we need to numerically calculate first and second-order derivatives of $f$ in the entire domain. In particular, we require time derivatives at $\tau=1$. For reliable numerical results, we must ensure that the functions are at least differentiable at this hypersurface. To this end, we explicitly enforce the known behaviour at the cylinder with the decomposition
\beq
\label{eq:2+1_FuncDecp}
f(\rho, \theta, \tau) = f_0(\theta, \tau) + \rho\, f_1(\theta, \tau) + \rho^2 {\cal F}(\rho, \theta, \tau),
\eeq
which replaces $f(\rho,\theta,\tau)$ by the new unknown ${\cal F}(\rho,\theta,\tau)$ and the additional functions $f_0(\theta,\tau)$, $f_1(\theta,\tau)$.
Taking into account the previous considerations, namely the study of Eq.~\eqref{eq:Eq_Cylider_mode_nl} with solution \eqref{eq:Sol_n0} at order $n=0$, we explicitly specify
\beq\label{eq:f0singlel}
f_0(\theta, \tau) = \sum_{\ell} \alpha_{0\ell}\, P_\ell(\cos\theta)\, P_\ell(\tau).
\eeq
The other additional function $f_1$ needs to satisfy Eq.~\eqref{eq:Eq_Cylider_mode_nl} at order $n=1$. Therefore, we numerically solve this equation together with the equation for ${\cal F}(\rho, \theta, \tau)$ that results from substituting \eqref{eq:2+1_FuncDecp} into the wave equation \eqref{eq:CWE}. 

In order to solve these coupled equations, we use the $(2+1)$-dimensional version of the fully pseudospectral code that was developed in~\cite{MacedoAnsorg2014}. A short summary of the method is given in \ref{sec:SpecCode_2+1}. 

For our numerical experiments, we again choose initial data that only contain a single $\ell'$-mode. This allows us to study the excitation of other modes in the solution.
We denote the corresponding solutions by ${\cal F}_{\ell'}(\rho, \theta, \tau)$, and we will particularly investigate the behaviour of the projections
\begin{equation}
\label{eq:LegendreProj}
 \Psi_{\ell, \ell'}(\rho, \tau) = \frac{2\ell+1}{2} \langle {\cal F}_{\ell'}(\rho, \theta, \tau), P_{\ell}(\cos\theta)\rangle
\end{equation}
of $\mathcal{F_{\ell'}}$ into the Legendre polynomial basis. 

\subsubsection{Mode coupling.}

We start by considering single $\ell'$-mode ID that reduce to the data from subsection \ref{sec:1+1coupling} at the cylinder, i.e.\ in the limit $\rho\to0$. 
It follows from Eq.~\eqref{eq:ID_1+1D} that the exact solution for $f_0$ is
\beq
 f_0(\theta, \tau) = P_{\ell'}(\cos\theta)\, P_{\ell'}(\tau),
\eeq
i.e.\ we choose $\alpha_{0\ell}=\delta_{\ell\ell'}$ in \eqref{eq:f0singlel}.
Next we construct ID for $f_1$ and ${\cal F}$. For that purpose, we first observe that \eqref{eq:ID_1+1D} [together with \eqref{eq:1+1ID}, \eqref{eq:1+1ID2} and \eqref{eq:2+1_FuncDecp}] implies
\bea
 \dot{f}_1(\theta, 0) = 10\, P_{\ell'}(\cos\theta), \quad 
 \dot{\cal{F}}_1(0,\theta, 0) = - P_{\ell'}(\cos\theta).
\eea
Then we use the regularity conditions \eqref{eq:RegCond_n1l0}-\eqref{eq:RegCond_n2l4} to obtain $f_1(\theta, 0)$ and ${\cal F}(0,\theta, 0)$. This specifies the ID at $\rho=0$. To obtain ID for all $\rho$, we could choose any functions ${\cal F}(\rho, \theta, 0)$ and $\dot{\cal F}(\rho,\theta,0)$ that take on the above values at $\rho=0$. Here, we make the probably simplest choice and extend ${\cal F}$ and $\dot{\cal F}$ as \emph{constant} functions to all $\rho$, i.e.\ we choose
\bea
 {\cal{F}}_1(\rho,\theta, 0) = {\cal{F}}_1(0,\theta, 0),\quad
 \dot{\cal{F}}_1(\rho,\theta, 0) = \dot{\cal{F}}_1(0,\theta, 0).
\eea

To exploit the code's capability of handling the entire $\kappa$-parameter space, we construct solutions for $\kappa=1/2$ and $\kappa=1$ with spectral resolution $N_{\rho}=20$,  $N_\theta = 11$ and $N_{\tau}=30$, where the angular part is again implemented in terms of $x=\cos\theta$. The outer boundary is fixed at $\rho_{\rm f} = 1/10$. 

\begin{figure}[ht!]\centering
\includegraphics[width=0.49\textwidth]{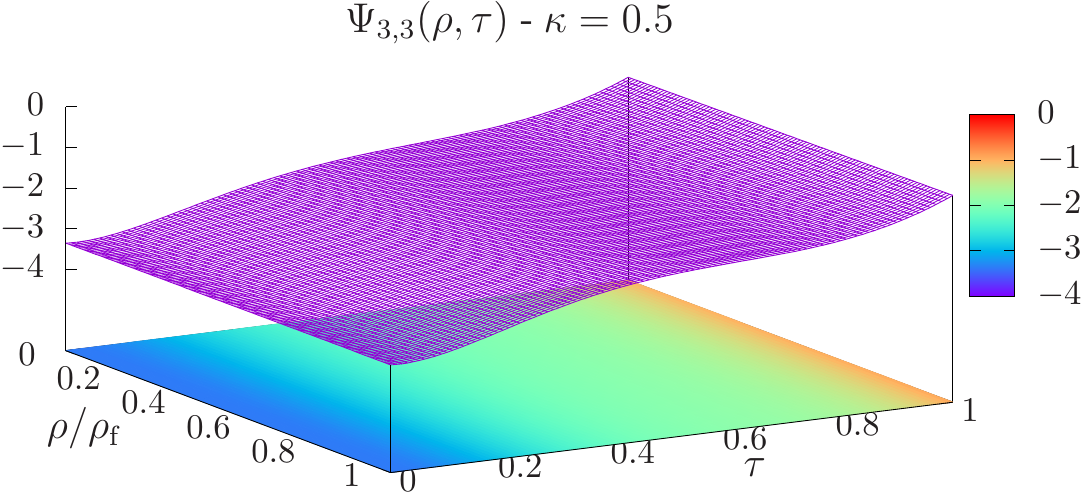}
\includegraphics[width=0.49\textwidth]{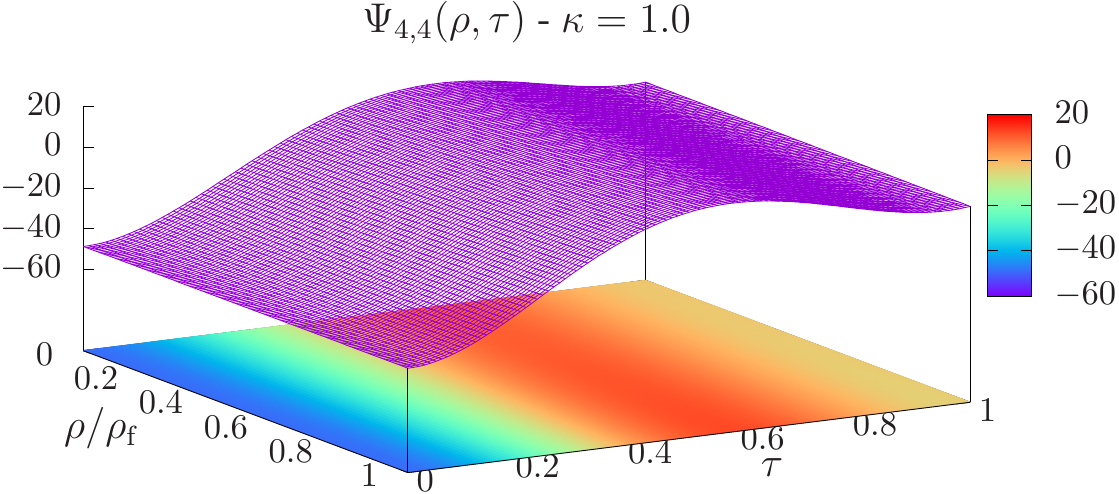}
 \caption{Time evolution around the cylinder at spacelike infinity in the Kerr background. Focus on dynamics of modes prescribed within the ID, i.e., on the angular projections  $\Psi_{\ell,\ell'}$ with $\ell = \ell'$ --- cf.~\eqref{eq:LegendreProj}. {\em Left Panel:} Spin parameter $\kappa=0.5$ and angular mode $\ell=\ell'=3$. {\em Right Panel:} Spin parameter $\kappa=1$ and angular mode $\ell=\ell'=4$.}
 \label{fig:2+1_l3_l4}
\end{figure}

Fig.~\ref{fig:2+1_l3_l4} shows results for ID with the single angular modes $\ell'=3$ ($\kappa = 1/2$) and $\ell'=4$ ($\kappa = 1$). Here, we focus on the solutions' projections into the dominant modes, i.e. $\Psi_{3, 3}(\rho, \tau)$ and $\Psi_{4, 4}(\rho, \tau)$. As expected, these functions turn out to agree with those from the previous $1+1$ studies on the cylinder $\rho=0$.
 
\begin{figure}[ht!]\centering
\includegraphics[width=0.49\textwidth]{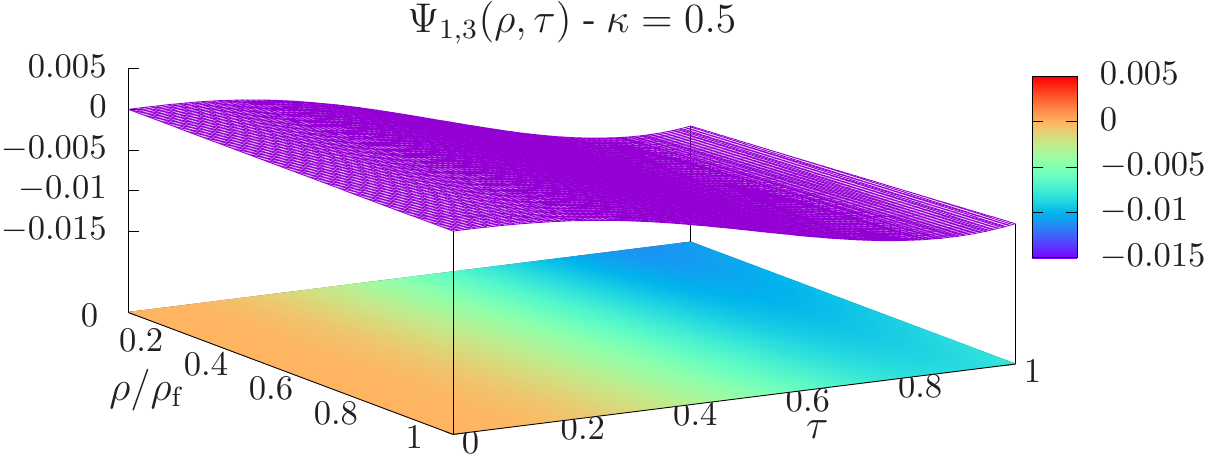}
\includegraphics[width=0.49\textwidth]{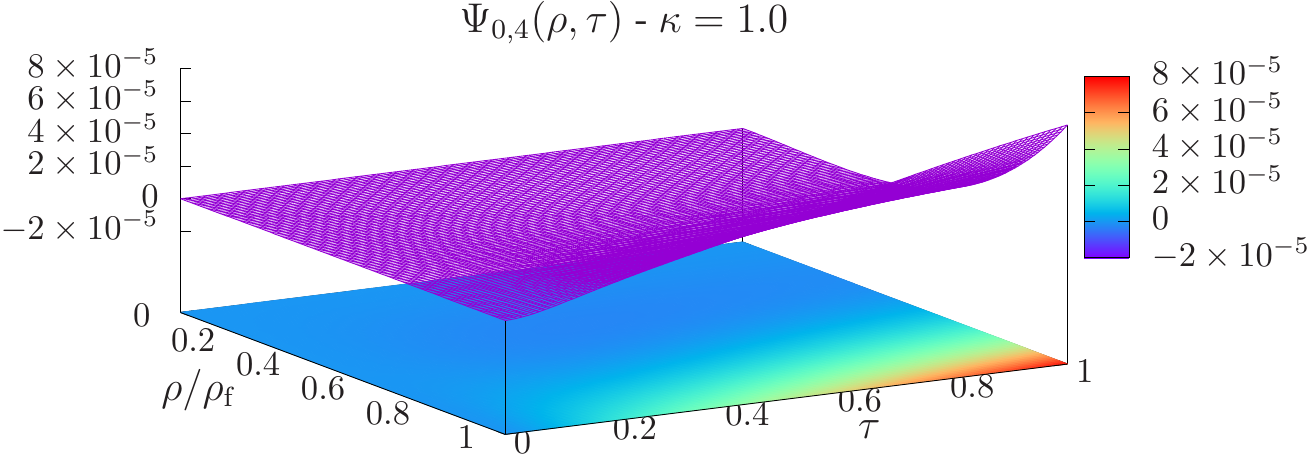}
\\[2ex]
\includegraphics[width=0.49\textwidth]{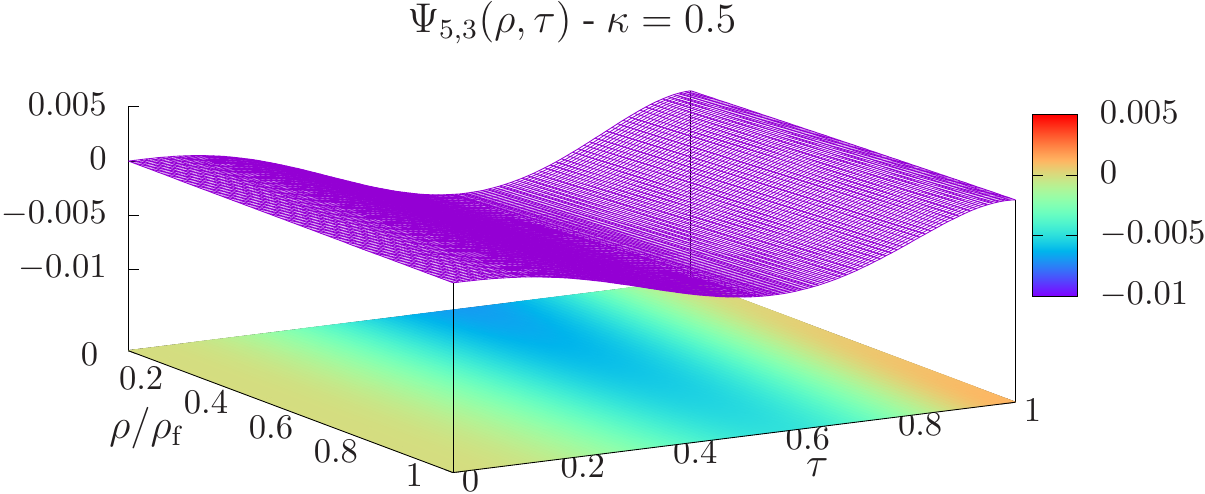}
\includegraphics[width=0.49\textwidth]{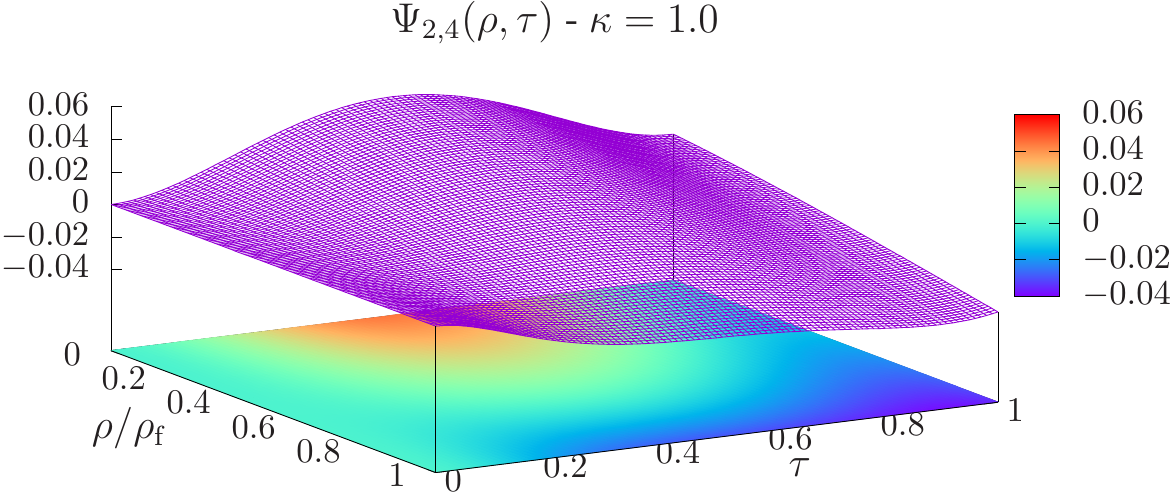}
\\[2ex]
\includegraphics[width=0.49\textwidth]{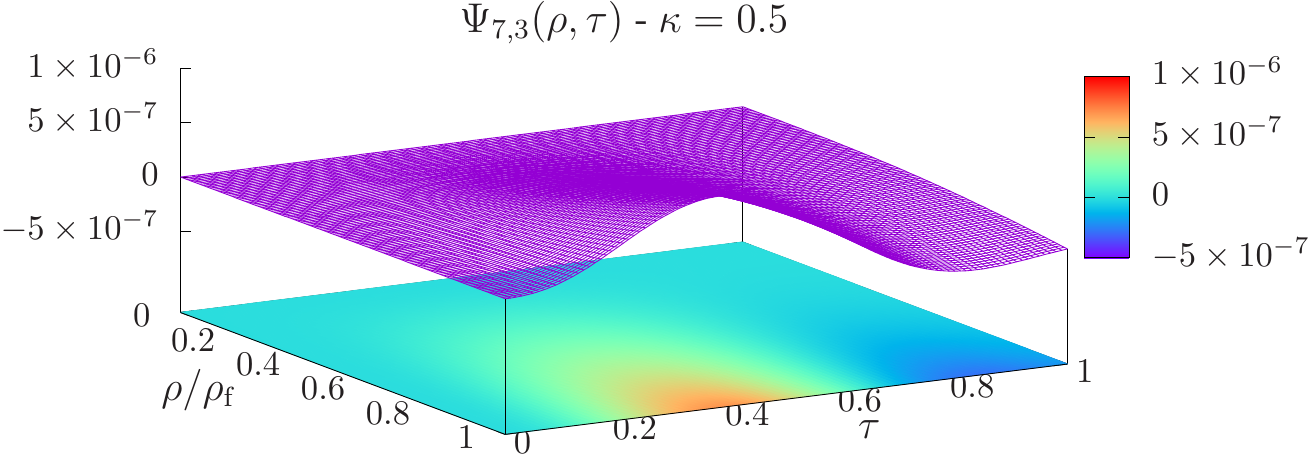}
\includegraphics[width=0.49\textwidth]{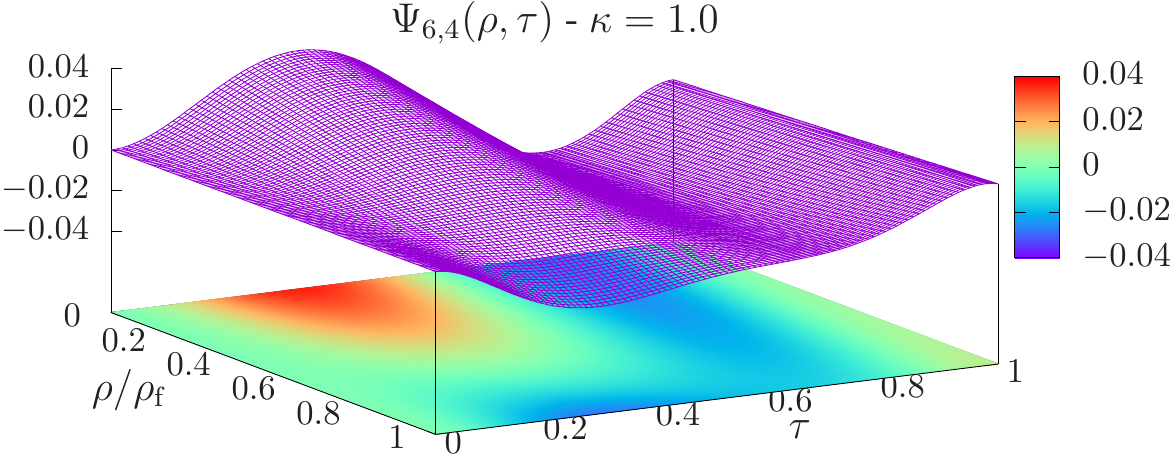}
\\[2ex]
\includegraphics[width=0.49\textwidth]{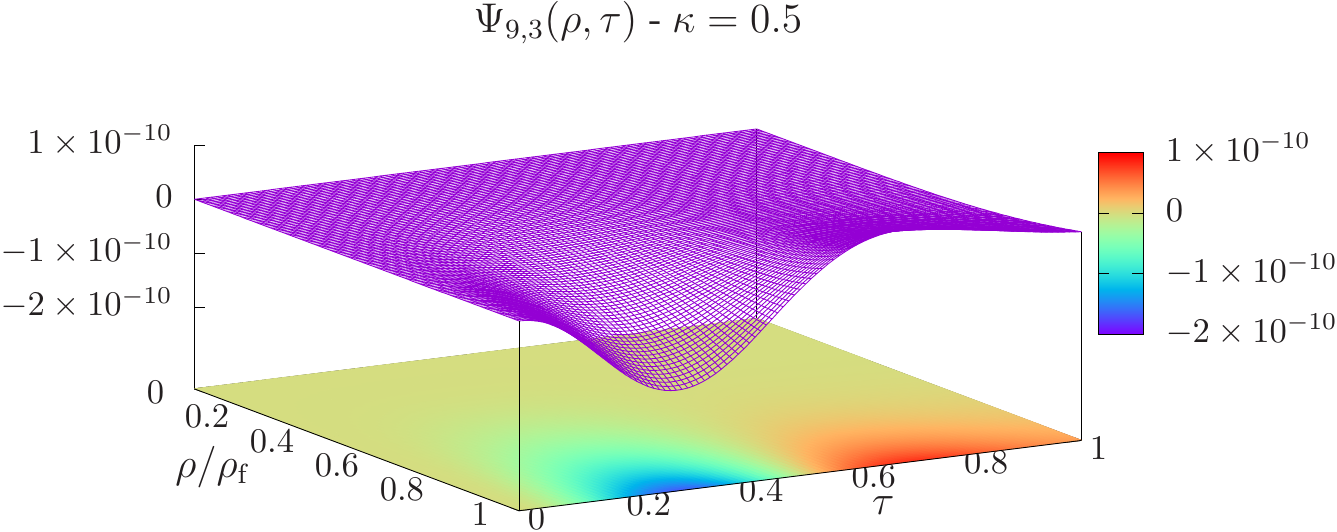}
\includegraphics[width=0.49\textwidth]{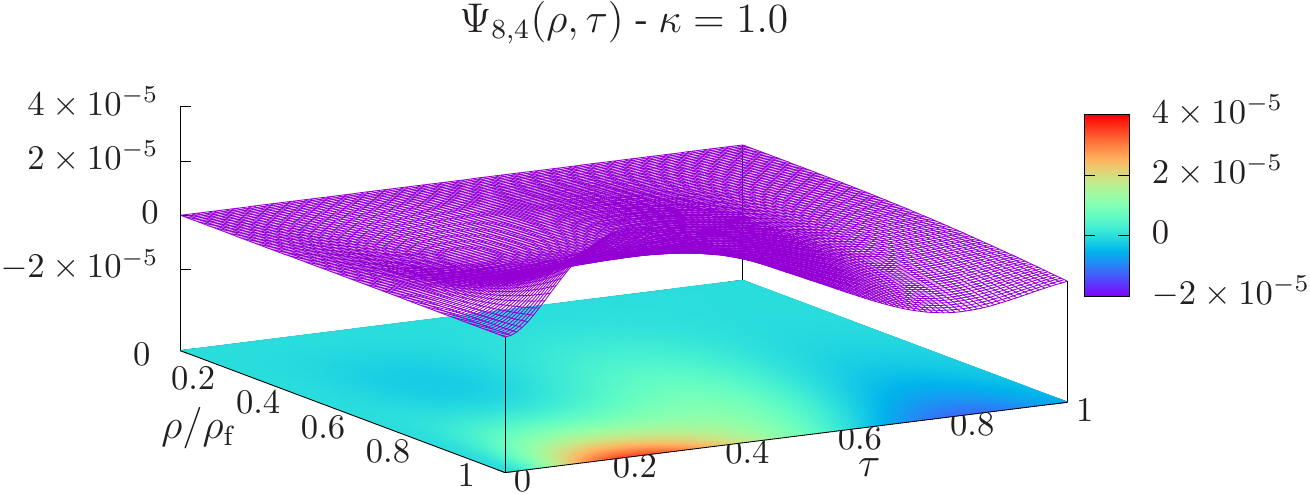}
\\[2ex]
\includegraphics[width=0.49\textwidth]{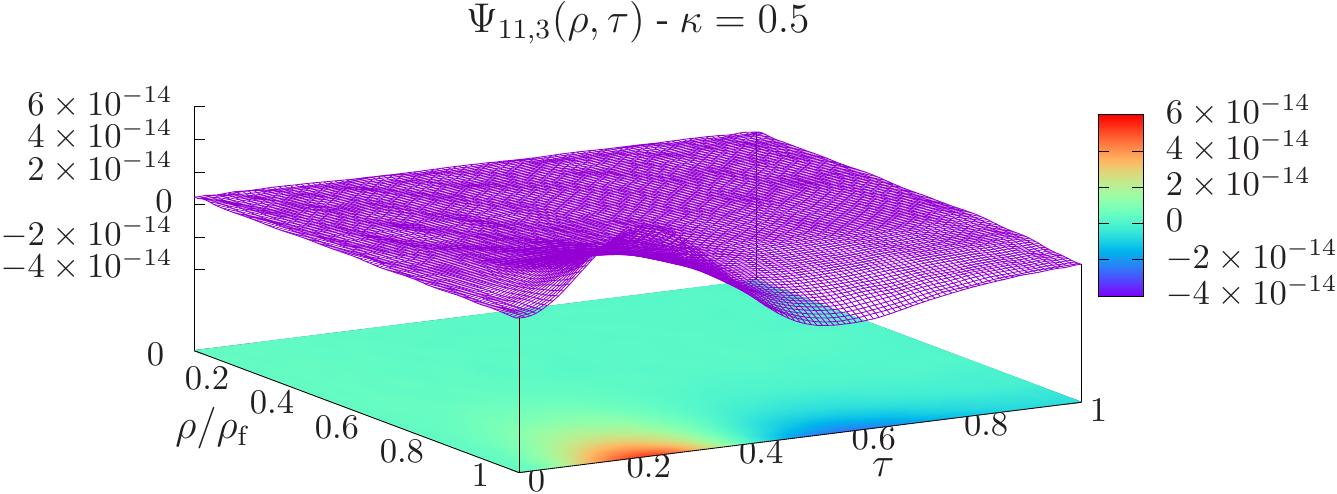}
\includegraphics[width=0.49\textwidth]{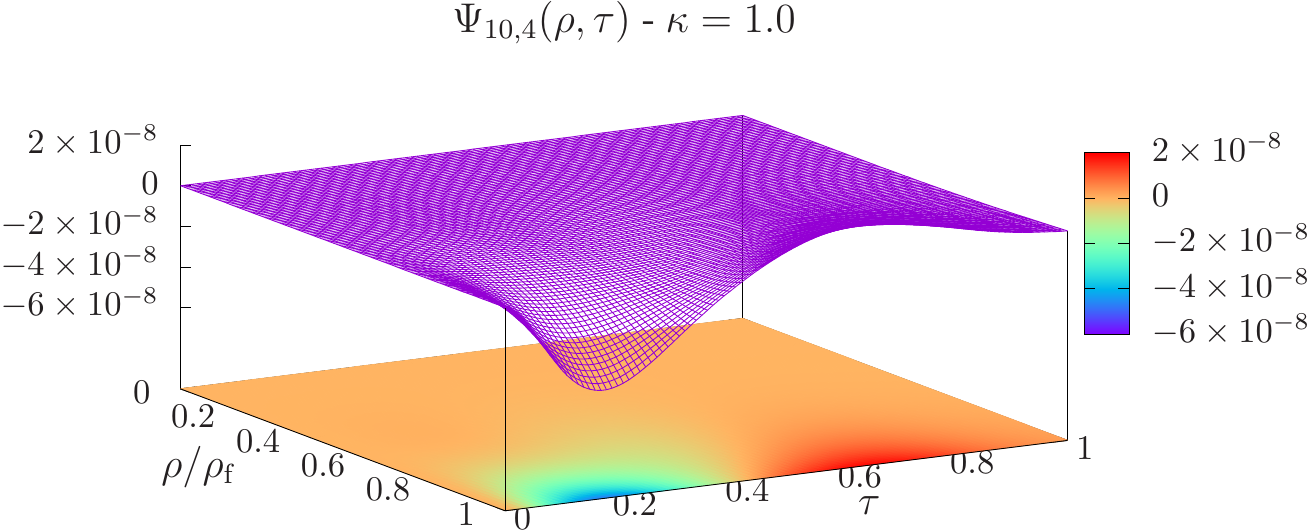}
  \caption{Time evolution around the cylinder at spacelike infinity in the Kerr background. Focus on the the angular projections  $\Psi_{\ell,\ell'}$ with $\ell \neq \ell'$, i.e. the excitation of modes absent in the ID. {\em Left Panel:} Spin parameter $\kappa=0.5$ and ID angular mode $\ell'=3$. Mode coupling excites the projections with $\ell = 1, 5, 7, 9$ and $11$ (from top to bottom). {\em Right Panel:} Spin parameter $\kappa=1$ and ID angular mode $\ell'=4$. Mode coupling excites the projections with $\ell = 0, 2, 6, 8$ and $10$ (from top to bottom).
  }
  \label{fig:2+1_modecouple}
\end{figure}

In our previous investigations of the equations intrinsic to the cylinder, we found that an $\ell'$-mode at order $n$ can only excite the additional modes $\ell=\ell'\pm2$ at order $n+1$. Since $\rho\neq0$ away from the cylinder, all functions $f_n$ in \eqref{eq:expansion} are simultaneously present, and hence we cannot restrict our attention to a single order $n$. Therefore, the full chain of excitations leads to all possible modes of the same parity as $\ell'$, i.e.\ all modes $\ell\geq0$ with $\ell = \ell' \pm 2k$, $k=1,2, 3,\dots$ 

This is illustrated with some examples in Fig.~\ref{fig:2+1_modecouple}.
The left panels show the results for $\ell'=3$ ($\kappa=1/2$) and the projections $\ell=1,5,7,9,11$. Similarly, the right panels show the modes $\ell=0,2,6,8,10$ for data with $\ell'=4$ ($\kappa=1$). Note that the amplitudes of the excited modes are several orders of magnitude smaller than the dominant one, and they decrease exponentially as we move away from the dominant mode. It is also apparent that only the additional modes $\ell=\ell'\pm2$ are present at $\rho=0$, while the other ones only have non-vanishing values away from the cylinder.

\subsubsection{Regularity of the solutions.}\label{sec:2+1_SolReg}

Next we study the regularity of the solutions and particularly investigate how ${\cal F}$ behaves either as a function of $\rho$ or of $\tau$.  Throughout this section, we fix the spin parameter to $\kappa=1/2$, and we choose the resolution $N_\rho=30$, $N_\theta=11$ and $N_\tau=100$.

For the next numerical calculations, we modify the ID considered in the previous section by including a nontrivial dependence on the radial coordinate. Specifically, we keep the same ID for the functions $f_0$ and $f_1$ as before, but for ${\cal F}$ we choose
\bea
\label{eq:ID_F_rho}
{\cal F}(\rho, \theta, 0 ) = f_2(\theta, 0) + \rho\, \cos(2\pi \rho/\rho_{\rm f})\,\ee^{-\rho/\rho_{\rm f}} P_{\ell'}(\cos\theta), \\
\dot{\cal F}(\rho, \theta, 0 ) = \dot{f}_2(\theta, 0) \nn
\eea
with $f_2(\theta, 0)$ and $\dot{f}_2(\theta,0)$ satisfying the regularity conditions.
Moreover, we consider single $\ell'$-mode ID with $\ell'=1,2,4$. 

In the case of $\ell'=1$-data, the corresponding regularity conditions were previously derived up to order $n=3$ in the radial expansion about the cylinder. For ID satisfying all conditions up to this order, we expect the behaviour $\Psi_{1,1}(\rho, \tau) \sim (1-\tau)^4\ln(1-\tau)$ near $\tau=1$. On the other hand, we have not established the regularity conditions for the mode $\ell'=2$ at order $n=3$. Nonetheless, following the logic of Eqs.~\eqref{eq:RegCond_n1l0} and \eqref{eq:RegCond_n2l1}, where we observe that no conditions are necessary for the regularity of the modes $\ell=0$ and $\ell=1$ at orders $n=1$ and $n=2$, respectively, it is plausible to assume that the mode $\ell=2$ requires no conditions at order $n=3$. (We already know that this is certainly correct in the Schwarzschild limit $\kappa=0$ discussed in \cite{FrauendienerHennig2018}.) Thus, it is interesting to check whether we observe the expected behaviour $\Psi_{2,2}(\rho, \tau) \sim (1-\tau)^4\ln(1-\tau)$ as $\tau\to1$ within the numerical solution. We also have not derived the regularity conditions of the mode $\ell'=4$ at order $n=3$. Thus this mode was chosen in order to test the numerical code against the expected behaviour $\Psi_{4,4}(\rho, \tau) \sim (1-\tau)^3\ln(1-\tau)$. 

In addition to these data, we also consider slightly modified single mode ID with $\ell'=5$. This is particularly interesting, because we have not derived regularity conditions for this mode at any order. Precisely because of this reason, we keep the ID for ${\cal F}(\rho, \theta, \tau)$ as in \eqref{eq:ID_F_rho}, but we work with vanishing initial values for the other auxiliary fields, i.e.\ we choose $f_0(\theta, \tau) = f_1(\theta, \tau) = 0$. In this case, we expect $\Psi_{5,5}(\rho, \tau) \sim (1-\tau)^2\ln(1-\tau)$.

Apart from the above discussed main modes $\psi_{\ell,\ell'}$ with $\ell=\ell'$, we will again observe additional excited modes. For these, the singular behaviour was not covered in the theoretical discussion, whence one aim of the numerical investigations is to explore their behaviour. 

Finally, while the described lower regularity of the solutions refers to their behaviour as functions of $\tau$ as $\tau\to1$ (i.e.\ near $\Scri^+$), we expect the solutions to be regular along the radial direction. This hypothesis will also be numerically tested. But first we investigate the behaviour in the $\tau$-direction.

\begin{figure}[ht!]\centering
\includegraphics[width=0.45\textwidth]{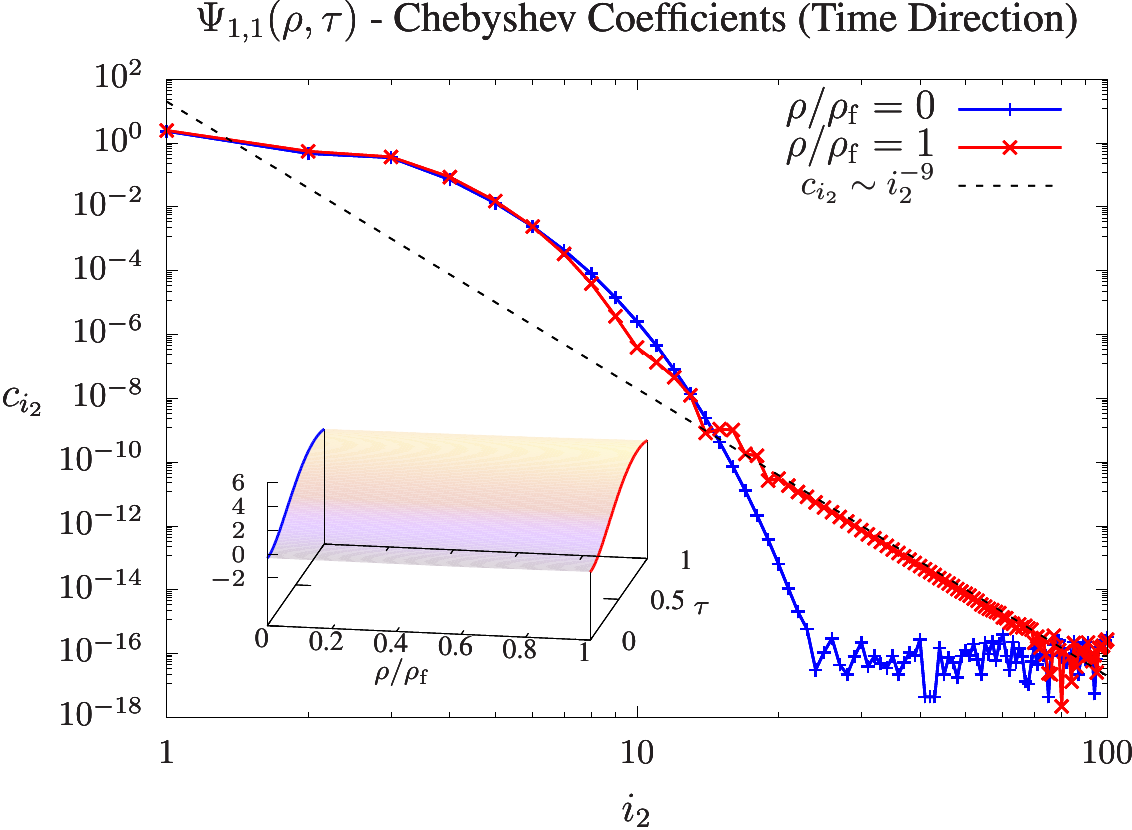}
\includegraphics[width=0.45\textwidth]{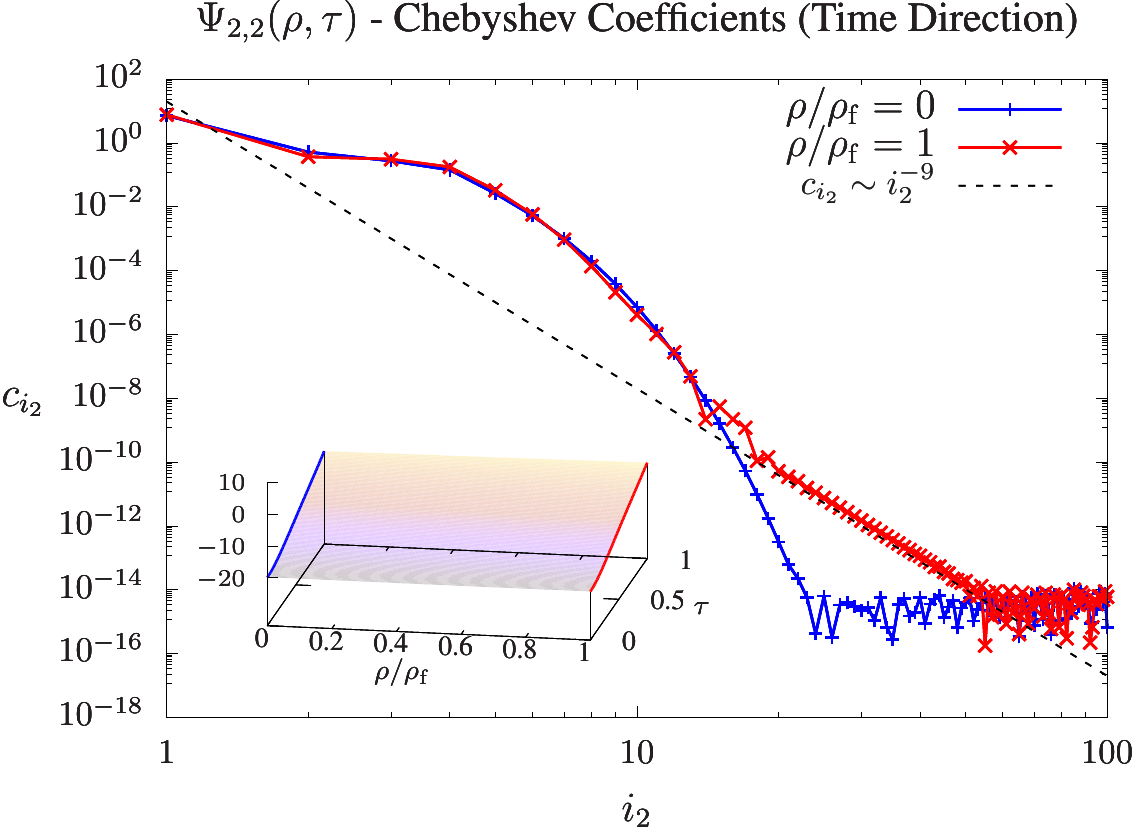}
\\[2ex]
\includegraphics[width=0.45\textwidth]{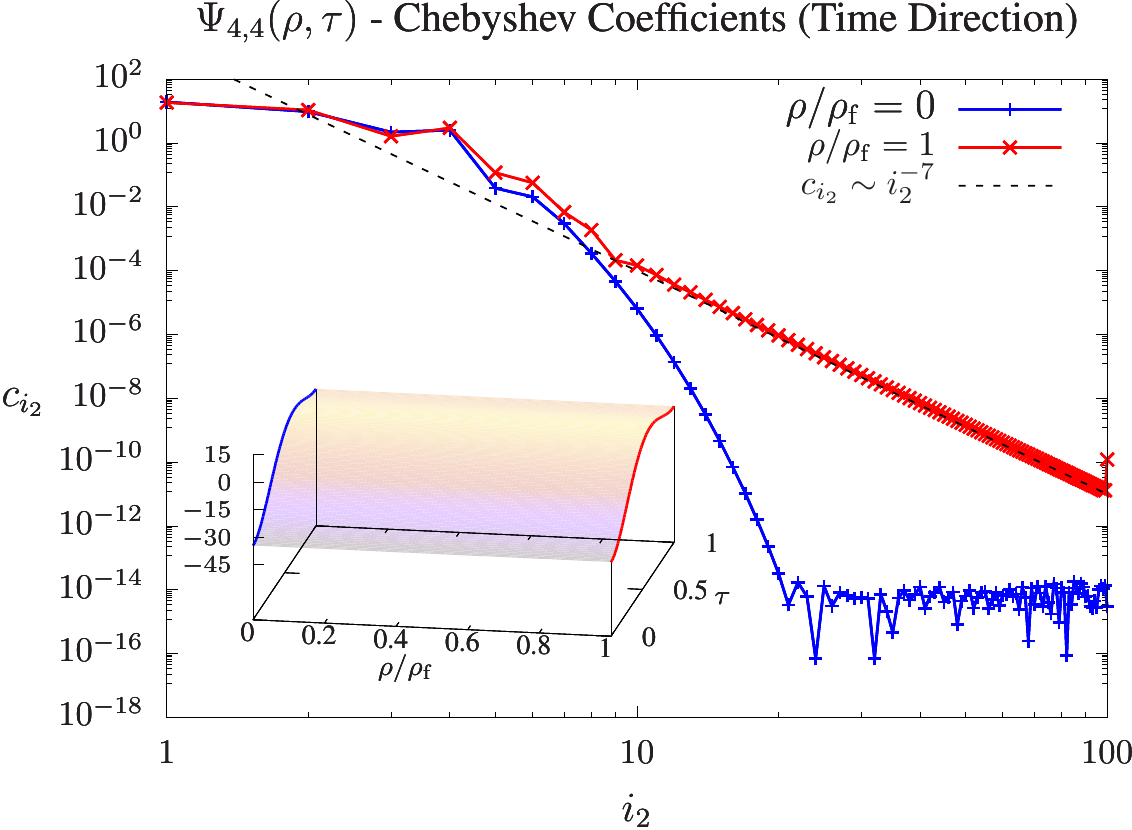}
\includegraphics[width=0.45\textwidth]{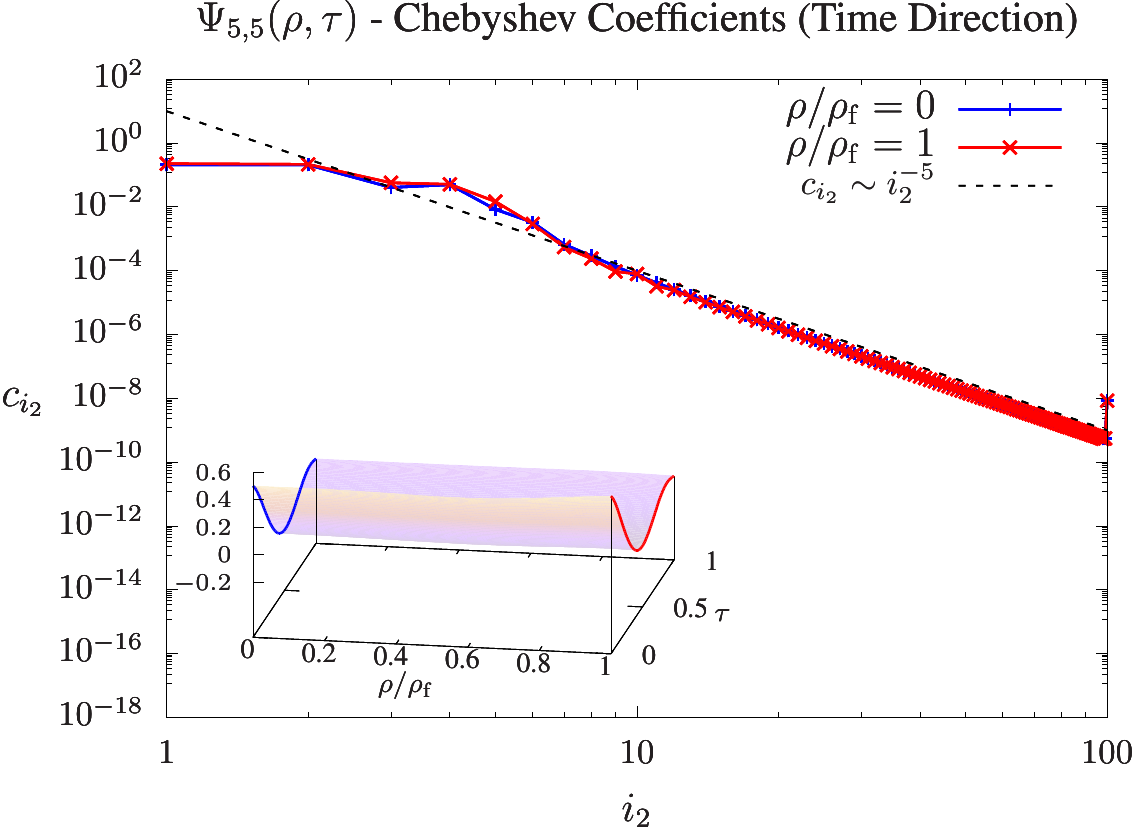}
    
 \caption{Chebyshev coefficients $c_{i_2}$ along the time direction (at a fixed radius $\rho$) associated to $\Psi_{\ell,\ell'}$ with $\ell = \ell'$. The insets display the time evolutions and delineate the curves $\rho/\rho_{\rm f} = 0$ (blue) and $\rho/\rho_{\rm f} = 1$ (red) along which the coefficients are calculated. Results are shown for a spin parameter $\kappa=1/2$. ID regularity conditions are enforced at $\rho=0$ for the angular modes $\ell'=1, 2$ and $4$. {\em Top panel:} Coefficients for $\ell=\ell'=1$ (left) and $\ell=\ell'=2$ (right) display an exponential decay at $\rho=0$ confirming the solutions' regularity at the cylinder. At $\rho=\rho_{\rm f}$, coefficients decay as $c_{i_2}\sim i_{i_2}^{-9}$ reflecting the singular behaviour $\sim(1-\tau)^4 \ln(1-\tau)$. {\em Bottom left panel:} Similar behaviour for the coefficients for $\ell=\ell'=4$. Here, the algebraic decay is $c_{i_2}\sim i_{i_2}^{-7}$ indicating a singularity in the form $\sim(1-\tau)^3 \ln(1-\tau)$. {\em Bottom right panel:} Since ID regularity conditions are not enforced for $\ell=\ell'=5$, the coefficients decay algebraically as $c_{i_2}\sim i_{i_2}^{-5}$ both at $\rho=0$ and $\rho=\rho_{\rm f}$, confirming the behaviour $\sim(1-\tau)^2 \ln(1-\tau)$.
 }
 \label{fig:cheb_time_main}
\end{figure}

\medskip
\noindent
{\bf Time direction --- main modes:} 
We study the Chebyshev coefficients along the $\tau$-direction for the main modes $\Psi_{\ell,\ell'}(\rho, \tau)$ with $\ell=\ell'=1,2,4,5$ at $\rho=0$ and $\rho=\rho_{\rm f}$. The results are shown in Fig.~\ref{fig:cheb_time_main}. 

For the three modes with $\ell=\ell'=1, 2, 4$,  we observe that the solutions are regular at $\rho=0$ as expected (because the ID satisfy the regularity conditions), which is confirmed by the exponential decay of the coefficients. On the other hand, at $\rho=\rho_{\rm f}$, we find an algebraic decay $c_{i_2}\sim i_2^{\varkappa}$, with $\varkappa=2\eta+1$, where $\eta$ is the exponent in the singular behaviour of the type $\sim(1-\tau)^\eta \ln(1-\tau)$. The numerical results show that $\varkappa = 9$ for $\ell'=1,2$, and $\varkappa=7$ for $\ell'=4$. This confirms the expected behaviour, namely singularities $\sim(1-\tau)^4 \ln(1-\tau)$ for $\ell'=1,2$ and $\sim(1-\tau)^3 \ln(1-\tau)$ for $\ell'=4$. 

For the mode $\ell=\ell'=5$, we also identify the expected behaviour. In this case, the coefficients decay algebraically with the exponent $\varkappa=5$ \emph{both} at $\rho=0$ and at $\rho=1$. This indicates that the solution everywhere has a singularity $\sim(1-\tau)^2 \ln(1-\tau)$.

\medskip
\noindent
{\bf Time direction --- coupled modes:} 
We now look at the additionally excited modes. For these, we again calculate the Chebyshev coefficients along the $\tau$-direction. By construction, the coupled modes are either absent or regular at $\rho=0$. Therefore, we only calculate the Chebyshev coefficients for two positive $\rho$-values, namely at $\rho=0.5\rho_{\rm f}$ and $\rho=\rho_{\rm f}$.

\begin{figure}[ht!]\centering
\includegraphics[width=0.49\textwidth]{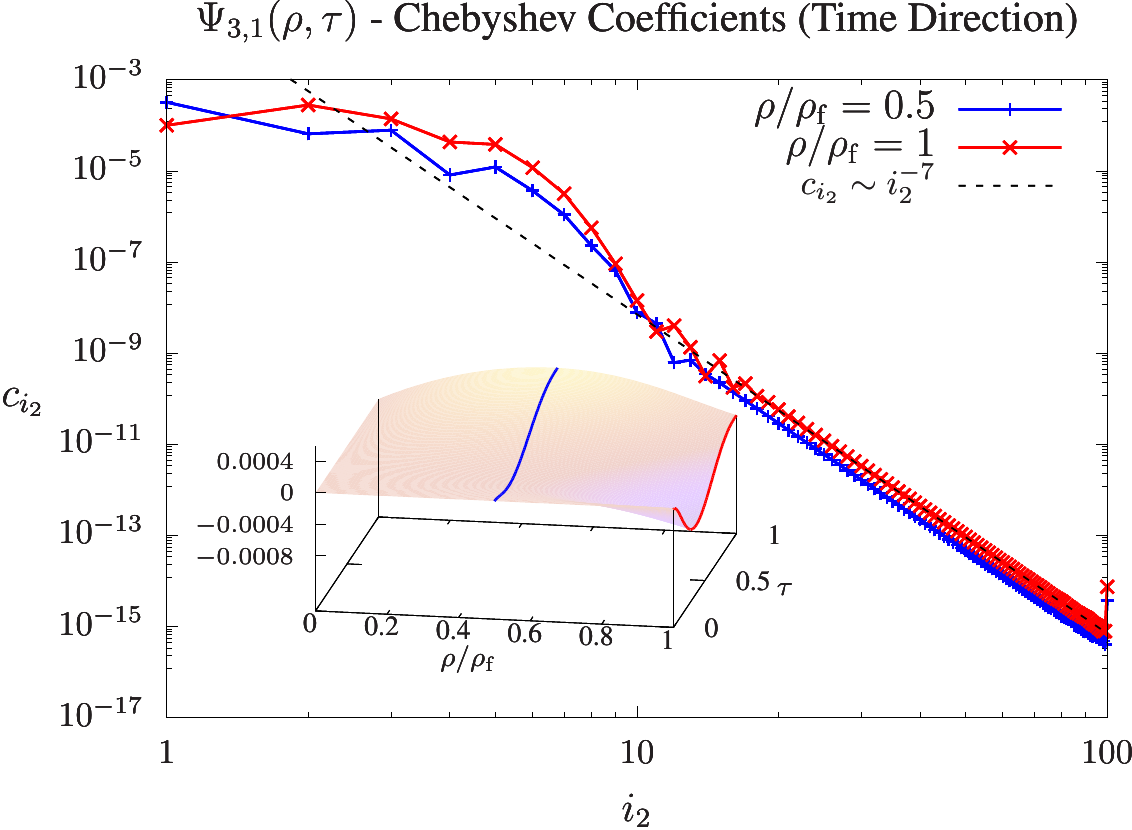}
\includegraphics[width=0.49\textwidth]{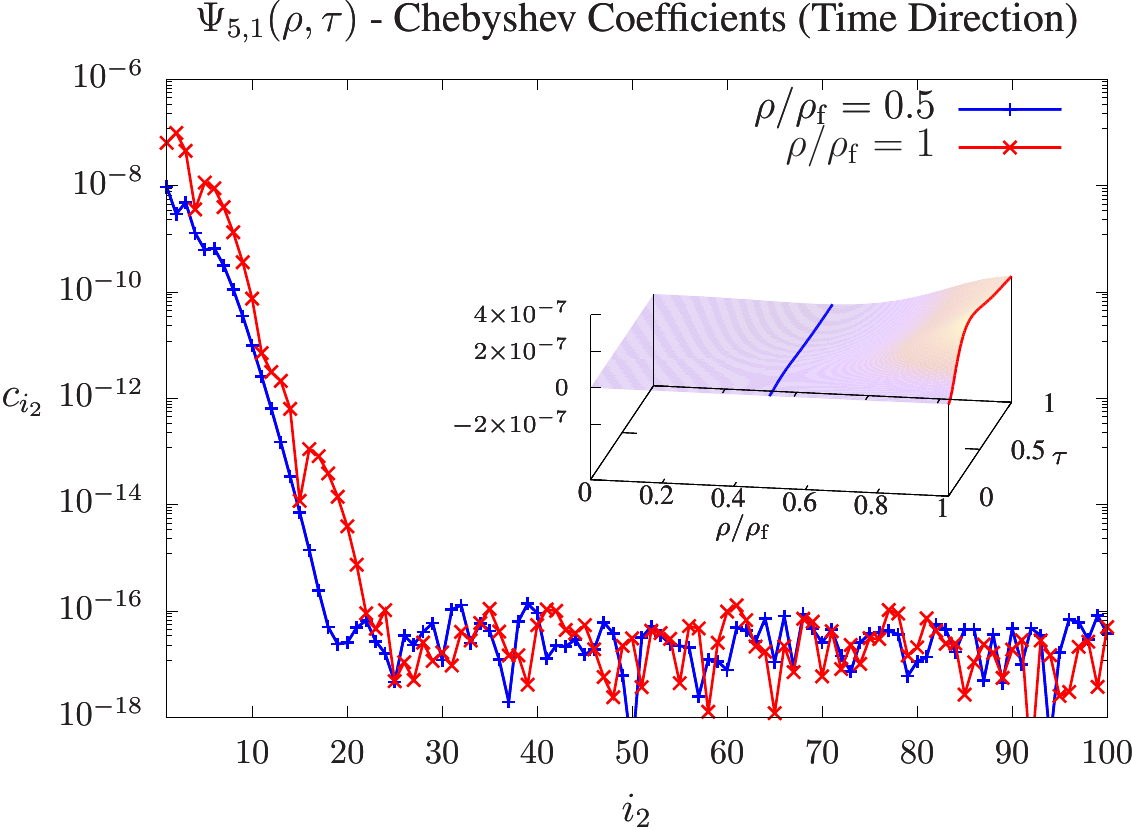}
\caption{Chebyshev coefficients $c_{i_2}$ along the time direction (at a fixed radius $\rho$) associated to the coupled modes $\Psi_{\ell,\ell'}$ with $\ell \neq \ell'$ (ID with mode $\ell'=1$). The insets display the time evolutions and delineate the curves $\rho/\rho_{\rm f} = 1/2$ (blue) and $\rho/\rho_{\rm f} = 1$ (red) along which the coefficients are calculated. Results are shown for a spin parameter $\kappa=1/2$. {\em Left panel:} Coefficients for mode $\ell=3$ decay as $c_{i_2}\sim i_{i_2}^{-7}$ reflecting the singular behaviour $\sim(1-\tau)^3 \ln(1-\tau)$. {\em Right panel:} Coefficients for mode $\ell=5$. No algebraic decay is identified above the numerical noise at $\sim10^{-16}$.
 }
 \label{fig:cheb_time_coupled_l1}
\end{figure}

\begin{figure}[ht!]\centering
\includegraphics[width=0.49\textwidth]{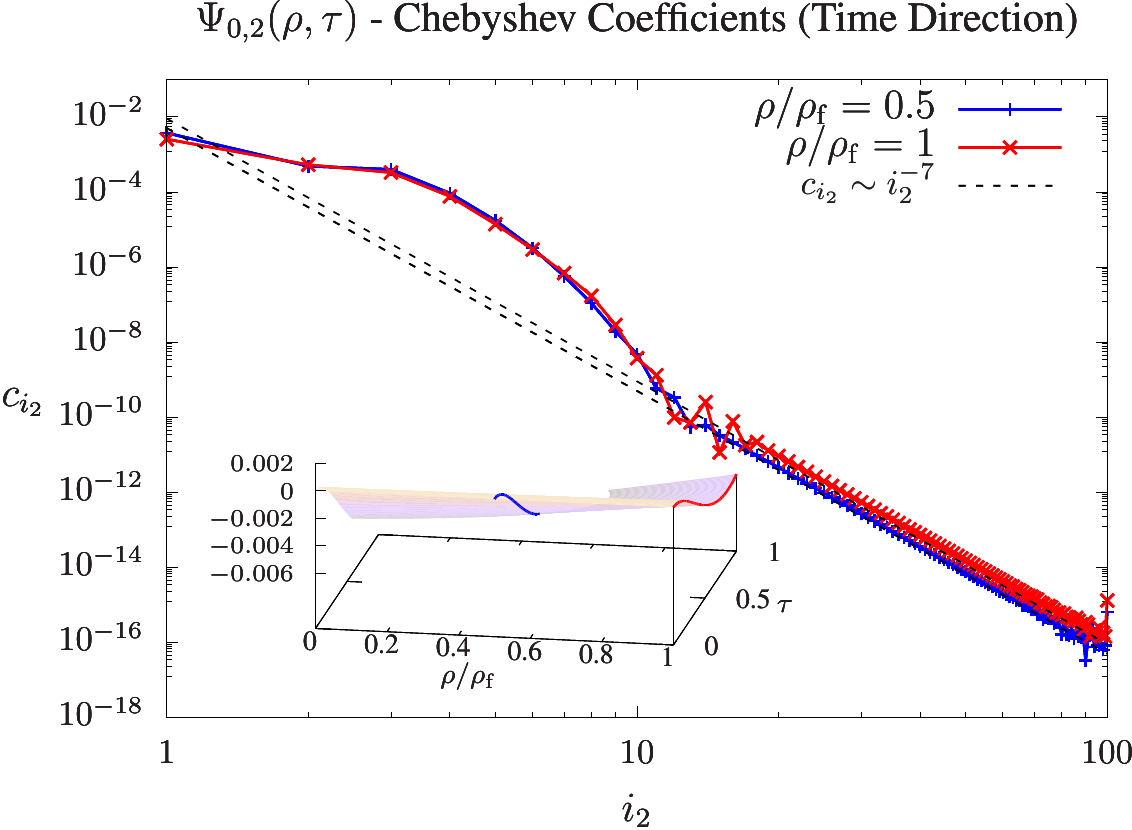}
\includegraphics[width=0.49\textwidth]{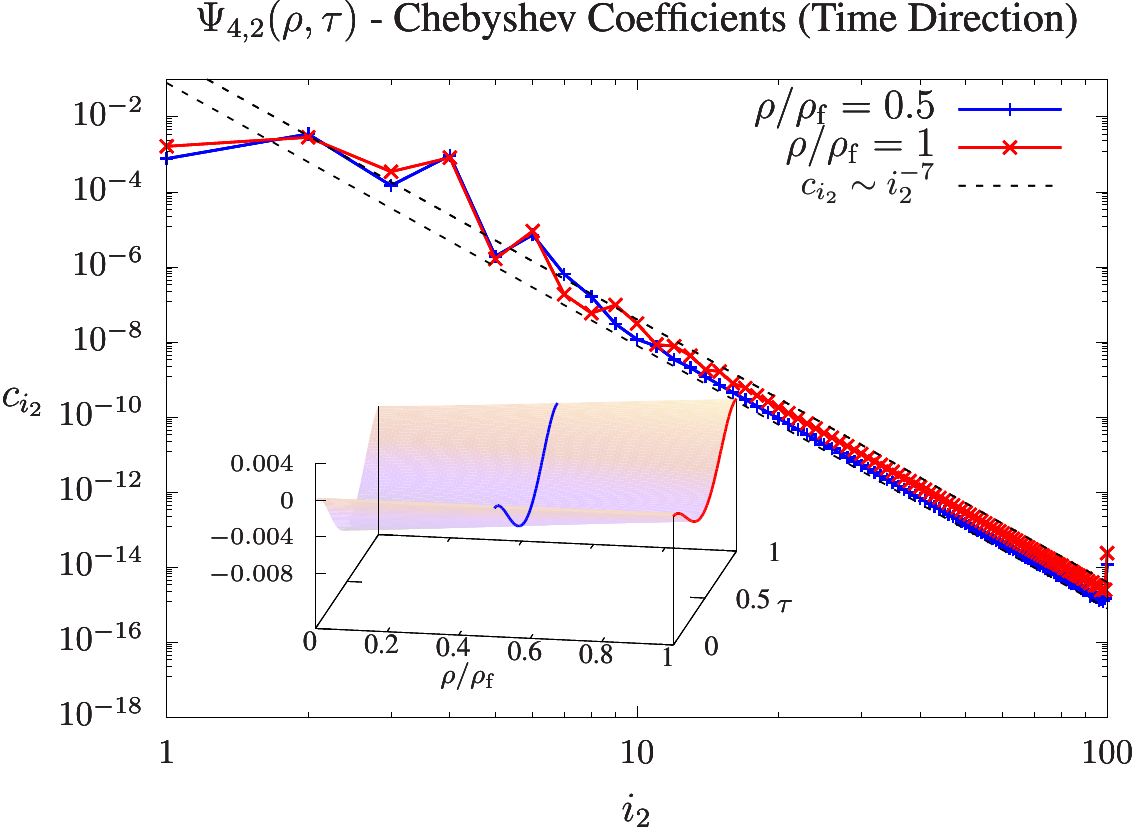}
  \\[2ex]
\includegraphics[width=0.49\textwidth]{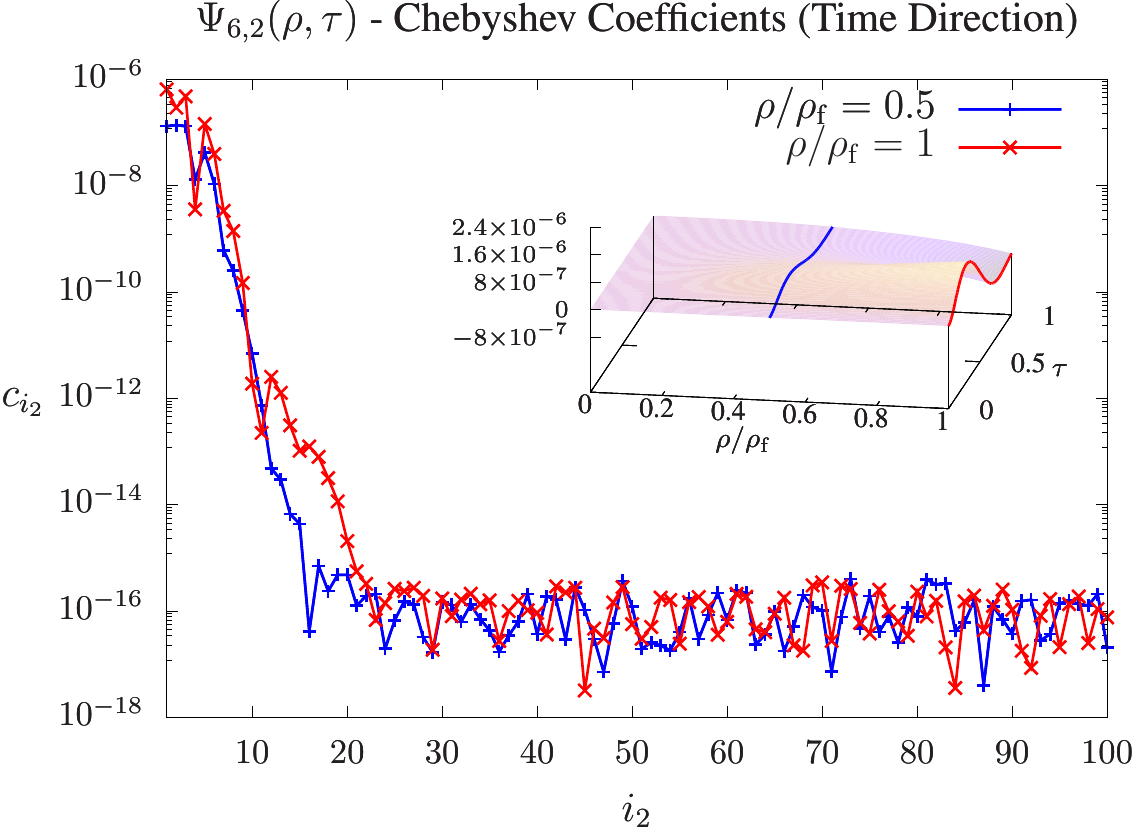}
\includegraphics[width=0.49\textwidth]{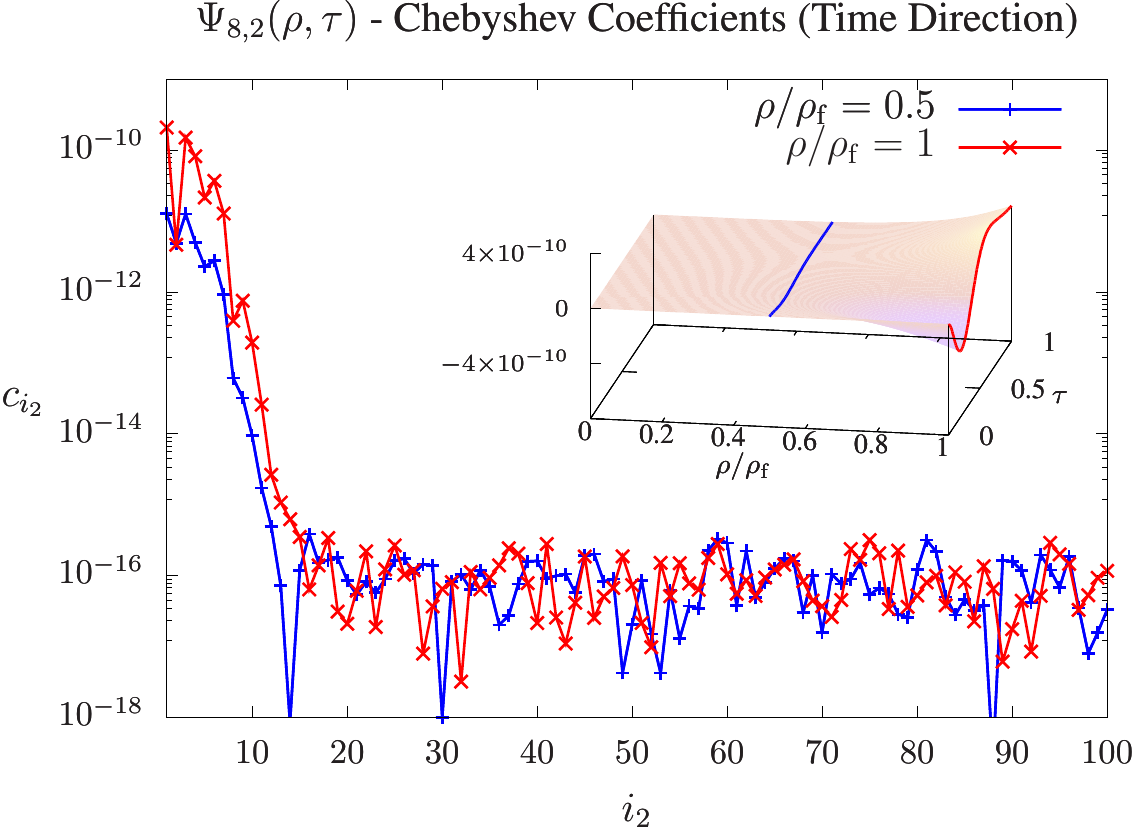}
 \caption{Chebyshev coefficients $c_{i_2}$ along the time direction (at a fixed radius $\rho$) associated to the coupled modes $\Psi_{\ell,\ell'}$ with $\ell \neq \ell'$ (ID with mode $\ell'=2$). The insets display the time evolutions and delineates the curves $\rho/\rho_{\rm f} = 1/2$ (blue) and $\rho/\rho_{\rm f} = 1$ (red) along which the coefficients are calculated. Results are shown for a spin parameter $\kappa=1/2$. {\em Top panels:} Coefficients for mode $\ell=0$ (left) and $\ell=4$ (right) decay as $c_{i_2}\sim i_{i_2}^{-7}$ reflecting the singular behaviour $\sim(1-\tau)^3 \ln(1-\tau)$. {\em Bottom panels:} Coefficients for mode $\ell=6$ (left) and $\ell=8$ (right). No algebraic decay is identified above the numerical noise at $\sim10^{-16}$.
 }
 \label{fig:cheb_time_coupled_l2}
\end{figure}

\begin{figure}[ht!]\centering
 \includegraphics[width=0.49\textwidth]{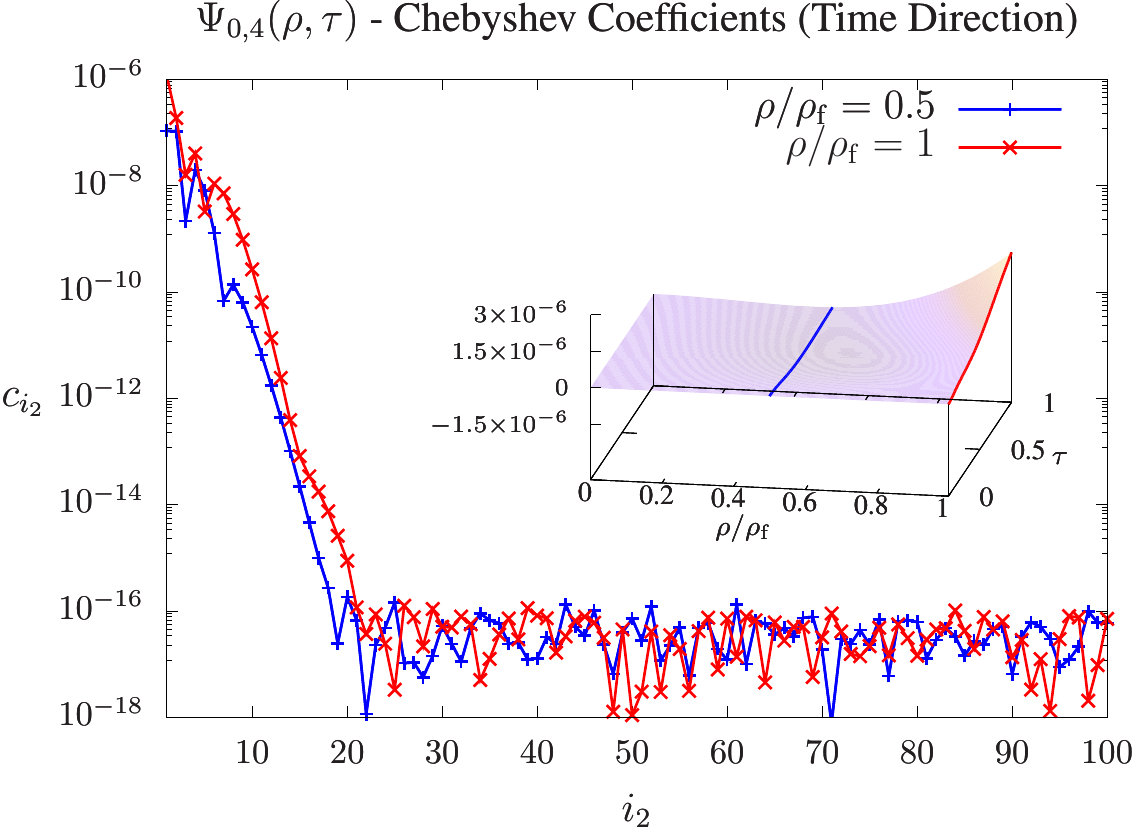}
 \includegraphics[width=0.49\textwidth]{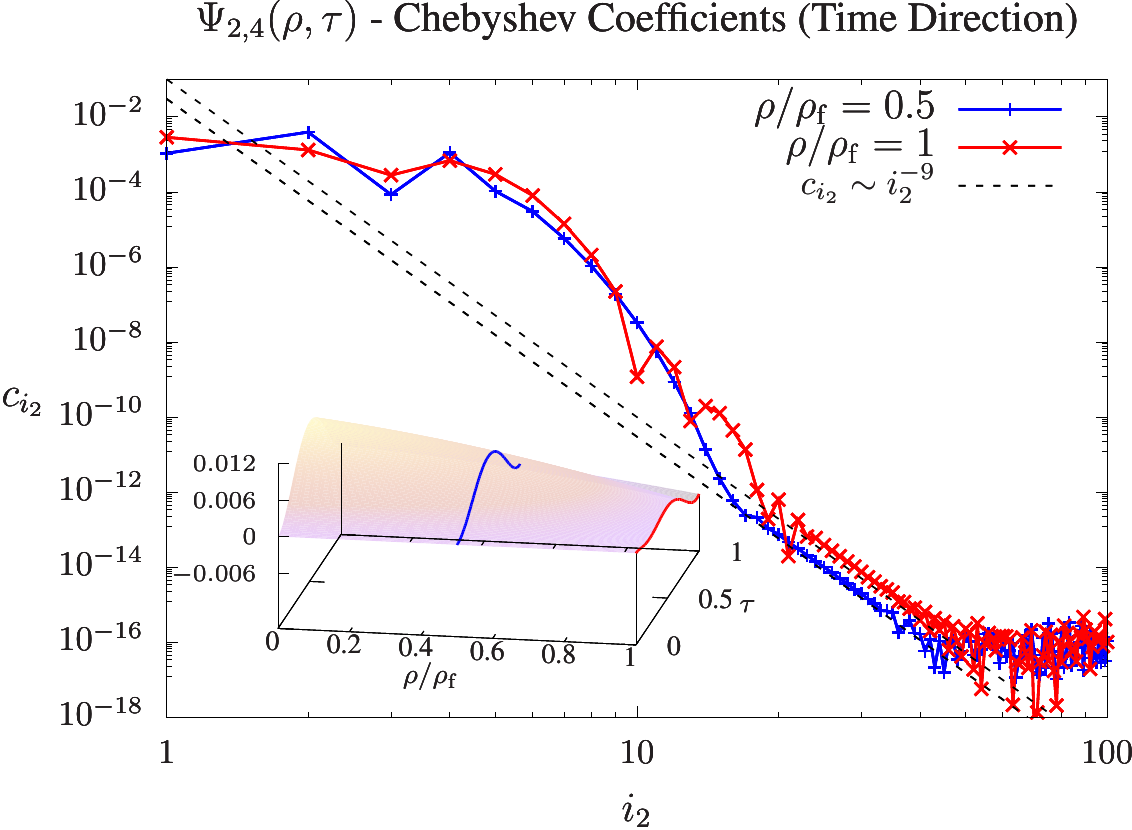}
  \\[2ex]
 \includegraphics[width=0.49\textwidth]{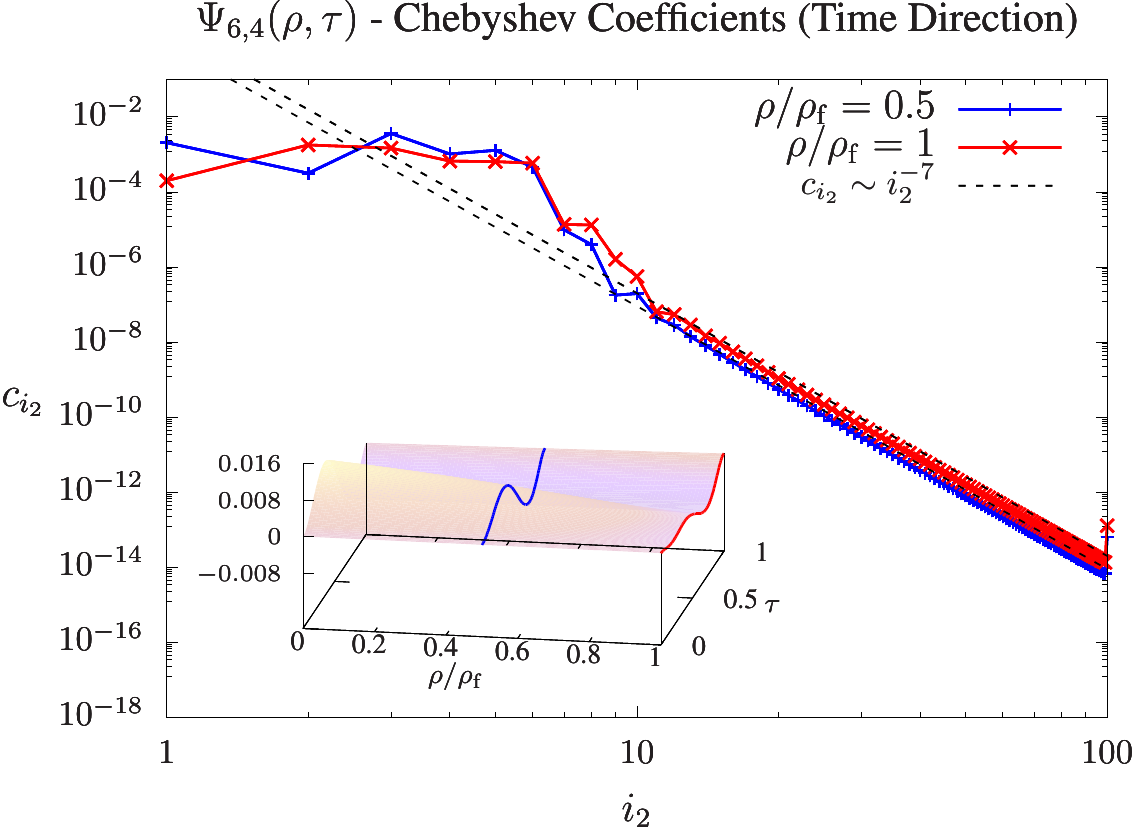}
 \includegraphics[width=0.49\textwidth]{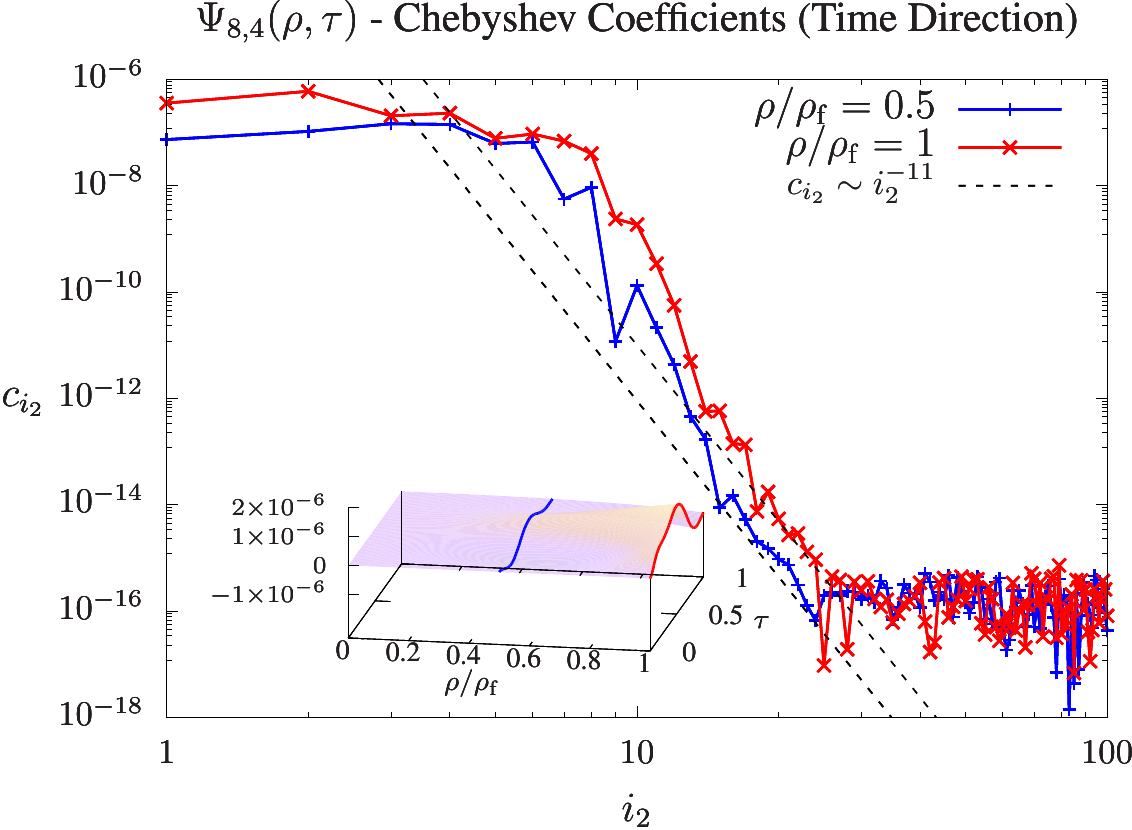}
 \caption{Chebyshev coefficients $c_{i_2}$ along the time direction (at a fixed radius $\rho$) associated to the coupled modes $\Psi_{\ell,\ell'}$ with $\ell \neq \ell'$ (ID with mode $\ell'=4$). The insets display the time evolutions and delineates the curves $\rho/\rho_{\rm f} = 1/2$ (blue) and $\rho/\rho_{\rm f} = 1$ (red) along which the coefficients are calculated. Results are shown for a spin parameter $\kappa=1/2$. {\em Top left panel:} Coefficients for mode $\ell=0$. No algebraic decay is identified above the numerical noise at $\sim10^{-16}$. {\em Top right panel:} Coefficients for mode $\ell=2$ decay as $c_{i_2}\sim i_{i_2}^{-9}$ reflecting the singular behaviour $\sim(1-\tau)^4 \ln(1-\tau)$. {\em Bottom left panel:} Coefficients for mode $\ell=6$ decay as $c_{i_2}\sim i_{i_2}^{-7}$ indicating the behaviour $\sim(1-\tau)^3 \ln(1-\tau)$. {\em Bottom right panel:} Coefficients for mode $\ell=8$ show a faint decay $c_{i_2}\sim i_{i_2}^{-11}$, which would correspond to the singular behaviour $\sim(1-\tau)^5 \ln(1-\tau)$. Such a decaying tendency is contaminated by limitations on the internal roundoff error.
 }
 \label{fig:cheb_time_coupled_l4}
\end{figure}

\begin{figure}[ht!]\centering
\includegraphics[width=0.49\textwidth]{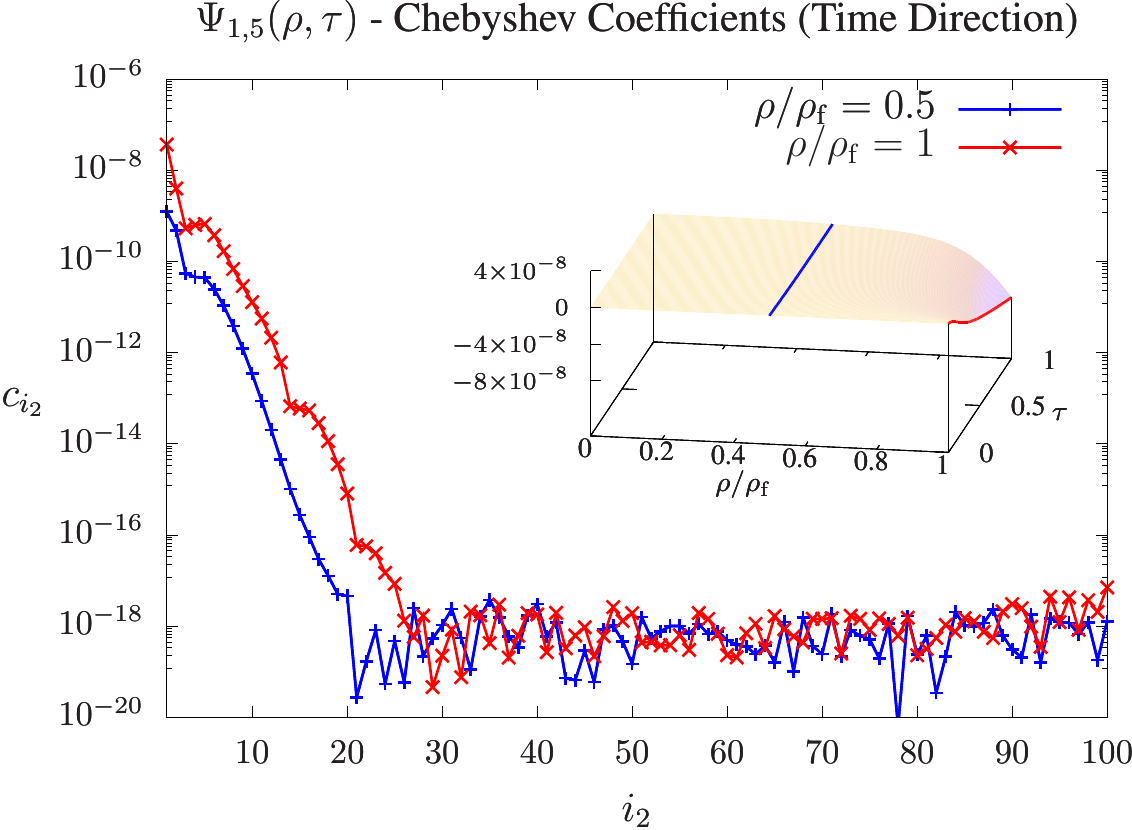}
\includegraphics[width=0.49\textwidth]{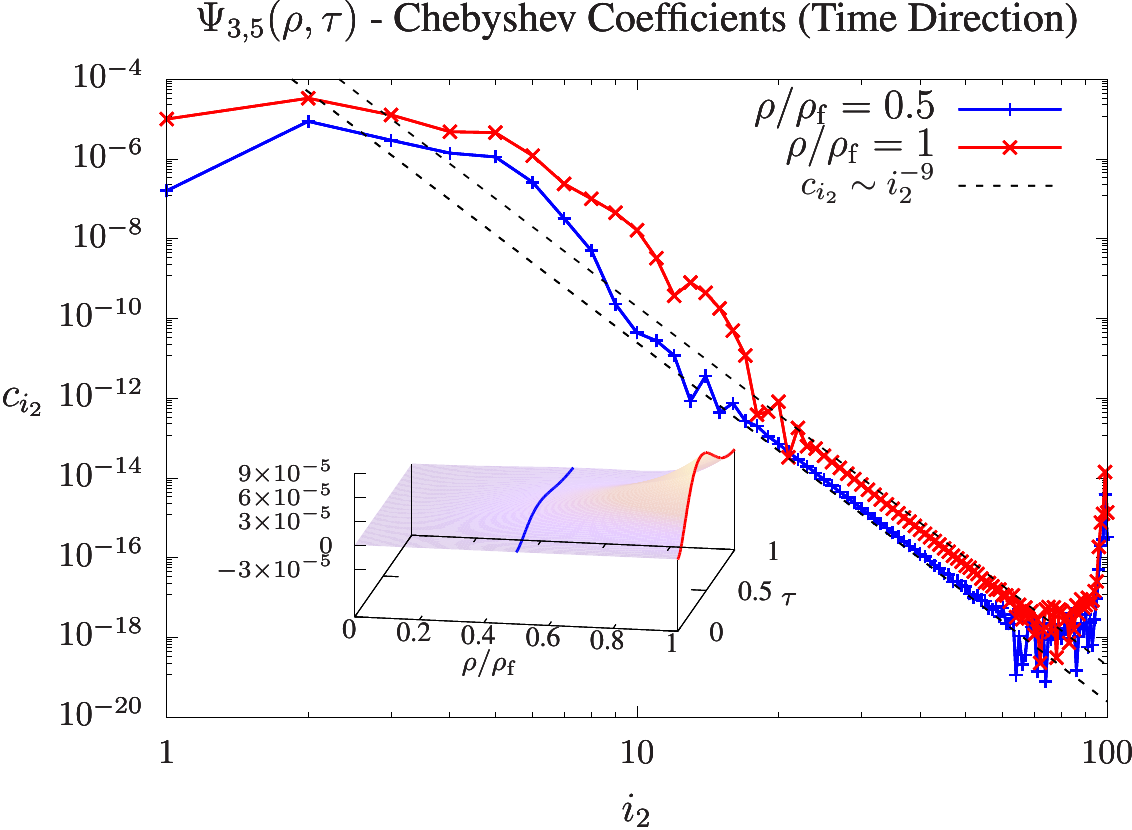}
  \\[2ex]
\includegraphics[width=0.49\textwidth]{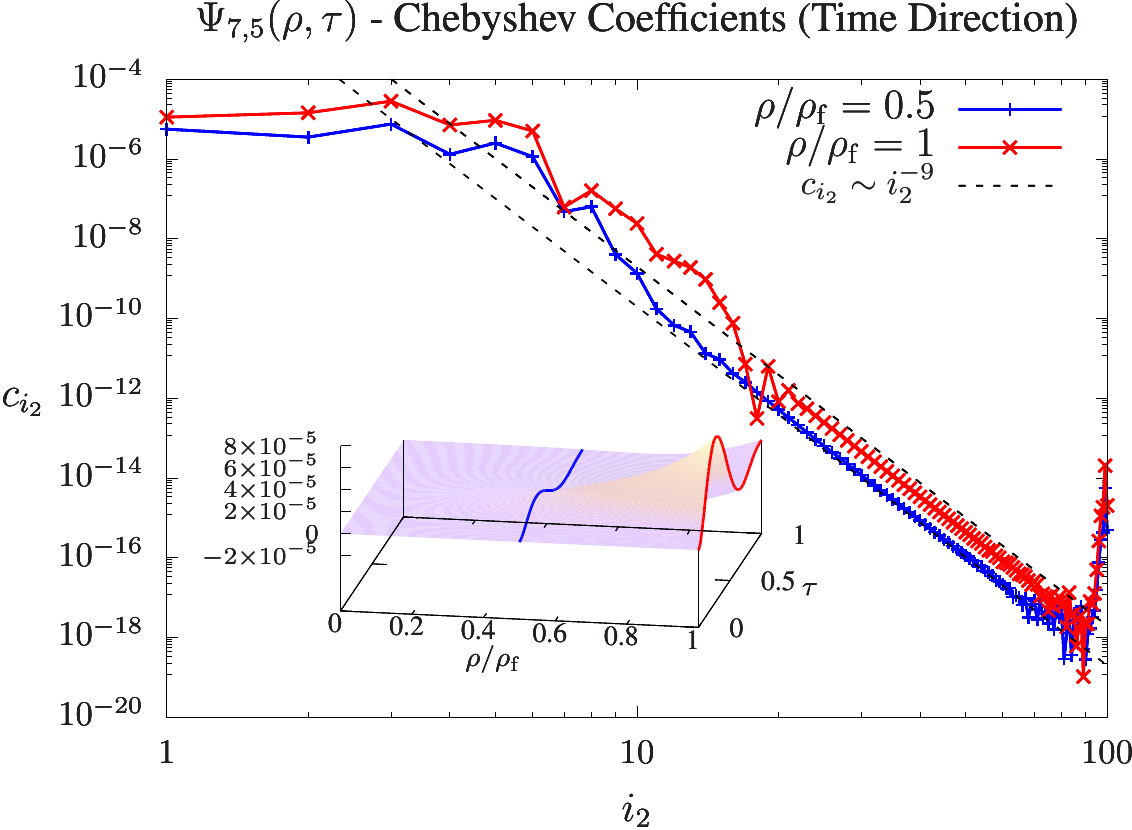}
\includegraphics[width=0.49\textwidth]{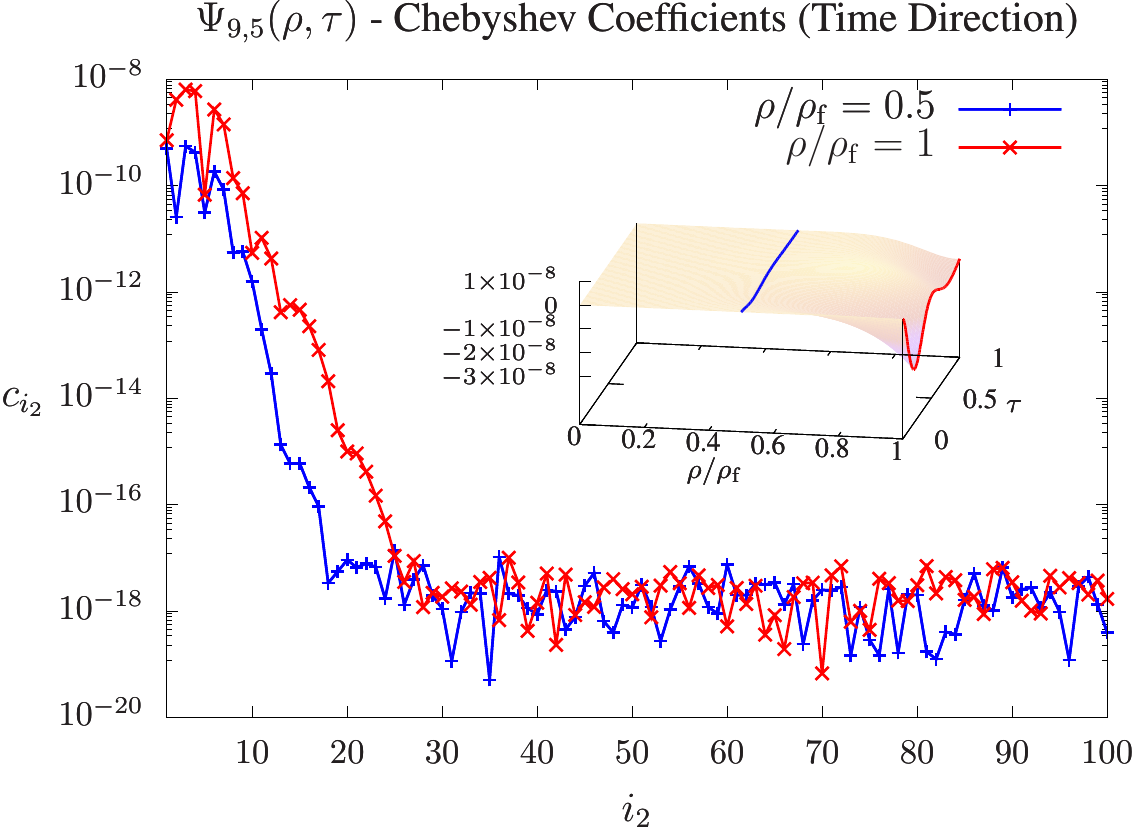}
 \caption{
 Chebyshev coefficients $c_{i_2}$ along the time direction (at a fixed radius $\rho$) associated to the coupled modes $\Psi_{\ell,\ell'}$ with $\ell \neq \ell'$ (ID with mode $\ell'=5$). The insets display the time evolutions and delineates the curves $\rho/\rho_{\rm f} = 1/2$ (blue) and $\rho/\rho_{\rm f} = 1$ (red) along which the coefficients are calculated. Results are shown for a spin parameter $\kappa=1/2$. No algebraic decays are identified above the numerical noise at $\sim10^{-16}$ for $\ell=1$ (top left panel) and $\ell=9$ (bottom right panel). {\em Top right panel:} Coefficients for mode $\ell=3$ decay as $c_{i_2}\sim i_{i_2}^{-9}$ reflecting the singular behaviour $\sim(1-\tau)^4 \ln(1-\tau)$. {\em Bottom left panel:} Coefficients for mode $\ell=7$ decay as $c_{i_2}\sim i_{i_2}^{-9}$ indicating the behaviour $\sim(1-\tau)^4 \ln(1-\tau)$.}
 \label{fig:cheb_time_coupled_l5}
\end{figure}

For our single mode ID with $\ell'=1,2,4,5$, we report on the exponent $\varkappa$ for the algebraic decay $c_{i_2}\sim i_0^{\varkappa}$ and the corresponding value $\eta$ for the singularity of the form $\sim(1-\tau)^\eta \ln(1-\tau)$. The Chebyshev coefficients are shown in Figs.~\ref{fig:cheb_time_coupled_l1}-\ref{fig:cheb_time_coupled_l5}.

It turns out that we can read off $\varkappa$ with confidence for $\ell=\ell'\pm 2$. 
On the other hand, for the (very weak) modes with $\ell > \ell' + 2$ or $\ell < \ell' - 2$, the analysis is not conclusive. In those cases, the Chebyshev coefficients show an apparent exponential decay. However, in a more conservative interpretation, we refrain from confirming that the solution is regular. Indeed, we could not identify the exponents $\varkappa$ for these higher modes due to the limit set by machine round-off error for double-precision floating-point operations. 
For instance, the coefficients for the mode $\Psi_{8,4}(\tau,\rho)$ show a behaviour resembling $\varkappa=11$ for a short range of values of the index $i_2$, but then quickly approach zero, see the last panel in Fig.~\ref{fig:cheb_time_coupled_l4}. Since the interval with algebraic decay is quite small, we cannot conclude that a decay with $\varkappa=11$ is the correct overall behaviour of this mode.

Nevertheless, the following results are obtained with confidence:
\bit
\item $\ell'=1$ (Fig.~\ref{fig:cheb_time_coupled_l1}): The excited mode with $\ell=3$ shows $\varkappa = 7$ ($\eta=3$). 

\item $\ell'=2$ (Fig.~\ref{fig:cheb_time_coupled_l2}): The excited modes with $\ell=0$ and $\ell=4$ both show $\varkappa = 7$ ($\eta=3$).

\item $\ell'=4$ (Fig.~\ref{fig:cheb_time_coupled_l4}): The excited mode with $\ell=2$ shows $\varkappa = 9$ ($\eta=4$), and for the mode with $\ell=6$ we find $\varkappa = 7$ ($\eta=3$). 

\item $\ell'=5$ (Fig.~\ref{fig:cheb_time_coupled_l5}): The excited modes with $\ell=3$ and $\ell=7$ both show $\varkappa = 9$ ($\eta=4$).
\eit

\begin{figure}[ht!]
\includegraphics[width=0.49\textwidth]{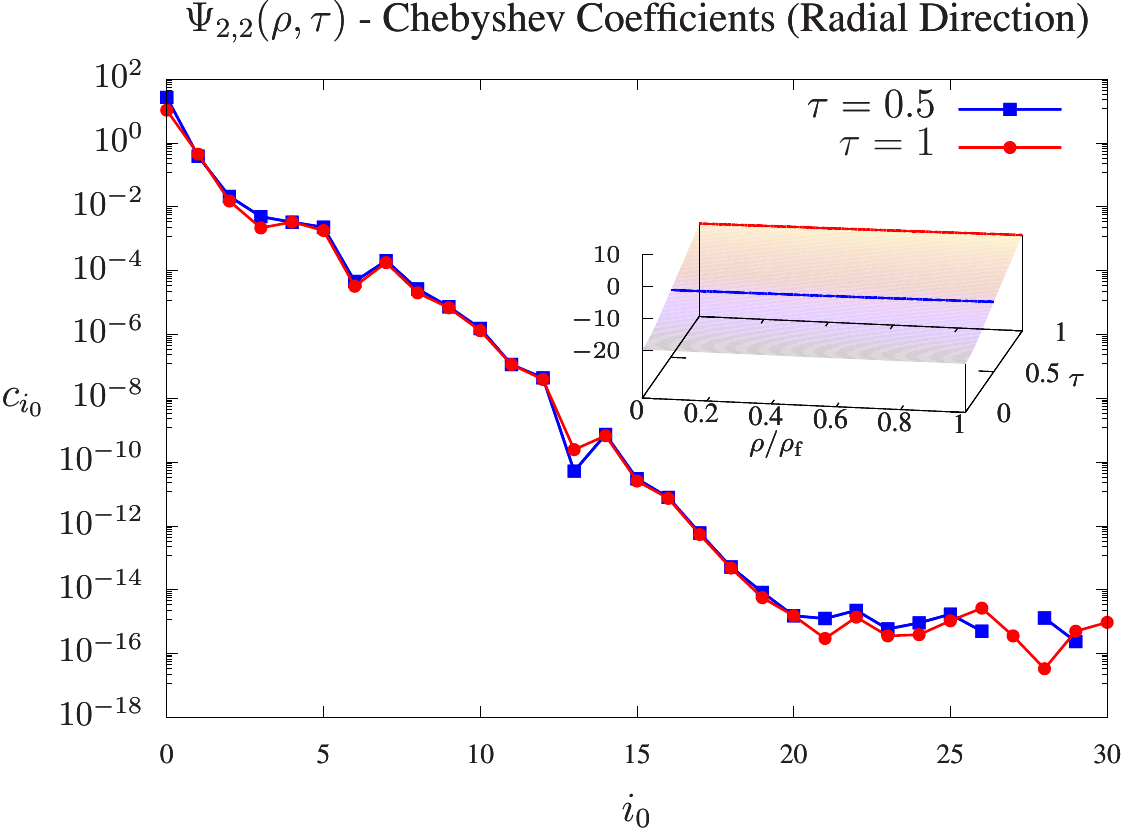}
\includegraphics[width=0.49\textwidth]{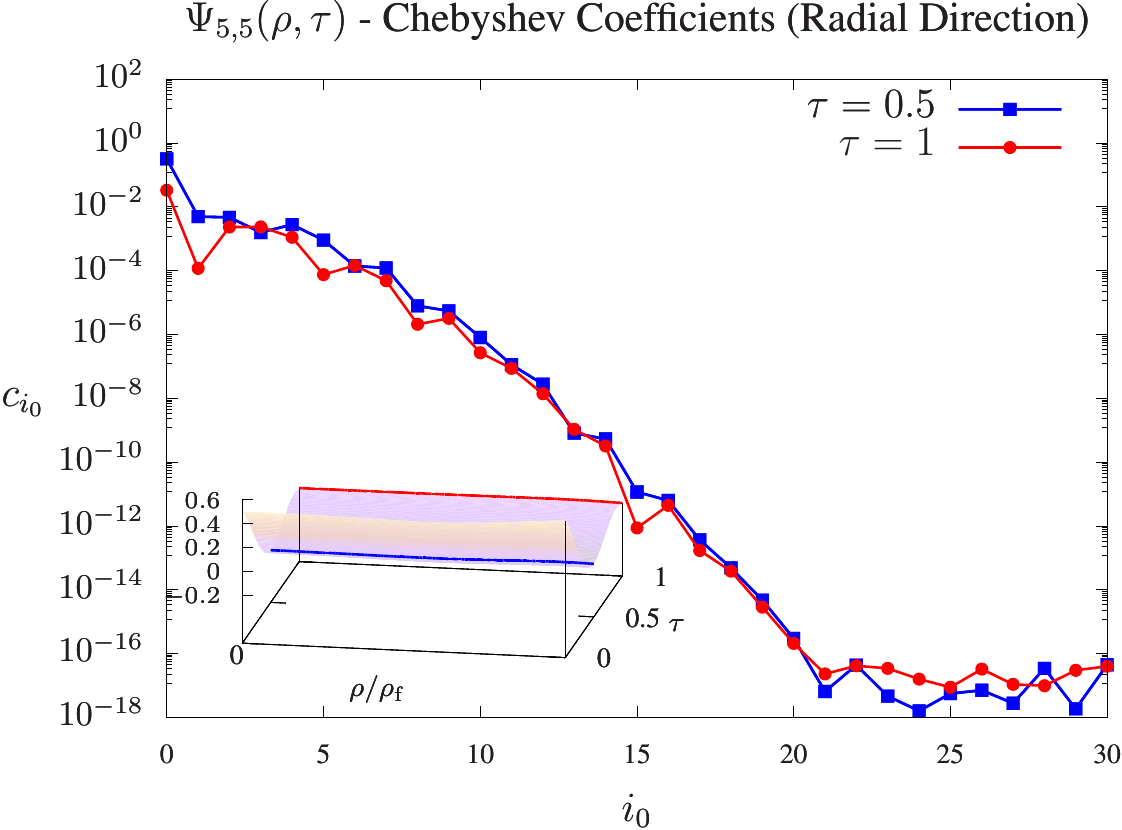}
\\[2ex]
\includegraphics[width=0.49\textwidth]{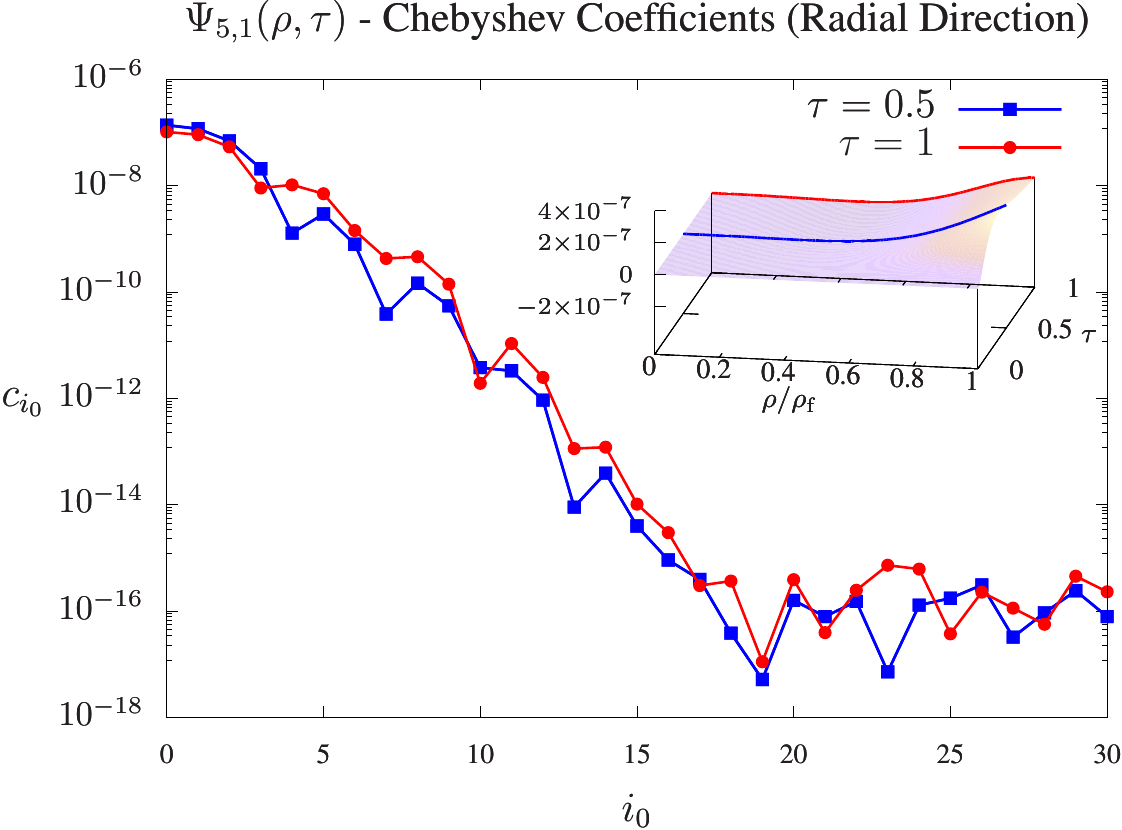}
\includegraphics[width=0.49\textwidth]{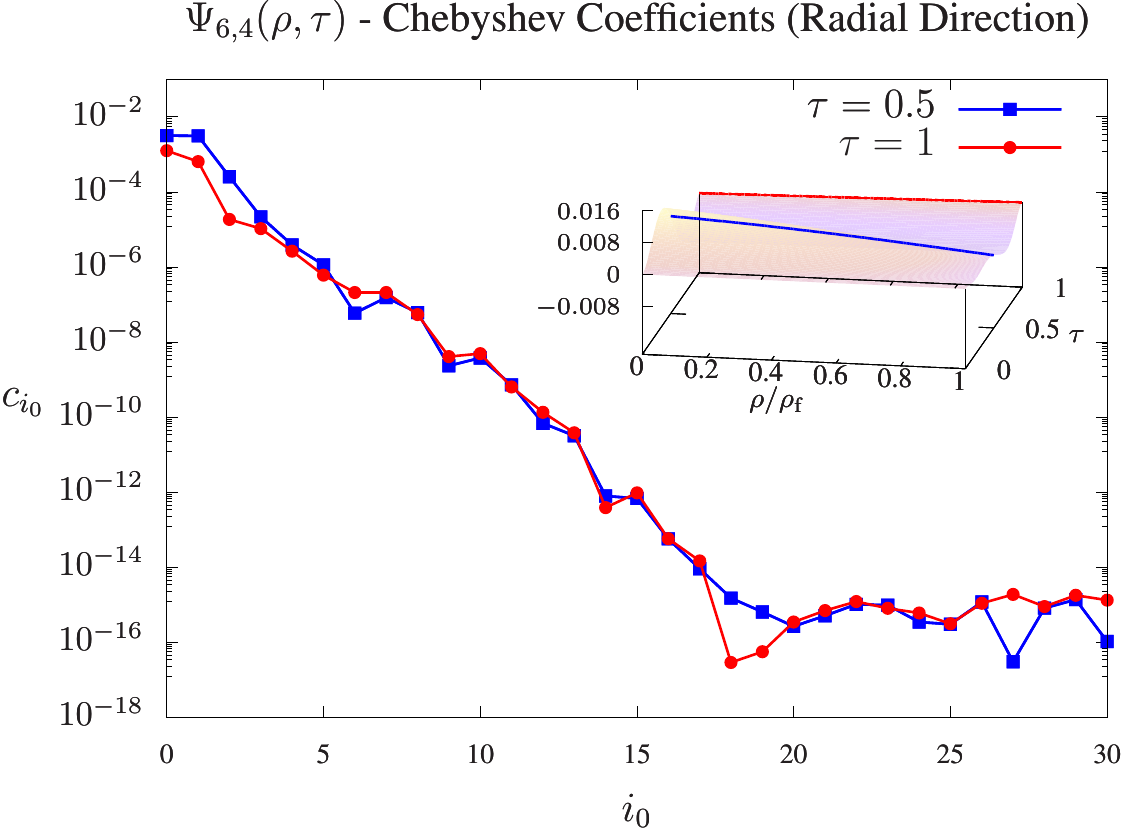}
\caption{Chebyshev coefficients $c_{i_2}$ along the radial direction (at a fixed time $\tau$) associated to different modes $\Psi_{\ell,\ell'}$. The insets display the time evolutions and delineates the curves $\tau = 1/2$ (blue) and $\tau = 1$ (red) along which the coefficients are calculated. Results are shown for a spin parameter $\kappa=1/2$. All coefficients decay exponentially indicating the solutions' regularity with respect to the coordinate $\rho$. {\em Top panels:} projections with main modes $\ell=\ell'$ for ID with $\ell' = 2$ (left) and $\ell'=5$ (right).  {\em Bottom panels:} Coupled modes with $\ell\neq \ell'$. ID $\ell'=1$ exciting mode $\ell=5$ (left) and ID $\ell'=4$ exciting mode $\ell=6$ (left).
  }
  \label{fig:cheb_radial}
\end{figure}

\medskip
\noindent
{\bf Radial direction:} Finally, we investigate the regularity of the solutions along the radial direction, where we do not expect any logarithmic terms. Some  representative examples for modes $\Psi_{\ell,\ell'}(\rho, \tau)$ are given Fig.~\ref{fig:cheb_radial}, which shows two main modes ($\ell=\ell'$) and two excited modes ($\ell\neq\ell'$). 

The top panels display the Chebyshev coefficients in the $\rho$-direction for $\Psi_{2,2}$ and $\Psi_{5,5}$ both at $\tau=0$ and $\tau=1$. The former reproduces the feature of our ID with a single $\ell'$-mode, whereas the latter shows that the same degree of regularity is kept during the whole evolution. 

The figure also displays the excited modes $\Psi_{5,1}$ and $\Psi_{6,4}$ in the bottom panels. Since these modes are absent at $\tau=0$, the Chebyshev coefficients in the $\rho$-direction are calculated at $\tau=0.5$ and $\tau=1$. The exponential decay of the coefficients again confirms the regularity of the solution.

\section{Discussion}\label{sec:discussion}

We have constructed a conformal compactification of the Kerr spacetime that includes a part of future null infinity $\Scri^+$ and that resolves spacelike infinity by blowing it up to Friedrich's cylinder representation. Then we have studied the conformally invariant wave equation on this background. Thereby, our focus was on the behaviour of the solutions on and near the cylinder. Due to the similar structure of characteristics of the wave equation and the conformal field equations, this analysis provides a useful model and a first numerical test for behaviour that can be expected for solutions to the full Einstein equations as well.

A main point of our investigations was to find out how the spin parameter $a$ in the Kerr solution influences the behaviour of solutions to the wave equation as compared to the situation on a Schwarzschild background ($a=0$), which was studied in \cite{FrauendienerHennig2017, FrauendienerHennig2018}. Another important point was to demonstrate that that the fully pseudospectral numerical scheme, which was able to produce highly accurate numerical solutions to the effectively $(1+1)$-dimensional problem on a Schwarzschild background, can also be applied to the much more intricate $(2+1)$-dimensional simulation of axisymmetric solutions to the Kerr wave equation. Once again, we observe that highly accurate numerical solutions can be obtained.

As far as the qualitative behaviour of the solutions is concerned, both our analytical investigations on the cylinder and our numerical simulations show that main features of the solutions are independent of the spin parameter $a$: There are special solutions that are regular in the entire spacetime, but apart from these, generic solutions are singular to a certain degree (in the sense of a finite $C^k$-regularity). The analytic considerations on the cylinder show that, exactly as in the Schwarzschild case, solutions contain logarithmic terms of the form $(1-\tau)^\eta\ln(1-\tau)$, with some $\eta\ge0$, as the critical set at $\tau=1$ is approached. However, by suitably restricting the initial data, singular terms can be removed at arbitrarily  many orders in a Taylor expansion about the cylinder at $\rho=0$. For that purpose, an increasing number of regularity conditions need to be satisfied by the initial data, if more orders are required to be regular. Our numerical investigations of the fall-off behaviour of the Chebyshev coefficients also show that these singularities are not restricted to the cylinder. We find --- again in perfect agreement with the Schwarzschild case --- that singularities `spread' from the cylinder to $\Scri^+$ via the critical set. Consequently, the solutions are generally singular as $\Scri^+$ at $\tau=1$ is approached. On the other hand, solutions have no singularities in the region $\tau<1$, i.e.\ the singularities are concentrated on the critical set and $\Scri^+$.

Besides all these similarities of the wave equations on a Schwarzschild and a Kerr background, an important difference is the coupling of different angular modes~$\ell$. In the Schwarzschild case, if we choose initial data at $\tau=0$ that only contain certain $\ell$-modes, then the solution will contain exactly the same modes in the region $\tau>0$. In the Kerr case, however, there is a coupling between the modes, and the solutions generally develop additional modes. The analytical investigations on the cylinder show that, if the Taylor order $n$ contains a mode $\ell$, then the next order $n+1$ generally contains not only this mode $\ell$, but also the additional modes $\ell\pm2$. With increasing orders $n$, an entire chain of excitations produces many additional modes, as illustrated in Fig.~\ref{fig:ModeCouplingSpatialInft} above. If we move away from the cylinder, then all Taylor orders are simultaneously required. This means that initial data with a single mode $\ell'$ usually lead to solutions with all $\ell$-modes of the same parity. However, since the amplitude of the excited modes rapidly decreases with growing $|\ell-\ell'|$, the most important contributions to the solution  have $\ell$-values somewhere near those in the initial data.

We also observe that the fully pseudospectral time-evolution is a very powerful method for studies of the Kerr wave equation. We are able to obtain highly accurate numerical solutions in this $(2+1)$-dimensional setting, even when (sufficiently weak) logarithmic singularities are present. With a modification of the coordinates near $\Scri^+$, similarly to the considerations in the Schwarzschild case \cite{FrauendienerHennig2018}, it should also be possible in future studies to accurately resolve solutions with singularities at even lower orders than considered here. Another interesting future generalisation of our investigations could be an analysis of the wave equation on a portion of the Kerr spacetime that contains not only a part of $\Scri^+$ and the cylinder, but also a part of past timelike infinity $\Scri^-$. This would allow us to directly study the interaction between the `late ingoing waves' (originating from $\Scri^-$) and the `early outgoing waves' (ending at $\Scri^+$) and the role of the cylinder in this transition from past to future null infinity. 
Apart from that, further studies are required on more generic $3D$ evolutions on the Kerr background, in which the conformal scalar field depends on the azimuthal $\phi$ angle as well. Thanks to the existence of an axial Killing vector, a Fourier decomposition in the azimuthal  coordinate $\phi$ in the form $f( \rho,\theta, \phi, \tau) \sim f_m( \rho,\theta, \tau)\,\e^{\ii m \phi}$ should again reduce the problem to a $2+1$ evolution for each individual mode $m$. Thus, it is expected that more general regularity conditions are required to cover all possible degrees of freedom for initial data that produce solutions of a certain regularity. In the context of the mode coupling discussed in this article, we expect that initial data with a single mode $(\ell', m')$ will excite modes $(\ell, m)$ with different axial numbers $\ell\neq \ell'$, but still with the same azimuthal number $m'=m$.

Finally, since all previous work with the fully pseudospectral scheme on various linear and nonlinear problems has always been very successful (whenever the solutions are sufficiently well-behaved, or when locations and type of singularities are known for an appropriate treatment), we do not see any principle obstacles to considering a variety of more involved problems in the future. For instance, Ref.~\cite{Gray2020} shows the separability of conformally coupled scalar field equations in general rotating black-hole spacetimes, opening the possibilities for studies in higher dimensions. Another interesting application could be the investigation of gravitational waves in a cylindrically-symmetric gravitational collapse~\cite{Mena2015}.

\appendix
\section{Geometrical perspective on the conformal compactificaton}
\label{sec:null_tetrad}

We discuss the coordinate transformation introduced in Sec.~\ref{sec:compactification} above from the perspective of a null tetrad basis. 
A null tetrad $(\tilde{\vec l}, \tilde{\vec n}, \tilde{\vec m}, \bar{\tilde{\vec m}})$ adapted to the Kerr metric in Boyer--Lindquist coordinates $(\tilde r,\theta,\varphi,\tilde t)$ is~\cite{Teukolsky:1972my,Teukolsky:1973ha}
\bea
\fl
\tilde{\vec l} = \zeta\Bigg(\partial_{\tilde r} + \frac{a}{\tilde{\Delta}}\,\partial_{\varphi} + \frac{\tilde{\Sigma}_0}{\tilde\Delta}\,\partial_{\tilde t}\Bigg), \quad
\tilde{\vec n} = \zeta^{-1}\Bigg(-\frac{\tilde\Delta}{2 \tilde\Sigma}\, \partial_{\tilde r} + \frac{a}{2 \tilde\Sigma}\, \partial_{\varphi} + \frac{\tilde{\Sigma}_0}{2 \tilde\Sigma}\,\partial_{\tilde t} \Bigg), \\
\fl
\tilde{\vec m} = \frac{1}{\sqrt{2}(r+\ii a \cos\theta)} \Bigg(\partial_{\theta} + \frac{\ii}{\sin\theta}\,\partial_{\varphi} + \ii a \sin\theta \,\partial_{\tilde t} \Bigg),
\eea
where $\zeta$ corresponds to the boost freedom, which will be fixed later, and where 
$\tilde{\Sigma}_0 := \tilde\Sigma|_{\theta=0} = r^2+a^2$.
 
Our objective is to construct a conformal null tetrad basis $(\vec l, \vec n, \vec m, \bar{\vec m})$ adapted to a coordinate system $(\rho, \theta, \phi, \tau)$ such that the outgoing null vector $\vec l$ is aligned with the time coordinate vector $\partial_\tau$. 

A first simplification is achieved with the transformation $\varphi\mapsto\phi$ in \eqref{eq:phitrans}. This corresponds to the change of basis vectors 
$\partial_{\tilde r}+\frac{a}{\tilde\Delta}\partial_{\varphi}\to\partial_{\tilde r}$ and hence effectively removes the axial component of $\tilde{\vec l}$. In the new coordinates $(\tilde r,\theta,\phi,\tilde t$\,), the null tetrad reads
\bea
\tilde{\vec l} = \zeta\Bigg(\partial_{\tilde r} + \frac{\tilde{\Sigma}_0}{\tilde\Delta} \, \partial_{\tilde t}\Bigg), \quad
\tilde{\vec n} = \zeta^{-1}\Bigg( -\frac{\tilde\Delta}{2 \tilde\Sigma} \, \partial_{\tilde r} + \frac{a}{\tilde\Sigma} \, \partial_{\phi} + \frac{\tilde{\Sigma}_0}{2 \tilde\Sigma} \, \partial_{\tilde t} \Bigg), \\
\tilde{\vec m} = \frac{1}{\sqrt{2}(r+\ii a \cos\theta)} \Bigg( \partial_{\theta} + \frac{\ii}{\sin\theta} \,\partial_{\phi} + \ii a \sin\theta \, \partial_{\tilde t} \Bigg).
\eea

The next steps are the compactification of the radial coordinate with \eqref{eq:compact_dimensionless} and the reformulation in terms of the parameter $\kappa$ defined in \eqref{eq:Defkappa}. Afterwards, the conformal rescaling of the null tetrad via $\tilde{\vec l} = \Theta\vec l$, $\tilde{\vec n} = \Theta\vec n$, $\tilde{\vec m} = \Theta\vec m$ with the conformal factor \eqref{eq:ConformalFactor} yields
\bea
\label{eq:NullTetrad_ConfCompactDimensionless}
\vec l = \zeta\Bigg(-r  \, \partial_r + \frac{\Sigma_0}{r\Delta} \, \partial_t\Bigg), \quad
 \vec n = \zeta^{-1}\Bigg( r \frac{\Delta}{2 \Sigma} \, \partial_{r} + r \frac{\kappa}{\Sigma} \, \partial_\phi + \frac{\Sigma_0}{2 r\Sigma} \, \partial_t \Bigg), \\
\vec m = \frac{1}{\sqrt{2}(1+\ii \kappa r \cos\theta)} \Bigg(\partial_{\theta} + \frac{\ii}{\sin\theta} \, \partial_{\phi} + \ii\kappa \sin\theta \, \partial_t \Bigg),
\eea
where $\Sigma_0:=\frac{\tilde{r}_+^2}{r^2}\tilde\Sigma_0=1+\kappa^2r^2\equiv\Sigma|_{\theta=0}$.

The final step is to introduce coordinates $(\rho, \theta, \phi, \tau)$ via a coordinate transformation $r=r(\rho,\tau)$ and $t=t(\rho, \tau)$ which: (i) blows up spatial infinity $i^0$ into a cylinder, and (ii) aligns $\vec l$ with $\partial_\tau$. 
The resulting $\rho$ and $\tau$-components of $\vec l$ are
\bea
 l^{\rho} = \frac{1}{d} \left( \dot{t} \, l^r - \dot{r}  \, l^t \right), \quad 
 l^\tau = \frac{1}{d} \left( -t'  \, l^r + r' \, l^t \right), \quad d:=r' \dot{t} - \dot{r} t',
\eea
where $'$ and $\dot{}$ denote $\rho$ and $\tau$-derivatives, respectively.

Objective (i) can be achieved by specifying $r=r(\rho,\tau)$ such that 
\begin{equation}
 \lim_{\rho\rightarrow0} r(\rho,\tau) = \lim_{\tau\rightarrow 1} r(\rho,\tau) = 0.
\end{equation}
The simplest choice is $r=\rho(1-\tau)$ as in \eqref{eq:i0_blowup}. We also impose $l^\rho=0$ and $l^\tau=1$ to achieve objective (ii). The former condition fixes the transformation $t(\rho, \tau)$ via $\dot t = l^t \, \dot r/l^r $. Specifically, the second equation in \eqref{eq:i0_blowup} is recovered by imposing $\tau=0 \leftrightarrow t = 0$ and by identifying 
\begin{equation}\label{eq:F_from_l}
 F = -\frac{l^r}{l^t}. 
\end{equation}
This result, together with condition $l^\tau=1$, fixes the boost parameter to $\zeta = -\dot{r}/r$. 

The null tetrad basis in the coordinates $(\rho, \theta, \phi, \tau)$ obtained in this way is 
\bea
\fl
\label{eq:NullTetrad_ioBlow}
\vec l = \partial_\tau, \quad
 \vec n = \frac{F(\rho)}{\rho} \frac{\Sigma_0}{\Sigma} \partial_{\rho} +  \frac{\kappa \rho(1-\tau)^2}{\Sigma} \, \partial_{\phi} + \frac{(1-\tau)}{2 \Sigma}\left[ \frac{2\Sigma_0 F(\rho)}{\rho^2} - (1-\tau)\Delta \right] \partial_\tau, \\
\fl 
\vec m = \frac{1}{\sqrt{2}(1+\ii \kappa \rho(1-\tau) \cos\theta)} \Bigg( F(\rho) \, \ii \kappa \sin\theta\,\partial_\rho + \partial_{\theta} + \frac{\ii}{\sin\theta} \, \partial_{\phi} + \frac{F(\rho)(1-\tau)}{\rho} \ii\kappa \sin\theta \, \partial_\tau \Bigg). \nn
\eea

While we have already recovered the desired coordinates, we still have one remaining degree of freedom for the null tetrad basis that keeps $\vec l$ aligned with $\partial_\tau$, namely a null rotation
\beq
\vec l \rightarrow \vec l, \quad \vec n\rightarrow \vec n + \alpha \bar\alpha \vec l + \bar\alpha \vec m + \alpha \bar{\vec m}, \quad \vec m\rightarrow \vec m + \alpha \vec l.
\eeq
This can be used to remove the $\tau$-component of $\vec m$ by choosing 
\beq
\alpha= -\frac{\ii \kappa F(\rho)(1-\tau)\sin\theta}{\sqrt{2}\rho\left[1+\ii\kappa\rho(1-\tau)\cos\theta\right]}.
\eeq
The final form of the null tetrad is
\bea\fl
\label{eq:NullTetrad_ioBlow_2}
\vec l = \partial_\tau, \quad
\vec m = \frac{1}{\sqrt{2}(1+\ii \kappa \rho(1-\tau) \cos\theta)} \Bigg( F(\rho) \, \ii \kappa \sin\theta\partial_\rho + \partial_{\theta} + \frac{\ii}{\sin\theta} \, \partial_{\phi} \Bigg). \\
\fl
\vec n =\frac{F(\rho)}{\rho\, \Sigma} \left[\Sigma_0- \kappa^2 (1-\tau)F(\rho)\sin^2(\theta)\right] \partial_{\rho} -  \frac{\kappa (1-\tau)}{\rho\Sigma}\left[F(\rho) -\rho^2(1-\tau)\right] \, \partial_{\phi} \nn \\
+ \frac{1-\tau}{2\rho^2 \Sigma}\left[ 2\Sigma_0 F(\rho) - (1-\tau)\rho^2\Delta - \kappa^2 (1-\tau)F(\rho)^2\sin^2(\theta) \right] \partial_\tau .
\eea

\section{Reviewing the Schwarzschild case}\label{sec:Schwarzschild}

We compare the Schwarzschild limit $\kappa=0$ of the metric and wave equation derived in this paper with the approach that was introduced in~\cite{Friedrich2004} and used for the numerical calculations in~\cite{FrauendienerHennig2017,FrauendienerHennig2018}. Both considerations are based on slightly different coordinates. 

Here, the starting point is the Kerr metric \eqref{eq:Kerr_BL} in Boyer--Lindquist coordinates $(\tilde{r}, \theta, \varphi, \tilde{t})$, which in the non-rotating limit $a\to0$ reduces to the Schwarzschild metric in \emph{Schwarzschild coordinates},
\beq\label{eq:SchwarzschildMetric}
 \tilde g = \left(1-\frac{2M}{\tilde r}\right)^{-1}\dd\tilde r^2+ \tilde{r}^2 \dd\theta^2
             +\tilde r^2 \sin^2\theta\,\dd\varphi^2
            - \left(1-\frac{2M}{\tilde r}\right) \dd\tilde t^{\,2}.
\eeq
On the other hand, the conformal compactification in \cite{Friedrich2004} is based on a representation of the Schwarzschild solution in \emph{isotropic coordinates}.

The series of coordinate transformations~\eqref{eq:phitrans}, \eqref{eq:compact_dimensionless} and \eqref{eq:i0_blowup}, applied to \eqref{eq:SchwarzschildMetric}, results in the conformal metric
\beq
 \fl
  g = \frac{(1-\tau)\left[ 1+ \tau -\rho(1+2\tau-\tau^2)\right]}{\rho^2(1-\rho)^2}\, \dd\rho^2 - \frac{2}{\rho(1-\rho)}\,\dd\rho\,\dd\tau  + \dd\theta^2 
  + \sin^2\theta\,\dd\varphi^2,
\eeq
 which is precisely the limit $\kappa\to0$ of \eqref{eq:confmetric}. Accordingly, the wave equation for an angular mode $\ell$ reads
\bea\label{eq:Eq_Schwarzschild}
 \Big( (1-\tau)\left[1+\tau - \rho(1+2\tau-\tau^2)\right] f_{,\tau} + 2\rho(1-\rho) f_{,\rho}  \Big)_{,\tau}   \\ \nn 
 \quad + \left[\ell(\ell+1) + \rho(1-\tau) \right] f = 0.
\eea
This equation is simpler than the one studied in~\cite{FrauendienerHennig2017,FrauendienerHennig2018} in the sense that \eqref{eq:Eq_Schwarzschild} involves polynomials in $\rho$ and $\tau$ of lower degrees.

We also look at the null tetrad basis \eqref{eq:NullTetrad_ioBlow} in the limit $\kappa\to0$, where we obtain
\bea\label{eq:NullTetrad_ioBlow_Schwarzeschild}
 \vec l = \partial_\tau, \quad
 \vec n = \rho(1-\rho)  \partial_{\rho}  + \frac{1-\tau}{2}\left[ 1+\tau - \rho(1  +2\tau -\tau^2)  \right] \partial_\tau, \\
 \vec m = \frac{1}{\sqrt{2}} \Bigg(\partial_{\theta} +  \frac{\ii}{\sin\theta} \, \partial_{\phi}  \Bigg). \nn
\eea
Next we now introduce a new function $\psi$ adapted to the tetrad vector $\vec n$ via $\psi = 2 n^a \nabla_a f$,  in order to obtain the following first-order reformulation of the wave equation,
\bea
\label{eq:Eq_Schwarzschild_FirstOrderRed_k}
 (1-\tau)\left[1+\tau - \rho(1+2\tau-\tau^2)\right] f_{,\tau} + 2\rho(1-\rho)f_{,\rho} - \psi = 0, \\
 \psi_{,\tau} + \left[\ell(\ell+1) + \rho(1-\tau) \right] f = 0.
\eea

Finally, we derive the relation between our coordinates $(\rho, \theta, \phi, \tau)$ and Friedrich's coordinates $(\bar\rho, \theta, \phi, \bar\tau)$. Using the transformation from isotropic coordinates $(\tilde{\overline{r}}, \theta, \phi, \tilde{\overline t})$ to Schwarzschild coordinates $(\tilde{r}, \theta, \phi, \tilde{t})$,
\beq
 \tilde{r} = \tilde{ \overline{r}} \left(1+\frac{\overline{r}_+}{\tilde{ \overline{r}}}\right)^2, \quad  \overline{r}_+ =\frac{M}{2},
\eeq
we can systematically find relations between the subsequent coordinate transformations. Similarly to Eq.~\eqref{eq:compact_dimensionless}, the first step consists in introducing a compactified radial coordinate and dimensionless coordinates $(\overline{r}, \theta, \phi, \overline{t})$ via
\beq
 \tilde{\overline{r}}=\frac{\overline{r}_+}{\overline{r}}, \quad
 \tilde{\overline{t}}=\overline{r}_+ \overline{r},
 \eeq
which implies the relations
\begin{equation}
 r = \frac{4 \overline{r}}{(1+\overline{r})^2}, \quad
 t=\frac{\overline{t}}{4}.
\end{equation}
In a next step, the blow-up of $i^0$ to the cylinder, equivalently to \eqref{eq:func_t} and \eqref{eq:i0_blowup}, is achieved by the transformation
\beq
\label{eq:i0_blowup_isotropic}
 \overline{r} = \bar{\rho}(1-\bar{\tau}), \quad 
 \overline{t} = \int_{\bar{\tau}}^{\bar{\rho}} \frac{\dd s}{\bar{F}(s)}, \quad 
 \bar{F}(\overline{r}):=\frac{\overline{r}^2(1-\overline{r})}{(1+\overline{r})^3}.
\eeq
Note that $\bar F$ and the function $F$ in the transformation \eqref{eq:i0_blowup} are related by $\displaystyle \bar{F}(\overline{r}) = F(r(\overline{r}))\, \frac{\dd r}{\dd \overline{r}}$, which reflects the definition of this function in terms of the radial component of the outgoing null vector $\vec l$, cf.\ \eqref{eq:F_from_l}. A comparison between the explicit form of the transformation \eqref{eq:i0_blowup_isotropic} --- which corresponds to Eq.~(7) in \cite{FrauendienerHennig2017} --- and \eqref{eq:i0_blowup} finally yields
\beq
 \rho = \frac{4\bar\rho}{(1+\bar\rho)^2}, \quad \tau = \bar\tau \, \frac{1-\bar\rho^2(1-\bar\tau)}{[1+\bar\rho(1-\bar\tau)]^2}.
\eeq

 \section{$2+1$ fully spectral code}
 \label{sec:SpecCode_2+1}
We review the main features of the fully spectral code for hyperbolic equations in $2+1$ dimensions that was developed in~\cite{MacedoAnsorg2014}. We begin with a first-order reduction in time of the unknown functions%
\footnote{
The decomposition \eqref{eq:2+1_FuncDecp} actually introduces the unknown function $f_1(\theta, \tau)$. However, the analysis at $\rho=0$ in Sec.~\ref{sec:BehaviourCylinder} shows that the time evolution can be written as $f_1(\theta, \tau) = f_1^{\ell'}(\tau) P_{\ell'}(\cos\theta)$, i.e.\ there is no mode coupling at this order, and the angular dependence in the time evolution is fixed by the individual mode $\ell'$ of the initial data. For simplicity, we omit the scripts $\ell'$ and $1$ of $f_1^{\ell'}(\tau)$ here.} 
$f(\tau)$ and ${\cal F}(\rho, \theta, \tau)$ by defining
\bea
{\cal E} = \frac{\partial}{\partial\tau} {\cal F}(\rho, \theta, \tau), \quad e = \frac{\partial}{\partial\tau}   f(\tau).
\eea
Then we incorporate the initial data ${\cal F}(\theta,\rho, 0)$, $\dot{\cal F}(\theta,\rho, 0)$, $f(0)$ and $\dot{f}(0)$ into the system by introducing auxiliary fields $ {\cal A}$,  ${\cal B}$, $\alpha$ and $\beta$ via
\bea
{\cal F}(\rho, \theta, \tau) 
 &=& {\cal F}(\rho, \theta, 0)  + \tau \, {\cal A}(\rho, \theta, \tau), \quad 
     f(\tau) = f(0) + \tau \, \alpha(\tau),\\
{\cal E}(\rho, \theta, \tau)  &=& \dot{\cal F}(\rho, \theta, 0) + \tau \, 
{\cal B}(\rho, \theta, \tau) , \quad e(\tau) = \dot{f}(0) + \tau \, \beta(\tau).
\eea
The numerical grid is obtained with a discretisation in terms of \emph{Chebyshev Gauss} collocation points for the time direction $\tau\in[0,1]$, and \emph{Chebyshev Lobatto} points in both spatial directions $\rho\in[0,\rho_{\rm f}]$ and $\theta\in[0,\pi]$. Specifically, for prescribed numbers $N_\rho$, $N_\theta$ and $N_\tau$, the grid points $(\rho_a,\tau_b,\theta_c)$ read
\bea
\rho_{a} = \frac{\rho_{\rm f}}{2}(1 + x_a), \quad &x_a=\cos\left(  \pi\frac{a}{N_\rho}\right), \quad &a = 0,\dots, N_\rho,\nn  \\
\label{eq:CollocationPoints}
\theta_{b} = \arccos\left( x_b \right), \quad &x_b=\cos\left(  \pi\frac{b}{N_\theta}\right), \quad &b = 0,\dots, N_\theta,\\
\tau_{c} =\frac{1}{2}(1 + x_c), \quad &x_c=\cos\left(  \pi\frac{c+\frac12}{N_\tau+1}\right), \quad &c = 0,\dots, N_\tau. \nn
\eea
At this stage, a comparison with the work \cite{MacedoAnsorg2014} is in order, where a \emph{Chebyshev Radau} grid is employed along the time direction. Such strategy is preferred whenever the time direction is not compact, because the discretisation of a given time domain $\tau\in[\tau_{\rm i}, \tau_{\rm f}]$ includes the final time slice $\tau=\tau_{\rm f}$ in the numerical grid. As a consequence, one can use the numerical solution at such a slice as initial data for the subsequent time domain. Here, however, the time direction is compact, with the surface $\tau=1$ representing future null infinity. Since we are not interested in the solution for $\tau>1$, we have the freedom to opt for the Chebyshev Gauss grid in the time direction. This grid does not include the singular surface $\tau=1$ and still retains all the desired properties of a spectral solver. Evaluating the solution at $\tau=1$ is then straightforward, since the expansions \eqref{eq:expansion1}, \eqref{eq:expansion2} below can easily be applied at any $\tau\in[0,1]$.

Following the standard procedure of pseudospectral schemes, we combine the values of the unknown functions ${\cal A}(\rho, \theta, \tau)$, ${\cal B}(\rho, \theta, \tau)$, $\alpha(\tau)$ and $\beta(\tau)$ at all collocation points \eqref{eq:CollocationPoints} to form the vector
\beq
\vec{X} = \Bigg( {\cal A}(\rho_a, \theta_b, \tau_c), \,  {\cal B}(\rho_a, \theta_b, \tau_c), \, \alpha(\tau_c), \, \beta(\tau_c) \Bigg).
\eeq
From this vector, Chebyshev coefficients ${\rm c}^{\cal A}_{i_0 \, i_1 \, i_2}$ and ${\rm c}^{\alpha}_{i_2}$ of the fields ${\cal A}(\theta, \rho, \tau)$ and $\alpha(\tau)$ --- and analogous for ${\cal B}(\theta, \rho, \tau)$ and $\beta(\tau)$ --- can be computed by inverting the equations
\bea\label{eq:expansion1}
{\cal A}(\rho_a, \theta_b, \tau_c) = \sum_{i_0=0}^{N_\rho} \sum_{i_1=0}^{N_\theta} \sum_{i_2=0}^{N_\tau} {\rm c}^{\cal A}_{i_0 \, i_1 \, i_2} T_{i_0}(x_a) T_{i_1}(x_b) T_{i_2}(x_c),\\
\label{eq:expansion2}
\alpha(\tau_c) = \sum_{i_2=0}^{N_\tau}  {\rm c}^{\alpha}_{i_2} T_{i_2}(x_a),
\eea
 where $T_{\ell}(\xi)$ denotes Chebyshev polynomials of the first kind, $T_\ell(\xi)=\cos[\ell \arccos(\xi)]$, $\xi[-1,1]$. For the pseudospectral solution procedure, we compute the approximation of first and second time and spatial derivatives of $\{{\cal A}, {\cal B}, \alpha, \beta\}$ at all grid points \eqref{eq:CollocationPoints} from the vector $\vec{X}$ with the help of the corresponding spectral differentiation matrices. Finally, evaluating the differential equations at the points \eqref{eq:CollocationPoints} yields a linear system of algebraic equations of order 
\beq
n_{\rm total} = 2\left[  1 + (N_\rho+1)(N_\theta+1)   \right] (N_\tau+1)
\eeq
for the entries of the vector $\vec X$. The linear system can straightforwardly be solved with an LU decomposition. However, it is well-known that the LU algorithm scales as $\sim n_{\rm total}^3$. Assuming that $N_\tau \sim N_\rho \sim N_\theta \sim N$, the size of problem grows as $n_{\rm total} \sim N^3$, which implies a prohibitive LU solver scaling as $\sim N^9$.
 
To overcome this difficulty, Ref.~\cite{MacedoAnsorg2014} introduces a BiCGStab-SDIRK scheme, i.e.\ an iterative spectral solver based on the so-called BiCGStab (biconjugate gradient stabilised) method, supplied with a pre-conditioner resulting from a Singly Diagonal Implicit Runge Kutta method (SDIRK). The SDIRK method requires the spectral solution of elliptical equations at each $\tau=$ constant slice. On such slices, we can again resort to an LU decomposition algorithm to invert the matrices representing the $2$D elliptical equations. Thus, each  BiCGStab iteration step scales as $\sim N_\tau \times n_{\rm 2D}^3 \sim N^7$. 
 
Finally, we can introduce another layer of refinement on the  BiCGStab-SDIRK scheme in order to enhance the speed of the elliptic solver at each $\tau=$ constant slice. Using (pseudo)spectral methods for solving elliptic equations is a standard strategy, in particular in GR~\cite{Grandclement:2007sb}. A typical approach also employs the iterative BiCGStab method, supplied with a pre-conditioner based on the finite difference (FD) representation of the derivatives matrices~\cite{Meinel:2008kpy}. Incorporating this algorithm into the fully pseudospectral solver yields the faster BiCGStab-SDIRK/FD solver.  
 
\begin{figure}\centering
 \includegraphics[width=9cm]{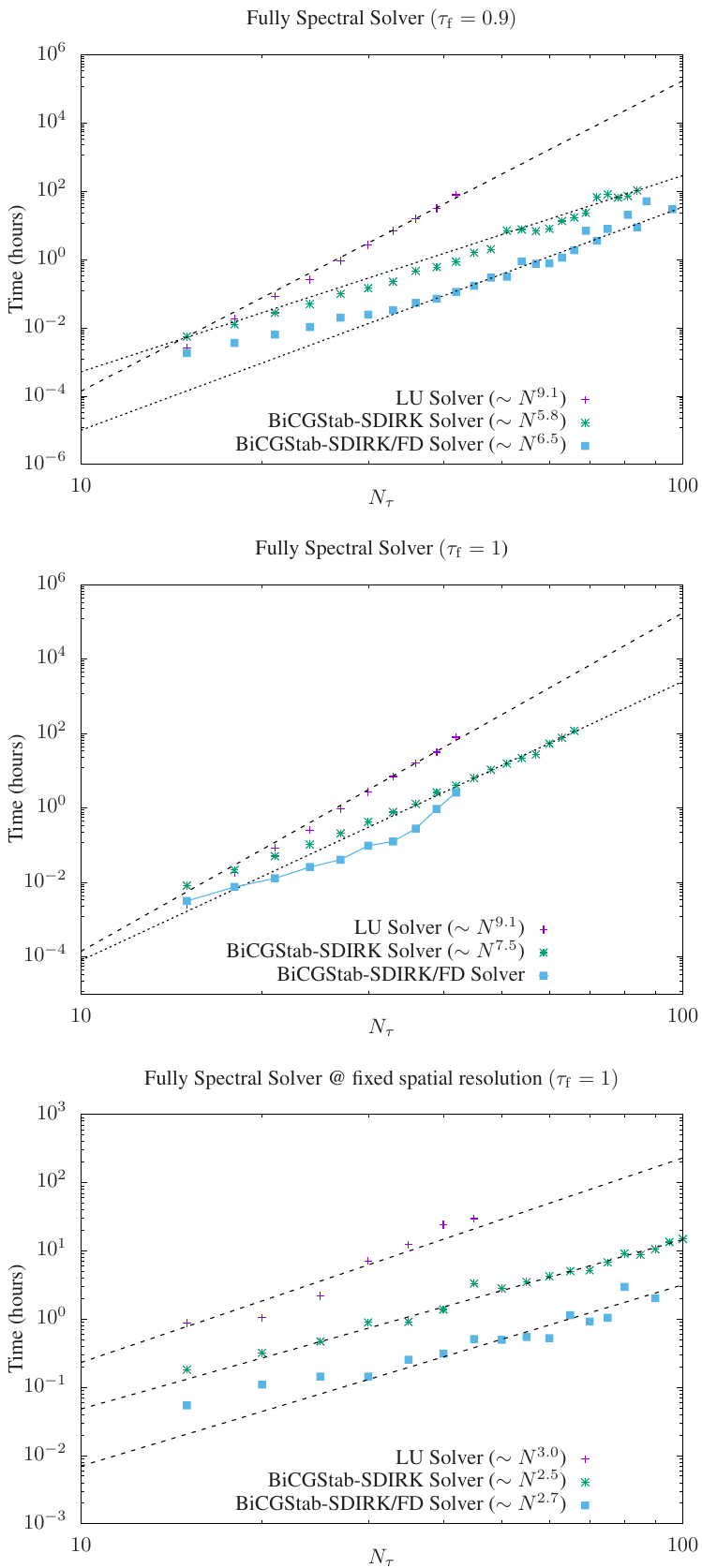}
 \caption{Comparison of the running times of the pseudospectral code for the three different methods for solving the linear equations described in the text.}
 \label{fig:SolverTime}
\end{figure}

The time complexity of the code is illustrated in Fig.~\ref{fig:SolverTime} for the single mode initial data from section~\ref{sec:2+1_SolReg} with $\ell'=3$ and $\kappa=1/2$. The resolutions scale as $N_\rho=2N$, $N_\theta=N$, and $N_\tau=3N$. The top panel shows the results when the equations are solved up to $\tau_{\rm f} = 0.9$, i.e.\ future null infinity at $\tau=1$ is not reached and the solution is regular. The predicted scaling for the LU solver is apparent, confirming that this methods becomes unproductive already for low values of $N$. The figure also shows how the use of the iterative BiCGStab-SDIRK scheme mitigates the problem, with considerable further improvement when the final BiCGStab-SDIRK/FD solver is used.
 
The middle panel shows the corresponding results when the integration includes $\tau=1$. Though prohibitive for moderate $N$, an advantage of the LU method is that it is utterly oblivious to the singular character of the equation at $\tau=1$. On the other hand, the BiCGStab-SDIRK scheme is slightly less efficient when the integration reaches $\tau=1$ because the solver requires a few more iterations to solve the linear system to a given accuracy. Moreover, the BiCGStab-SDIRK/FD becomes unstable at moderate $N$ (here  $N_\rho=30$, $N_\theta=15$, and $N_\tau=45$).

This instability arises, because we are rescaling the resolution in all dimensions at the same time. For our example, the spatial resolution is already saturated at $N_\theta=15$, $N_\rho=30$, i.e.\ any further increase in $N$ does not yield a more accurate solution. On the contrary, it only adds numerical noise to the system. The accumulation of this round-off error for high values of $N$ triggers instabilities in the finite difference preconditioner, which then undermines the scheme's efficiency. Indeed, the bottom panel of Fig.~\ref{fig:SolverTime} shows the time for integrating the equations up to $\tau=1$ against the resolution $N_\tau$ when the spatial resolution is fixed (here $N_\rho=25$, $N_\theta=10$). Once again, the efficiency of each solver becomes evident.

\section*{Acknowledgments}
We thank Juan A. Valiente Kroon for a careful reading of the manuscript.
RPM would like to thank the warm hospitality of the Department of Mathematics and Statistics at the University of Otago. Part of this research was supported by the University of Otago through a Division of Sciences Strategic Seeding Grant. The work was also partially supported by the the European Research Council
Grant No.\ ERC-2014-StG 639022-NewNGR ``New frontiers in numerical general relativity". The project utilised Queen Mary's Apocrita HPC facility, supported by QMUL Research-IT.




\section*{References}

\begin{thebibliography}{99}
 
\bibitem{AcenaKroon2011}
 Ace\~na, A.~E., Valiente Kroon, J.~A.,
 \emph{Conformal extensions for stationary spacetimes},
  Class.\ Quantum Grav.\ {\bf 28}, 225023 (2011)  

\bibitem{Alcubierre2009}
Alcubierre, M.,
\emph{Introduction to $3+1$ numerical relativity},
Cambridge University Press (Cambridge, 2009)

\bibitem{AnsorgHennig2011}
 Ansorg, M.\ and Hennig, J.,
 \emph{The interior of axisymmetric and stationary black holes: numerical and analytical studies},
 J.~Phys.: Conf.\ Ser. {\bf 314}, 012017 (2011)

\bibitem{BeyerDoulis2012}
 Beyer, F., Doulis, G., Frauendiener, J., Whale, B., 
 \emph{Numerical space-times near space-like and null infinity. The spin-2 system on Minkowski space}, 
 Class.\ Quantum Grav.\ {\bf 29}, 245013 (2012)

\bibitem{BeyerDoulis2014}
 Beyer, F., Doulis, G., Frauendiener, J., Whale, B., 
 \emph{The spin-2 equation on Minkowski background}, 
 Springer Proc.\ Math.\ Stat.\ {\bf 60}, 465 (2014) 

\bibitem{BeyerFrauendienerHennig2020}
 Beyer, F., Frauendiener, J., and Hennig, J.,
 \emph{Explorations of the infinite regions of spacetime},
 Int. J. Mod. Phys. D 29, 2030007 (2020)
 
\bibitem{Boyd}
Boyd, J.~P.,
\emph{Chebyshev and Fourier spectral methods},
Dover publications, Mineola, (2001)

\bibitem{Canuto}
Canuto, C., Hussaini, M., Quarteroni, A., Zang, T.: Spectral Methods:
\emph{Fundamentals in Single Domains},
Scientific Computation. Springer Berlin Heidelberg (2007).
 
\bibitem{DoulisFrauendiener2013}
 Doulis, G., Frauendiener, J., 
 \emph{The second order spin-2 system in flat space near space-like and null-infinity}, 
 Gen.\ Relativ.\ Gravit.\ {\bf 45}, 1365 (2013) 
 
\bibitem{Frauendiener2004}
 Frauendiener, J.,
 \emph{Conformal infinity},
 Living Rev.\ Relativity {\bf 7}, 1  (2004)
 
\bibitem{FrauendienerHennig2014}
 Frauendiener, J.\ and Hennig, J.,
 \emph{Fully pseudospectral solution of the conformally invariant wave equation near the cylinder at spacelike infinity},
 Class.\ Quantum Grav.\ {\bf 31}, 085010 (2014)
  
\bibitem{FrauendienerHennig2017}
  Frauendiener, J.\ and Hennig, J.,
  \emph{Fully pseudospectral solution of the conformally invariant wave equation near the cylinder at spacelike infinity. II: Schwarzschild background}, 
  Class.\ Quantum Grav. {\bf 34}, 045005 (2017)

 \bibitem{FrauendienerHennig2018}
  Frauendiener, J.\ and Hennig, J.,
  \emph{Fully pseudospectral solution of the conformally invariant wave equation near the  cylinder at spacelike infinity. III: Nonspherical Schwarzschild waves and singularities at null infinity},
  Class.\ Quantum Grav.\ {\bf 35}, 065015 (2018)
  
\bibitem{Friedrich1998}
 Friedrich, H.,
 \emph{Gravitational fields near space-like and null infinity},
 J.\ Geom.\ Phys.\ {\bf 24}, 83 (1998)
    
\bibitem{Friedrich2004}
  Friedrich, H.,
  \emph{Smoothness at null infinity and the structure of initial data}
  in The Einstein equations and the large scale behavior of gravitational fields,
  edited by Chru\'sciel, P.\ and Friedrich, H.\
  (Springer, Basel, 2004) 
  
  \bibitem{Gray2020}
Gray, F.\ {\em et al.},
\emph{Conformally coupled scalar in rotating black hole spacetimes},
Phys.\ Rev.\ D \textbf{101}, 084031 (2020)
  
 \bibitem{Grandclement:2007sb}
  Grandclement, P.\ and Novak, J.,
  \emph{Spectral methods for numerical relativity},
  Living Rev.\ Rel.\ \textbf{12}, 1 (2009)
  
  \bibitem{HennigAnsorg2009}
  Hennig, J.\ and Ansorg, M.,
  \emph{A fully pseudospectral scheme for solving singular hyperbolic equations on conformally compactified space-times},
  J.~Hyperbolic Differ.\ Equ. {\bf 6}, 161 (2009)
 
 \bibitem{Hennig2013}
  Hennig, J.,
  \emph{Fully pseudospectral time evolution and its application to $1+1$ dimensional physical problems},
  J.~Comput.\ Phys.\ {\bf 235}, 322 (2013)
  
 \bibitem{Kroon2016}
  Kroon, J.~A.~V.,
  \emph{Conformal methods in general relativity},
  Cambridge University Press (Cambridge, 2016) 
   
 \bibitem{Meinel:2008kpy}
 Meinel, R., Ansorg, M., Kleinw\"achter, A., Neugebauer, G., and Petroff, D.,
 \emph{Relativistic Figures of Equilibrium},
 Cambridge University Press (Cambridge 2008)
  
 \bibitem{MacedoAnsorg2014}
 Macedo, R.~P.\ and Ansorg, M.,
 \emph{Axisymmetric fully spectral code for hyperbolic equations},
 J.~Comput.\ Phys.\ {\bf 276}, 357 (2014)
 
\bibitem{MacedoValienteKroon2018}
 Macedo, R.~P.\ and Valiente Kroon, J.\ A.,
 \emph{Spectral methods for the spin-2 equation near the cylinder at spatial infinity},
 Class.\ Quantum Grav.\ {\bf 35}, 125007 (2018)
 
\bibitem{Macedo2020}
 Macedo, R.~P.,
\emph{Hyperboloidal framework for the Kerr spacetime},
Class.\ Quant.\ Grav.\ \textbf{37}, 065019 (2020)

\bibitem{Mena2015}
Mena, F.~C.,
\emph{Cylindrically symmetric models of gravitational collapse to black holes: A short review},
Int.\ J.\ Mod.\ Phys.\ D \textbf{24}, 1542021 (2015)

\bibitem{Paetz:2018nbd}
T.~T.~Paetz,
\emph{On the smoothness of the critical sets of the cylinder at spatial infinity in vacuum spacetimes},
J.\ Math.\ Phys.\ \textbf{59}, 102501 (2018)

\bibitem{Paetz:2019sjd}
T.~T.~Paetz,
\emph{On the choice of a conformal Gauss gauge near the cylinder representing spatial infinity},
J.\ Math.\ Phys.\ \textbf{60}, 072501 (2019)
 
\bibitem{Penrose1963}
 Penrose, R.,
 \emph{Asymptotic properties of fields and space-times},
 Phys.\ Rev.\ Lett.\ {\bf 10}, 66 (1963)

\bibitem{Penrose1964a}
 Penrose, R.,
 \emph{Conformal treatment of infinity},
 in \emph{Relativity, groups and topology}, ed.\ by B.\ deWitt and C.\ deWitt,
 (Gordon and Breach, New York, London, 1964), p.\ 565; 
 republished in Gen.\ Relativ.\ Gravit.\ {\bf 43}, 901 (2011)
 
\bibitem{Penrose1964b}
 Penrose, R.,
 \emph{The light cone at infinity}, in {Relativistic theories of gravitation},
 ed.\ by Infeld, L.\ (Pergamon Press, Oxford, 1964)

\bibitem{Penrose1965}
 Penrose, R.,
 \emph{Zero rest-mass fields including gravitation: asymptotic behaviour},
 Proc.\ Roy.\ Soc.\ London A {\bf 284}, 159 (1965)  
 
\bibitem{Teukolsky:1972my}
Teukolsky, S.~A.,
\emph{Rotating black holes --- separable wave equations for gravitational and electro\-magnetic perturbations},
Phys.\ Rev.\ Lett.\ \textbf{29}, 1114 (1972)
 
\bibitem{Teukolsky:1973ha}
Teukolsky, S.~A.,
\emph {Perturbations of a rotating black hole. 1. Fundamental equations for gravitational electromagnetic and neutrino field perturbations},
Astrophys.\ J.\ \textbf{185}, 635 (1973)

\bibitem{ValienteKroon2007}
 Valiente Kroon, J.~A.,
 \emph{The Maxwell field on the Schwarzschild spacetime: behaviour near spatial infinity},
 Proc.\ Roy.\ Soc.\ Lond.\ A {\bf 463}, 2609 (2007)

\bibitem{ValienteKroon2008}
 Valiente Kroon, J.~A.,
 \emph{Estimates for the Maxwell field near the spatial and null infinity of the Schwarzschild spacetime},
 J.\ Hyp.\ Diff.\ Eqns.\ {\bf 6}, 229 (2009)
%
\end{thebibliography}
\end{document}